\documentclass[pre,aps,amsmath,amssymb,amsfonts,floatfix,superscriptaddress,showpacs,twocolumn,nofootinbib]{revtex4-1}
\usepackage{graphicx}
\usepackage{color}
\usepackage{bbm}
\usepackage[utf8]{inputenc}
\usepackage{dcolumn}
\usepackage{hyperref}
\usepackage{textcomp}
\usepackage[english]{babel}

\hypersetup{colorlinks=true,breaklinks,linkcolor=blue,urlcolor=blue,citecolor=blue}
\usepackage[caption=false]{subfig}

\usepackage{comment}
\usepackage{tikz}
\usetikzlibrary{calc}
\usetikzlibrary{decorations.pathmorphing}
\usetikzlibrary{decorations.pathreplacing}
\usetikzlibrary{decorations.markings}

\tikzset{->-/.style={decoration={
  markings,
  mark=at position #1 with {\arrow{>}}},postaction={decorate}}}
\tikzset{-<-/.style={decoration={
  markings,
  mark=at position #1 with {\arrow{<}}},postaction={decorate}}}

\newcommand{\cpxi}{{\ensuremath{i}}}
\newcommand\Let{\mathrel{\mathop:\!\!=}}
\newcommand\teL{\mathrel{=\!\!\mathop:}}
\newcommand{\iom}{{\ensuremath{\cpxi\omega}}}
\newcommand{\inu}{{\ensuremath{\cpxi\nu}}}
\newcommand{\kv}{\ensuremath{\mathbf{k}}}
\newcommand{\rv}{\ensuremath{\mathbf{r}}}
\newcommand{\KV}{\ensuremath{\mathbf{q}}}
\newcommand{\qv}{\ensuremath{\mathbf{q}}} 

\newcommand{\abs}[1]{\ensuremath{\lvert#1\rvert}}

\newcommand{\av}[1]{\ensuremath{\left\langle #1 \right\rangle}}

\newcommand{\up}{\ensuremath{\uparrow}}
\newcommand{\dn}{\ensuremath{\downarrow}}

\def \Im {\mathop {\rm Im}}

\begin{document}

\title{Beyond extended dynamical mean-field theory: Dual boson approach to the two-dimensional extended Hubbard model}

\author{Erik G. C. P. van Loon}
\affiliation{Radboud University Nijmegen, Institute for Molecules and Materials, NL-6525 AJ Nijmegen, The Netherlands}

\author{Alexander I. Lichtenstein}
\affiliation{I. Institut f\"ur Theoretische Physik, Universit\"at Hamburg, Jungiusstra\ss e 9, D-20355 Hamburg, Germany}

\author{Mikhail I. Katsnelson}
\affiliation{Radboud University Nijmegen, Institute for Molecules and Materials, NL-6525 AJ Nijmegen, The Netherlands}

\author{Olivier Parcollet}
\affiliation{Institut de Physique Th\'eorique (IPhT), CEA, CNRS, 91191 Gif-sur-Yvette, France}

\author{Hartmut Hafermann}
\affiliation{Institut de Physique Th\'eorique (IPhT), CEA, CNRS, 91191 Gif-sur-Yvette, France}

\date{\today}

\begin{abstract}
The dual boson approach [Ann. Phys. {\bf 327}, 1320 (2012)] provides a means to construct a diagrammatic expansion around the extended dynamical mean-field theory (EDMFT). In this paper, we present the numerical implementation of the approach and apply it to the extended Hubbard model with nearest-neighbor interaction $V$. 
We calculate the EDMFT phase diagram and study the effect of diagrams beyond EDMFT on the transition to the charge-ordered phase.
Including diagrammatic corrections to the EDMFT polarization shifts the EDMFT phase boundary to lower values of $V$. The approach interpolates between the random phase approximation in the weak coupling limit and EDMFT for strong coupling.
Neglecting vertex corrections leads to results reminiscent of the EDMFT + $GW$ approximation. We however find significant deviations from the dual boson results already for small values of the interaction, emphasizing the crucial importance of fermion-boson vertex corrections. 
\end{abstract}


\maketitle

\section{Introduction}

The description of correlated electron systems is theoretically challenging. Such systems are  characterized by an intricate interplay between the kinetic energy and the strong Coulomb repulsion. A minimal model to capture this competition is the Hubbard model. It consists of a hopping term, which describes the electron motion, and a local interaction term. For narrow bands, one may expect the intra-atomic matrix elements of the long-range Coulomb interaction to dominate. Hubbard proposed to restrict the interaction to these elements~\cite{Hubbard63}.
In a seminal paper, Anderson conjectured that the model captures the essential features of the high-temperature cuprate superconductors~\cite{Anderson87}.

There may be cases however where nonlocal interaction parameters are sizable. Adatom systems on semiconductor surfaces have been found to exhibit nonlocal interaction parameters with a magnitude reaching as much as 30\% of the on-site Coulomb interaction~\cite{Hansmann13}. Moreover, the nonlocal interaction decays slowly as $1/r$ with distance $r$, as determined by the static dielectric constant of the substrate, rendering even long-range contributions to the interaction important. 
The screening effect of the nonlocal interaction can make a material appear metallic, 
which would be on the verge of the insulating state if only the on-site Coulomb interaction were considered, as observed in graphene~\cite{Wehling11}. For graphene, benzene and silicene, the nonlocal terms were found to reduce the effective local interaction by more than a factor of $2$~\cite{Schuler13}. In metals and semiconductors, the long-range Coulomb interaction leads to plasmons and can induce charge-ordering transitions.

In the extended Hubbard model, nonlocal interaction terms are added to describe such physics, giving rise to the Hamiltonian
\begin{align}
\label{H}
H&=\sum_{ij\sigma}t_{ij} c^{\dagger}_{i\sigma}c_{j\sigma} + \sum_{i}U  n_{i\uparrow}n_{i\downarrow} + \frac{1}{2}\sum_{ij}V_{ij}n_{i}n_{j}.
\end{align}
Here Latin indices denote lattice sites, $\sigma=\uparrow,\downarrow$ label the spin projections and $n_{i\sigma}=c^{\dagger}_{i\sigma}c_{i\sigma}$ and $n_{i}=\sum_{\sigma}n_{i\sigma}$ are  density operators. The Hamiltonian depends on the electron hopping amplitudes $t_{ij}$, the local Hubbard repulsion $U$ and nonlocal interaction parameters $V_{ij}$, respectively~\cite{schubin1934}.

In the absence of a nonlocal Coulomb interaction $V_{ij}=0$, it reduces to the Hubbard model. Different approximations exist to treat this case in the interesting correlated regime. Many of them are based on quantum impurity models (QIMs), which provide a means to sum local contributions in a non-perturbative manner. Dynamical mean-field theory (DMFT)~\cite{Georges96,Kotliar06} maps the problem to a local QIM subject to a self-consistency condition. It has significantly increased our understanding of the Mott transition. Various extensions to DMFT have been developed, which aim to include the effects of spatial correlations neglected in the original approach. Cluster generalizations of DMFT~\cite{Maier05} treat short-range correlations. They have made it possible to address some important aspects of Mott physics, such as the nodal-antinodal dichotomy~\cite{Werner09,Ferrero09,Gull10} and superconductivity (see, e.g., Refs.~\onlinecite{Maier05-2,Gull13}).
A clear advantage of cluster methods is the presence of a control parameter (cluster size). In practice, however, it is not possible to converge the calculations with respect to this parameter in the physically interesting medium to low-temperature or doped regimes. It is therefore important to study diagrammatic extensions of DMFT, which provide a complementary viewpoint. In these methods, long-range dynamical spatial correlations are treated through a combination of numerical and analytical techniques. The work of Kusunose~\cite{Kusunose06}, the dynamical vertex approximation (D$\Gamma $A)~\cite{Toschi07}, the one-particle irreducible (1PI) approach~\cite{Rohringer13} and the dual fermion (DF) method~\cite{Rubtsov08,Rubtsov09} belong to this category.

The extended Hubbard model, on the other hand, is much less studied and fewer methods are available. One can account for screening by deriving reduced effective on-site Coulomb interaction parameters~\cite{Schuler13}. This way, one may use above mentioned methods, but one neglects important effects due to dynamical and nonlocal screening. Cluster extensions of DMFT can treat the nonlocal interaction within the cluster~\cite{Bolech03,Poteryaev04,Tong05,Merino07,Aichhorn07,Hassan10}. In the weakly correlated regime, on the other hand, plasmons and the dielectric screening are described by the  random phase approximation (RPA), while the self-energy can be computed from the screened interaction $W$ in the so-called $GW$-~approximation~\cite{Aryasetiawan98}.

Extended dynamical mean-field theory (EDMFT)~\cite{Si96,Kajueter96,Smith00,Chitra00,Chitra01} provides a means to address the effects of nonlocal Coulomb interaction when correlations are strong. As in DMFT, the lattice problem is mapped to a QIM supplemented with a self-consistency condition. The screening effect of the nonlocal interaction leads to a local retarded interaction which is determined from an additional self-consistency condition on the bosonic bath. The effect of nonlocal corrections has been included by combining EDMFT with the $GW$ approximation~\cite{Sun02}. EDMFT~+~$GW$ has recently been reexamined systematically~\cite{Ayral12,Ayral13} and applied to aforementioned adatom systems on surfaces within a first-principles description~\cite{Hansmann13}.

The dual boson (DB) approach~\cite{Rubtsov12} is a diagrammatic extension of EDMFT which aims to address the fermionic degrees of freedom and the collective bosonic excitations, such as plasmons~\cite{vanLoon14,Hafermann14-2}, on equal footing. In can be applied to correlated lattice fermion models with local and nonlocal interaction.
Strong local correlations are accounted for on the impurity level, while spatial correlations and nonlocal collective excitations are treated diagrammatically. This separation is similar to the DF method.

In this paper, we present an efficient numerical implementation of the DB approach and apply it to the extended Hubbard model with nearest-neighbor Coulomb interaction.
The paper is organized as follows:
In Sec.~\ref{sec:dualboson} we derive the approach for the extended Hubbard model and discuss its formal relation to EDMFT. The computational scheme is discussed in Sec.~\ref{sec:compscheme}, followed by a short summary of the implementation details in Sec.~\ref{sec:impl}.
To set the stage for the discussion of the DB results, we first discuss some numerical results obtained within EDMFT in Sec~\ref{sec:edmftresults}. In particular, we show the phase diagram in the $U$--$V$ plane and discuss the behavior of the self-energy, local susceptibility and the three-leg fermion-boson vertex at some marked points therein. These quantities enter the dual perturbation theory. How the phase diagram is modified through the DB diagrammatic corrections is investigated in Sec.~\ref{dbresults}. In Sec.~\ref{sec:edmftgw} we consider a simplified, computationally less demanding approximation, obtained by systematically neglecting vertex corrections. This allows us to relate the DB approach to EDMFT~+~$GW$ and to elucidate the role of vertex corrections. In Sec.~\ref{sec:summary} we summarize our findings.
A discussion of the technical aspects underlying the formalism and its implementation as well as detailed derivations are provided in the appendixes.

\section{Dual Boson Formalism}
\label{sec:dualboson}

The DB approach was introduced in Ref.~\onlinecite{Rubtsov12}. Its derivation relies on a decoupling of the long-range Coulomb-~interaction via a Hubbard-Stratonovich transformation. Here we provide the derivation of the formalism for the specific case of the extended Hubbard model. Instead of the transformation based on complex fields for the decoupling used in the original work, we employ a decoupling based on real fields. This approach is similar to the derivation of EDMFT in Refs.~\onlinecite{Sun02,Ayral13} and therefore more clearly reveals the relation of these methods.

We seek the solution of the extended Hubbard model \eqref{H}, giving rise to the imaginary-time action
\begin{align}
\label{slat}
S_{\text{latt}}[c^{*},c]=&-\sum_{i\nu\sigma} c^{*}_{i\nu\sigma}[\inu+\mu]c_{i\nu\sigma} + U\sum_{\qv\omega}n_{\qv\omega\uparrow}n_{-\qv,-\omega\downarrow}\notag\\
&+ \sum_{\kv\nu\sigma}\varepsilon_{\kv}c^{*}_{\kv\nu\sigma}c_{\kv\nu\sigma} + \frac{1}{2}\sum_{\KV\omega}V_{\KV}n_{\KV\omega}n_{-\KV-\omega}.
\end{align}
Here $c^{*}$ and $c$ denote Grassmann variables. The Fourier transforms of the hopping amplitudes and nonlocal interaction are denoted by $\varepsilon_{\kv}$ and $V_{\KV}$. The fermionic and bosonic Matsubara frequencies are $\inu_{n}=(2n+1)\pi/\beta$ and $\iom_{m}=2m\pi/\beta$, respectively, where $\beta=1/T$ is the inverse temperature.

In EDMFT, the lattice problem is mapped to a QIM with a hybridization function $\Delta(\tau-\tau')$ and a local retarded interaction $\Lambda(\tau-\tau')$. These functions are determined through self-consistency conditions, which leads to a dynamical mean-field description of the model. In the DB approach,  the QIM serves as the starting point of the perturbation expansion. To achieve this, we replace all sites with QIMs by formally adding and subtracting an arbitrary hybridization and retarded interaction at each lattice site. 
This leaves the original action unaltered and leads to
\begin{align}
\label{slat_rew}
S_{\text{latt}}[c^{*},c]=\sum_{i}S_{\text{imp}}[c_{i}^{*},c_{i}] &- \sum_{\kv\nu\sigma}c^{*}_{\kv\nu\sigma}(\Delta_{\nu\sigma}-\varepsilon_{\kv})c_{\kv\nu\sigma}\notag\\
&-\frac{1}{2}\sum_{\KV\omega\sigma}n_{\KV\omega}(\Lambda_{\omega}-V_{\KV})n_{-\KV-\omega}.
\end{align}
The impurity action $S_{\text{imp}}$ is given by
\begin{align}
\label{simp}
S_{\text{imp}}[c^{*},c]=&-\sum_{\nu\sigma} c^{*}_{\nu\sigma}[\inu+\mu-\Delta_{\nu\sigma}]c_{\nu\sigma}\notag\\
&+ U\sum_{\omega}n_{\omega\uparrow}n_{-\omega\downarrow} + \frac{1}{2}\sum_{\omega}n_{\omega}\Lambda_{\omega} n_{-\omega}.
\end{align}
We now decouple the QIMs by applying suitable Hubbard-Stratonovich transformations to the remainder of Eq.~\eqref{slat_rew}. This is similar to the DF approach. The first term is decoupled through the following identity for Grassmann variables:
\begin{align}
\label{hstfermion1}
&\int \prod_{k}df_{k}^{*}df_{k} e^{-f_{i}^{*}[\alpha^{f}D^{-1}\alpha^{f}]_{ij}f_{j} -c^{*}_{i}\alpha^{f}_{ij}f_{i}  - f^{*}_{i}\alpha^{f}_{ij}c_{i}}\notag\\ 
&= \det [\alpha^{f}D^{-1}\alpha^{f}] e^{c^{*}_{i}D_{ij}c_{j}},
\end{align}
where $\alpha^{f}$ and $D$ denote arbitrary matrices.
It is natural to decouple the density-density interaction term in the charge channel, in particular since we are interested in the charge fluctuations and screening effects induced by $V$. This is achieved through the following transformation based on real fields:
\begin{align}
\label{realhst}
&\int \frac{\prod_{i} d\phi_{i}}{\sqrt{(2\pi)^{N}}} e^{-\frac{1}{2}\phi_{i}[\alpha^{b}W^{-1}\alpha^{b}]_{ij}\phi_{j} \pm \phi_{i}\alpha^{b}_{ij}  n_{i}} \notag\\
&=\sqrt{\det [\alpha^{b} W^{-1}\alpha^{b}]}^{-1} e^{\frac{1}{2}n_{i}W_{ij}n_{j}}.
\end{align}
Here the matrix $W$ is assumed to be positive definite.\footnote{For the case that the matrix $W$ is not positive definite, see Appendix~\ref{app:hst}.} In the above, we choose the negative sign.\footnote{The choice of the sign only affects the sign of the electron-boson vertex and does not affect end results.}
\emph{A priori}, the couplings $\alpha^{f}$ and $\alpha^{b}$ in above equations are arbitrary, but it is important that the coupling is local. This allows us to integrate out the fermionic degrees of freedom locally, leading to a theory in terms of dual variables $f, f^{*}$ and $\phi$ only.
Note that a local coupling preserves the topological structure of the diagrams; diagrams describing processes between nearest-neighbor sites in terms of DFs correspond to the same kind of (i.e., nearest-neighbor) processes in terms of the physical fermions.

The equations take a particularly simple form by letting $\alpha^{f}\to g_{\nu\sigma}^{-1}$ and $\alpha^{b}\to\chi_{\omega}^{-1}$, where the impurity Green's function $g_{\nu\sigma}$ and charge susceptibility $\chi_{\omega}$ are diagonal matrices. They are defined as
\begin{align}
g_{\nu\sigma} &\Let -\av{c_{\nu\sigma} c^{*}_{\nu\sigma}},\\
\label{chidef}
\chi_{\omega} &\Let -\left(\av{n_{\omega}n_{-\omega}}-\av{n}\av{n}\delta_{\omega}\right),
\end{align}
where here and in the following $\av{\ldots}$ denotes the impurity average:
\begin{align}
\av{\dots}\Let\frac{1}{Z_{\text{imp}}}\int \mathcal{D}[c^{*},c] \ldots e^{-S_{\text{imp}}[c^{*},c]}.
\end{align}
Applying the Hubbard-Stratonovich transformations to the partition function $Z=\int\mathcal{D}[c^{*},c]\exp(-S_{\text{latt}}[c^{*},c])$ with $D=\Delta-\epsilon$ and $W=\Lambda - V$ and regrouping terms, we obtain
\begin{align}
\label{partitionfunction}
Z =  \int \mathcal{D}[f^{*},f;\phi] &\int  \mathcal{D}[c^{*},c]  e^{-\sum_{i}S_{\text{site}}[c_{i}^{*},c_{i};f_{i}^{*},f_{i},\phi_{i}]}\notag\\
&\times D_{f} e^{-\sum_{\kv\nu\sigma}f^{*}_{\kv\nu\sigma}g_{\nu\sigma}^{-1}(\Delta_{\nu\sigma}-\varepsilon_{\kv})^{-1}g_{\nu\sigma}^{-1}f_{\kv\nu\sigma}}
\notag\\
&\times D_{b} e^{-\frac{1}{2}\sum_{\KV\omega} \phi_{\KV\omega}\chi_{\omega}^{-1}(\Lambda_{\omega}-V_{\KV})^{-1}\chi_{\omega}^{-1}\phi_{\KV\omega}},
\end{align}
where
\begin{align}
\label{detf}
D_{f}  &= \det[g_{\nu\sigma}(\Delta_{\nu\sigma}-\varepsilon_{\kv})g_{\nu\sigma}],\\
\label{detb}
D_{b}^{-1} &= \sqrt{\det[\chi_{\omega}(\Lambda_{\omega}-V_{\KV})\chi_{\omega}]}
\end{align}
are the determinants arising from the integral transformation. While these are irrelevant for the calculation of expectation values, they are required to establish relations between the dual and the physical fermion propagators.

$S_{\text{site}}$ in Eq.~\eqref{partitionfunction} is the part of the action which is site diagonal,
\begin{align}
\label{s_site}
&S_{\text{site}}[c^{*},c;f^{*},f,\phi]=S_{\text{imp}}[c^{*},c] + S_{\text{cf}}[c^{*},c;f^{*},f,\phi],
\end{align}
where, in turn,
\begin{align}
S_{\text{cf}}[c^{*},c;f^{*},f,\phi] =& \sum_{\nu\sigma} \left(f^{*}_{\nu\sigma}g_{\nu\sigma}^{-1}c_{\nu\sigma}+c^{*}_{\nu\sigma}g_{\nu\sigma}^{-1}f_{\nu\sigma}\right)\notag\\
&+\sum_{\omega}\phi_{\omega}\chi_{\omega}^{-1}n_{\omega}.
\end{align}
In order to arrive at an action which depends on dual variables only, we formally integrate out the original fermionic degrees of freedom. It is possible to do this for each lattice site separately, because the coupling to the dual variables is local. Evaluation of the path integral means taking the average over the impurity degrees of freedom:
\begin{align}
\label{dualdef}
&\frac{1}{Z_{\text{imp}}}\int \mathcal{D}[c^{*},c] e^{-S_{\text{site}}[c^{*},c;f^{*},f,\phi]}
=\av{e^{-S_{\text{cf}}[c^{*},c;f^{*},f,\phi]}}.
\end{align}
We expand the generating functional in the sources $f,f^{*}$, and $\phi$. The result can be written in the following form:
\begin{align}
\label{Vdef}
\ln \av{e^{-S_{\text{cf}}[c^{*},c;f^{*},f,\phi]}}=&-\sum_{\nu\sigma}f_{\nu\sigma}^{*}g^{-1}_{\nu\sigma}f_{\nu\sigma}\notag\\
& - \frac{1}{2}\sum_{\omega}\phi_{\omega}\chi_{\omega}^{-1}\phi_{\omega} - \tilde{V}[f^{*},f;\phi].
\end{align}
This equation defines the dual interaction $\tilde{V}$. Expanding the logarithm yields the \emph{connected} correlation functions of the impurity model, which are coupled to dual variables. The leading terms of $\tilde{V}$ are given by
\begin{align}
\label{V}
&\tilde{V}[f^{*},f;\phi] = - \sum_{\omega}\phi_{\omega}\chi^{-1}_{\omega}\av{n}\delta_{\omega} + \sum_{\nu\omega\sigma}\lambda_{\nu\omega}^{\sigma} f^{*}_{\nu\sigma}f_{\nu+\omega,\sigma}\phi_{\omega}\notag\\
& - \frac{1}{4} \sum_{\nu\nu'\omega}\sum_{\sigma_{i}}\gamma^{\sigma_{1}\sigma_{2}\sigma_{3}\sigma_{4}}_{\nu\nu'\omega}f_{\nu\sigma_{1}}^{*}f_{\nu+\omega,\sigma_{2}}f^{*}_{\nu'+\omega,\sigma_{3}}f_{\nu'\sigma_{4}}.
\end{align}
We neglect higher order terms. The theory thus involves two types of vertex functions: a three-leg fermion-boson vertex $\lambda$ which mediates the coupling between the DFs and (charge) bosons and a four-leg fermion-fermion vertex $\gamma$. These are local and obtained from the impurity model correlation functions in the following manner:
\begin{align}
\label{lambdadef}
\lambda_{\nu\omega}^{\sigma} &\Let \frac{g_{\nu\omega}^{\sigma(3)} - \beta g_{\nu\sigma}\av{n}\delta_{\omega}}{g_{\nu\sigma}g_{\nu+\omega,\sigma}\chi_{\omega}},\\
\label{gammadef}
\gamma_{\nu\nu'\omega}^{\sigma\sigma'} & \Let \frac{g^{(4)\sigma\sigma'}_{\nu\nu'\omega} + \beta g_{\nu\sigma}g_{\nu+\omega\sigma}\delta_{\nu\nu'}\delta_{\sigma\sigma'} - \beta g_{\nu\sigma}g_{\nu'\sigma'}\delta_{\omega}
}{g_{\nu\sigma}g_{\nu+\omega,\sigma}g_{\nu'+\omega\sigma'}g_{\nu'\sigma'}}
\end{align}
(see Appendix~\ref{app:lambda} for details).
Here we use the short-hand notation $\gamma^{\sigma\sigma'}\Let \gamma^{\sigma\sigma\sigma'\sigma'}$. The three- and four-leg vertex functions are obtained from the impurity correlation functions
\begin{align}
\label{chi3def}
g_{\nu\omega}^{(3)\sigma}&\Let -\av{c_{\nu\sigma}c^{*}_{\nu+\omega,\sigma}n_{\omega}},\\
\label{chi4def}
g_{\nu\nu'\omega}^{(4)\sigma\sigma'} & \Let +\av{c_{\nu\sigma}c^{*}_{\nu+\omega,\sigma}c_{\nu'+\omega,\sigma'}c^{*}_{\nu'\sigma'}}.
\end{align}

It is more convenient to work with the original model \eqref{H} written in terms of density fluctuations, i.e. replacing $n_{i}\to \bar{n}_{i}=n_{i}- \av{n_{i}}$ so that $\av{\bar{n}_{i}}=0$. This eliminates the first term in \eqref{V} and is compensated by a shift in the chemical potential. It is further easy to see that the correlation functions \eqref{chidef} and \eqref{lambdadef} remain unchanged, since $\chi_{\omega} \to \bar{\chi}_{\omega}=-\av{\bar{n}_{\omega}\bar{n}_{-\omega}}\equiv\chi_{\omega}$ and similarly $\bar{\lambda}_{\nu\omega}^{\sigma}=\lambda_{\nu\omega}^{\sigma}$.

Combining Eqs.~\eqref{partitionfunction} and \eqref{Vdef}, we obtain the action in dual variables:
\begin{align}
\tilde{S}[f^{*},f;\phi] = &- \sum_{\kv\nu\sigma}f^{*}_{\kv\nu\sigma} \tilde{\mathcal{G}}_{\kv\nu\sigma}^{-1} f_{\kv\nu\sigma} - \frac{1}{2} \sum_{\KV\omega} \phi_{\KV\omega}\tilde{\mathcal{X}}_{\KV\omega}\phi_{\KV\omega}\notag\\ 
&+ \tilde{V}[f^{*},f,\phi].\label{dualaction}
\end{align}
The \emph{bare} dual propagators are denoted by calligraphic symbols. They are given by:
\begin{align}
\label{gdual}
\tilde{\mathcal{G}}_{\kv\nu\sigma} =&\left\{\left[g_{\nu\sigma}^{-1} + (\Delta_{\nu\sigma}-\varepsilon_{\kv})\right]^{-1}\right\} - g_{\nu\sigma},\\
\label{chidual}
\tilde{\mathcal{X}}_{\KV\omega} =& \left\{\left[\chi^{-1}_{\omega} + (\Lambda_{\omega}-V_{\KV})\right]^{-1}\right\}-\chi_{\omega}.
\end{align}
We use the following definitions of the propagators in terms of the fields
\begin{align}
\label{gdualdef}
\tilde{G}_{\kv\nu\sigma} &\Let -\av{f_{\kv\nu\sigma}f_{\kv\nu\sigma}^{*}},\\
\label{xdualdef}
\tilde{X}_{\qv\omega} &\Let -\av{\phi_{\qv\omega}\phi_{-\qv,-\omega}}.
\end{align}
The elements of the dual action are obtained numerically from the solution of the QIM. We have hence achieved a description in which the strong local correlation physics is determined by the QIM, while weaker nonlocal corrections to the (dual) fermionic and bosonic self-energies are obtained through a Feynman-type diagrammatic expansion in the dual interaction $\tilde{V}$. We discuss the perturbation theory in Sec.~\ref{perturbationtheory}. The diagrammatic rules of this expansion are provided in Appendix~\ref{app:feynmanrules}.

\subsection{Relation to EDMFT}
\label{sec:EDMTrel}

In order to establish the connection of the approach to EDMFT, we use relations between the dual and physical lattice propagators. These are obtained by equating the appropriate derivatives of the generating functional before and after introducing the dual particles~\cite{Rubtsov09,Hafermannphd}. The determinants of the transformations, Eqs.~\eqref{detf} and \eqref{detb}, have to be taken into account. 
A difference to the complex-field decoupling of Ref.~\onlinecite{Rubtsov12} is that here a factor $1/2$ appears in the interaction. At the same time, the determinant \eqref{detb} in the real-field decoupling enters with a square-root, so that its derivative also produces a factor $1/2$. As a result, this factor drops out and the result is the same as in the original paper:
\begin{align}
\label{grel}
G_{\kv\nu\sigma} = (\Delta_{\nu\sigma}-\epsilon_{\kv})^{-1} +& (\Delta_{\nu\sigma}-\!\epsilon_{\kv})^{-1}g_{\nu\sigma}^{-1}\tilde{G}_{\kv\nu\sigma}\notag\\
&g_{\nu\sigma}^{-1}(\Delta_{\nu\sigma} - \epsilon_{\kv})^{-1},\\
\label{chirel}
X_{\KV\omega} = (\Lambda_{\omega} - V_{\KV})^{-1} +& (\Lambda_{\omega}\!-\!V_{\KV})^{-1}\chi_{\omega}^{-1}\tilde{X}_{\KV\omega}\notag\\
&\chi_{\omega}^{-1}(\Lambda_{\omega} - V_{\KV})^{-1}.
\end{align}
Inserting the bare dual propagators, Eqs.~\eqref{gdual} and \eqref{chidual}, in place of $\tilde{G}_{\kv\nu\sigma}$ and $\tilde{\chi}_{\KV\omega}$, one obtains the EDMFT lattice Green's functions:
\begin{align}
\label{gedmft}
G^{\text{EDMFT}}_{\kv\nu\sigma} =&\left[g_{\nu\sigma}^{-1} + (\Delta_{\nu\sigma}-\varepsilon_{\kv})\right]^{-1},\\
\label{xedmft}
X^{\text{EDMFT}}_{\KV\omega} =& \left[\chi^{-1}_{\omega} + (\Lambda_{\omega}-V_{\KV})\right]^{-1}.
\end{align}
EDMFT can hence be obtained in a theory of noninteracting DFs and bosons. The same argument can be phrased differently:
Defining the dual fermionic and bosonic self-energies,
\begin{align}
\label{sigmadual}
\tilde{\Sigma}_{\kv\nu\sigma} &= \tilde{\mathcal{G}}_{\kv\nu\sigma}^{-1} - \tilde{G}_{\kv\nu\sigma}^{-1},\\
\label{pidual}
\tilde{\Pi}_{\KV\omega} &= \tilde{\mathcal{X}}_{\KV\omega}^{-1} - \tilde{X}_{\KV\omega}^{-1},
\end{align}
the transformation rules can also be written in the following form~\cite{Rubtsov12}:
\begin{align}
G^{-1}_{\kv\nu} &\!\!=\! (g_{\nu\sigma}+g_{\nu\sigma}\tilde{\Sigma}_{\kv\nu}g_{\nu\sigma})^{-1}+\Delta_{\nu\sigma}\!-\!\epsilon_{\kv},\label{gdtog} \\
X^{-1}_{\KV\omega} &\!\!=\! (\chi_\omega+\chi_{\omega}\tilde{\Pi}_{\KV\omega}\chi_{\omega})^{-1}+\Lambda_{\omega}\!-\!V_{\KV}.\label{xdtox}
\end{align}
We see again that for $\tilde{\Sigma}\equiv 0$ and $\tilde{\Pi}\equiv 0$, these are  identical to the EDMFT propagators. Hence EDMFT emerges as a zero-order approximation in DB. In this sense, the DB approach may be regarded as a diagrammatic extension of EDMFT. 

In the above, we have ignored the fact that an EDMFT solution corresponds to a specific value of the hybridization function and retarded interaction.  \emph{A priori}, these functions are arbitrary in the DB approach. In order to complete the relation to EDMFT, it remains to fix their values by imposing proper self-consistency conditions.
Identifying the terms in braces in Eqs.~\eqref{gdual} and \eqref{chidual} as the EDMFT single- and two-particle propagators, we see that the following conditions are equivalent:
\begin{alignat}{3}
\label{selfcG}
\frac{1}{N}\sum_{\kv}\tilde{\mathcal{G}}_{\kv\nu\sigma} &=  0\quad &\Leftrightarrow&\quad \frac{1}{N}\sum_{\kv} G_{\kv\nu\sigma}^{\text{EDMFT}}&=g_{\nu\sigma},\\
\label{selfcX}
\frac{1}{N}\sum_{\KV} \tilde{\mathcal{X}}_{\KV\omega} &= 0\quad  &\Leftrightarrow&\quad\frac{1}{N}\sum_{\KV}X_{\qv\omega}^{\text{EDMFT}} &= \chi_{\omega}.
\end{alignat}
The conditions on the right-hand side are the EDMFT self-consistency conditions.\footnote{Imposing the self-consistency condition on the two-particle Green's function corresponds to a particular formulation of EDMFT also used in Refs.~\onlinecite{Smith00,Chitra00}. In an alternative formulation, the self-consistency is imposed on the propagator of the bosonic field. The two formulations are equivalent, as noted in Ref.~\onlinecite{Sun02}.}
Therefore, enforcing the conditions on the dual propagators on the left-hand side and neglecting diagrammatic corrections reproduces EDMFT.
When diagrams are taken into account, we impose the analogous conditions on the corresponding renormalized propagators, i.e.,
\begin{align}
\label{eq:gsc}
\frac{1}{N}\sum_{\kv}\tilde{G}_{\kv\nu\sigma} &=  0,\\
\label{eq:xsc}
\frac{1}{N}\sum_{\KV} \tilde{X}_{\KV\omega} &= 0.
\end{align}
How this is achieved in practice is discussed in Appendix~\ref{app:impdetails:scloop}.

\subsection{Perturbation theory}
\label{perturbationtheory}

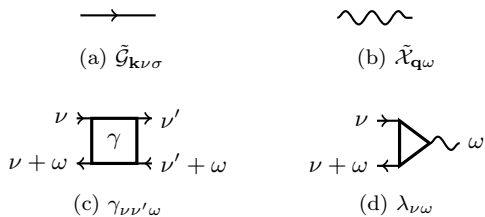
\begin{figure}
\subfloat[$\tilde{\mathcal{G}}_{\kv\nu\sigma}$]{
  \begin{tikzpicture}
    \coordinate (leftfermion) at (0,0) ; 
    \coordinate (rightfermion) at ($(leftfermion)+(1,0)$) ; 
    % Fermion line  
    \draw[thick,->-=0.5] (leftfermion) -- (rightfermion);
    
    % For alignment
    \node [left] at (leftfermion) {\phantom{$\nu+\omega$}} ;
    \node [right] at (rightfermion) {\phantom{$\nu'+\omega$}} ;
    
  \end{tikzpicture}
}\qquad
\subfloat[$\tilde{\mathcal{X}}_{\KV\omega}$]{
  \begin{tikzpicture}
    \coordinate (leftboson) at (0,0) ; 
    \coordinate (rightboson) at ($(leftboson)+(1,0)$) ; 
    % Boson line
    \draw[thick,decorate,decoration=snake] (leftboson) -- (rightboson) ;

    % For alignment
    \node [left] at (leftboson) {\phantom{$\omega$}} ;
    \node [right] at (rightboson) {\phantom{$\nu+\omega$}} ;
    
  \end{tikzpicture}
}\\
\subfloat[$\gamma_{\nu\nu'\omega}$]{
  \begin{tikzpicture}
    \coordinate (fermionnw) at (0,0) ; 

    \coordinate (squarenw) at ($(fermionnw)+(0.2,0)$) ; 
    \coordinate (squarene) at ($(squarenw)+(0.6,0)$) ; 
    \coordinate (squarese) at ($(squarene)+(0,-0.6)$) ; 
    \coordinate (squaresw) at ($(squarenw)+(0,-0.6)$) ; 

    \coordinate (fermionne) at ($(squarene)+(0.2,0)$) ; 
    \coordinate (fermionse) at ($(squarese)+(0.2,0.)$) ; 
    \coordinate (fermionsw) at ($(squaresw)+(-0.2,0)$) ; 

    % gamma
    \draw[very thick] (squarenw) -- (squarene) -- (squarese) -- (squaresw) -- cycle ;
    % Fermion line      
    \draw[thick,-<-=0.5] (fermionsw) -- (squaresw);
    % Fermion line  
    \draw[thick,-<-=0.5] (fermionne) -- (squarene);
    % Fermion line  
    \draw[thick,->-=0.5]  (fermionnw) -- (squarenw);
    % Fermion line  
    \draw[thick,->-=0.5]  (fermionse) -- (squarese);
    \node [left] at ($(fermionsw) $) {$\nu+\omega$} ;
    \node [left] at ($(fermionnw) $) {$\nu$} ;
    \node [right] at ($(fermionse) $) {$\nu'+\omega$} ;
    \node [right] at ($(fermionne) $) {$\nu'$} ;
    
    \coordinate (squarew) at ($(squarenw)!0.5!(squaresw)$) ;
    \coordinate (squaree) at ($(squarene)!0.5!(squarese)$) ;    
    \node at ($(squarew)!0.5!(squaree)$) {$\gamma$} ;
  \end{tikzpicture}
}\qquad
\subfloat[$\lambda_{\nu\omega}$]{
  \begin{tikzpicture}
    \coordinate (diagram1) at (0,0) ;

    \coordinate (leftboson) at ($(diagram1)$ ) ;  
    \coordinate (rightboson) at ($(leftboson)+(0.4,0)$) ;  

    \coordinate (nw) at ($(leftboson) + (-0.4,0.3)$) ;
    \coordinate (sw) at ($(leftboson) + (-0.4,-0.3)$) ;
    \coordinate (ne) at ($(rightboson) + (0.1,0.3)$) ;
    \coordinate (se) at ($(rightboson) + (0.1,-0.3)$) ;

    \coordinate (fermionn) at ($(nw)+(-0.3,0.0)$) ;
    \coordinate (fermions) at ($(sw)+(-0.3,0.0)$) ;
    
    % Fermion line  
    \draw[thick,->-=0.5] (fermionn) -- (nw);
    % Fermion line  
    \draw[thick,-<-=0.5]  (fermions) -- (sw);

    \draw[very thick] (leftboson) -- (nw) -- (sw) -- cycle;
%    \draw[very thick] (rightboson) -- (ne) -- (se) -- cycle;

    \draw[thick,decorate,decoration=snake] (leftboson) -- (rightboson);  

    \node [left] at ($(fermions) $) {$\nu+\omega$} ;
    \node [left] at ($(fermionn) $) {$\nu$} ;
    \node [right] at ($(rightboson) $) {$\omega$} ;
    
  \end{tikzpicture}
}
\caption{The building blocks of the dual perturbation theory and frequency conventions. (Top) The bare DF and boson propagators $\tilde{\mathcal{G}}$ (a) and $\tilde{\mathcal{X}}$ (b). (Bottom) Dual fermions interact via the vertex function $\gamma$. $\lambda$ is the fermion-boson interaction. $\nu$ denotes fermionic frequencies; $\omega$ denotes bosonic Matsubara frequencies.}
\label{fig:feynmanrules}
\end{figure}

\begin{figure}[t]
\subfloat[]{
  \begin{tikzpicture}
  \coordinate (diagram1) at (0,0) ;

\coordinate (leftend) at ($(diagram1)$ ) ;
\coordinate (leftfermion) at ($(leftend) + (-0.3,0)$) ;
\coordinate (toplefttriangle) at ($(leftend) + (0.3,0.4)$) ;
\coordinate (rightlefttriangle) at ($(leftend) + (0.6,0.)$) ;

\coordinate (leftrighttriangle) at ($(rightlefttriangle) + (0.5,0)$) ;
\coordinate (toprighttriangle) at ($(leftrighttriangle) + (0.3,0.4)$) ;
\coordinate (rightend) at ($(leftrighttriangle) + (0.6,0.)$) ;
\coordinate (rightfermion) at ($(rightend) + (0.3,0)$) ;

\draw[very thick] (leftend) -- (toplefttriangle) -- (rightlefttriangle) -- cycle;

\draw[very thick] (rightend) -- (toprighttriangle) -- (leftrighttriangle) -- cycle;

\draw[thick,->-=0.5] (leftfermion) -- (leftend) ;
\draw[thick,-<-=0.5] (rightfermion) -- (rightend) ;

\draw[thick,->-=0.5] (rightlefttriangle) -- (leftrighttriangle);
\draw[thick,decorate,decoration=snake] (toprighttriangle) -- (toplefttriangle);

\node[below] at ($(rightlefttriangle)!0.5!(leftrighttriangle)$) {$\begin{array}{c}\kv+\qv,\\ \nu+\omega\end{array}$} ;
\node[above] at ($(toplefttriangle)!0.5!(toprighttriangle)$) {$\qv, _{\phantom{0}}\omega_{\phantom{0}}$} ;
\end{tikzpicture}
}\qquad\qquad
\subfloat[]{
\begin{tikzpicture}
    \coordinate (diagram1) at (0,0) ;

    \coordinate (leftend) at ($(diagram1)$ ) ;
    \coordinate (leftboson) at ($(leftend) + (-0.3,0)$) ;
    \coordinate (toplefttriangle) at ($(leftend) + (0.4,0.3)$) ;
    \coordinate (bottomlefttriangle) at ($(leftend) + (0.4,-0.3)$) ;

    \coordinate (toprighttriangle) at ($(toplefttriangle) + (0.7,0)$) ;
    \coordinate (bottomrighttriangle) at ($(toprighttriangle) + (0,-0.6)$) ;
    \coordinate (rightend) at ($(toprighttriangle) + (0.4,-0.3)$) ;
    \coordinate (rightboson) at ($(rightend) + (0.3,0)$) ;

    \draw[very thick] (leftend) -- (toplefttriangle) -- (bottomlefttriangle) -- cycle;

    \draw[very thick] (rightend) -- (toprighttriangle) -- (bottomrighttriangle) -- cycle;

    \draw[thick,decorate,decoration=snake] (leftboson) -- (leftend) ;
    \draw[thick,decorate,decoration=snake] (rightboson) -- (rightend) ;

    \draw[thick,-<-=0.5] (bottomlefttriangle) -- (bottomrighttriangle);
    \draw[thick,-<-=0.5] (toprighttriangle) -- (toplefttriangle);

    \node[below] at ($(bottomlefttriangle)!0.5!(bottomrighttriangle)$) {$\begin{array}{c}\kv+\qv,\\ \nu+\omega^{\phantom{0}}\end{array}$} ;
    \node[above] at ($(toplefttriangle)!0.5!(toprighttriangle)$) {$\kv, \nu^{\phantom{0}}$} ;
  \end{tikzpicture}
}  
  \caption{Second-order diagrams contributing to the nonlocal fermionic  (a) and bosonic (b) DF self-energy $\tilde{\Sigma}_{\kv\nu\sigma}$ and $\tilde{\Pi}_{\qv\omega}$, respectively.}
  \label{fig:diagrams}
\end{figure}
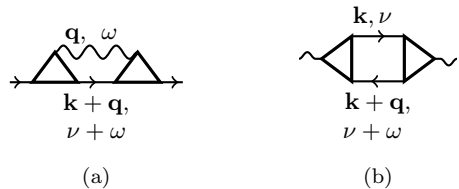

The elements of the perturbation theory are depicted in Fig.~\ref{fig:feynmanrules}. These are the DF and boson propagators and two types of vertices: a (``four-leg'') fermion-fermion vertex and a (``three-leg'') fermion-boson vertex $\lambda$.  The diagrammatic rules for the DB perturbation theory are an extension of those of the DF approach (see Refs.~\onlinecite{Hafermannphd,Hafermann12-2}). We state them explicitly in Appendix~\ref{app:feynmanrules}.

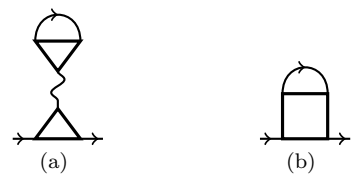
\begin{figure}[b]
\subfloat[]{
\begin{tikzpicture}
  \coordinate (diagram1) at (0,0) ;

\coordinate (leftend) at ($(diagram1)$ ) ;
\coordinate (leftfermion) at ($(leftend) + (-0.3,0)$) ;
\coordinate (downtriangle) at ($(leftend) + (0.3,0.4)$) ;
\coordinate (rightlefttriangle) at ($(leftend) + (0.6,0.)$) ;
\coordinate (rightfermion) at ($(rightlefttriangle) + (0.3,0)$) ;

\coordinate (toptriangle) at ($(downtriangle) + (0.,0.5)$) ;
\coordinate (toprighttriangle) at ($(toptriangle) + (0.3,0.4)$) ;
\coordinate (toplefttriangle) at ($(toptriangle) + (-0.3,0.4)$) ;

\draw[very thick] (leftend) -- (downtriangle) -- (rightlefttriangle) -- cycle;

\draw[very thick] (toptriangle) -- (toprighttriangle) -- (toplefttriangle) -- cycle;

\draw[thick,->-=0.5] (leftfermion) -- (leftend) ;
\draw[thick,-<-=0.5] (rightfermion) -- (rightlefttriangle) ;

\draw[thick,->-=0.5,bend left=90,looseness=2] (toplefttriangle) to (toprighttriangle);
\draw[thick,decorate,decoration=snake] (toptriangle) -- (downtriangle);
\end{tikzpicture}}
\qquad\qquad\qquad\subfloat[]{
\begin{tikzpicture}
    \coordinate (fermionin) at (0,0) ; 

    \coordinate (squaresw) at ($(fermionin)+(0.3,0)$) ; 
    \coordinate (squarenw) at ($(squaresw)+(0,0.6)$) ; 
    \coordinate (squarene) at ($(squarenw)+(0.6,0)$) ; 
    \coordinate (squarese) at ($(squarene)+(0,-0.6)$) ; 
    
    \coordinate (fermionout) at ($(squarese)+(0.3,0)$) ;
    
    % gamma
    \draw[very thick] (squarenw) -- (squarene) -- (squarese) -- (squaresw) -- cycle ;
    % Fermion line  
    \draw[thick,->-=0.5] (fermionin) -- (squaresw);
    \draw[thick,-<-=0.5] (fermionout) -- (squarese);

    \draw[thick,->-=0.5,bend left=90,looseness=2] (squarenw) to (squarene);
\end{tikzpicture}}
\caption{\label{fig:localloop}Second-order diagrams with a local dual fermion loop.}

\end{figure}

The dual self-energy $\tilde{\Sigma}$ is determined by the sum of all topologically distinct diagrams where one external fermion line enters and one exits and which are irreducible with respect to the dual propagators. An example of such a diagram is shown in Fig.~\ref{fig:diagrams} (a). This particular diagram describes the effect of renormalization of the fermionic degrees of freedom due to the bosonic (charge) excitations in the system. The diagram explicitly evaluates to
\begin{align}
\label{sigmaexample}
 \tilde{\Sigma}^{(2)}_{\kv\nu\sigma} &= -\frac{T}{N} \sum_{\KV\omega} \lambda^\sigma_{\nu\omega}\tilde{G}^\sigma_{\kv+\KV\nu+\omega} \tilde{X}_{\KV\omega} \lambda^\sigma_{\nu+\omega,-\omega}.
\end{align}
The DB self-energy $\tilde{\Pi}$ is given by all diagrams irreducible with respect to the propagators and with two external endpoints where bosonic lines can be attached. As an example, the second-order diagram is shown in Fig.~\ref{fig:diagrams} (b) and given by
\begin{align}
\label{piexample}
\tilde{\Pi}^{(2)}_{\KV\omega} &=  \frac{T}{N} \sum_{\kv\nu\sigma} \lambda^\sigma_{\nu+\omega,-\omega} \tilde{G}_{\kv\nu\sigma} \tilde{G}_{\kv+\KV\nu+\omega\sigma}\lambda^\sigma_{\nu\omega}.
\end{align}
Diagrams containing a local DF loop cancel because of the self-consistency condition \eqref{eq:gsc} and the locality of the vertices. Examples are shown in Fig.~\ref{fig:localloop}.

\subsection{Invariance with respect to the decoupling scheme}
\label{sec:invariance}

\emph{A priori}, the division of the interaction into $U$ and $V_{ij}$ in the lattice action \eqref{slat} is arbitrary. 
One may choose to decouple the $V$-term with or without the local interaction included. In the EDMFT~+~$GW$ approach, the former is referred to as the $UV$-decoupling scheme and the latter is the $V$-decoupling scheme.
The two schemes give the same result in EDMFT but different results in EDMFT~+~$GW$~\cite{Ayral13}. Using the identity 
\begin{align}
\label{nid}
n_{i\up} n_{i\dn} = \frac{1}{2} n_{i} n_{i} - \frac{1}{2} n_{i},
\end{align}
we can absorb the $U$- into the $V$-~term with a simultaneous shift of the chemical potential:
\begin{align}
\label{utov}
&U\sum_{i}n_{i\uparrow}n_{i\downarrow} + \frac{1}{2}\sum_{ij}V_{ij}n_{i}n_{j} -\mu\sum_{i}n_{i}\notag\\
&=\frac{1}{2}\sum_{ij}\overline{V}_{ij}n_{i}n_{j} -\overline{\mu}\sum_{i}n_{i}.
\end{align}
Here, $\overline{V}_{ij}=U\delta_{ij}+V_{ij}$ and $\overline{\mu}=\mu+U/2$. 

The DB theory is invariant under the choice of decoupling, as can be seen as follows: The dual action \eqref{dualaction} depends on $V$ only through the propagator $\tilde{\mathcal{X}}$. In addition, $V$ only appears in the combination $(\Lambda-V)$. As a result, all dual quantities are functions of $(\Lambda-V)$. Inspection of the transformation rules \eqref{grel},\eqref{chirel} or \eqref{gdtog},\eqref{xdtox} shows that this is also the case for physical quantities.

If we absorb the local $U$ term into $V$ according to \eqref{utov}, the underlying impurity action becomes
\begin{align}
S_{\text{imp}}[c^{*},c]=&-\sum_{\nu\sigma} c^{*}_{\nu\sigma}[\inu+\mu-\Delta_{\nu\sigma}]c_{\nu\sigma}\notag\\
&+ \frac{1}{2}\sum_{\omega}n_{\omega}\Lambda_{\omega} n_{-\omega} - \frac{U}{2}n_{\omega=0}.
\end{align}
Because the retarded interaction is an auxiliary quantity, we have the freedom to shift it by the same amount, i.e. $\overline{\Lambda}=\Lambda+U$. As a consequence, $\Lambda-V=\overline{\Lambda}-\overline{V}$ stays invariant.
Using the local version of the identity \eqref{nid}, we see that the second line in the equation above becomes
\begin{align}
&\frac{1}{2}\sum_{\omega}n_{\omega}\bar{\Lambda}_{\omega} n_{-\omega}-\frac{U}{2}n_{\omega=0}\notag\\
&=U\sum_{\omega}n_{\omega\uparrow}n_{\omega\downarrow} + \frac{1}{2}\sum_{\omega}n_{\omega}\Lambda_{\omega} n_{-\omega},
\end{align}
so that we recover the impurity action \eqref{simp}.
Hence the impurity model and thus all results of the theory remain invariant.
The deeper reason is of course that the definition of the auxiliary impurity is arbitrary and one may as well choose one which does not contain the local interaction $U$.
That EDMFT is invariant is evident from the fact that it appears as the zero-order approximation in our approach.

\subsection{Two-particle excitations}

There is a direct connection between the bosonic excitations of dual and physical fermions.
Rewriting the dual Dyson equation \eqref{pidual} in the form
\begin{align}
\tilde{X}_{\KV\omega} =  \frac{1}{\tilde{\mathcal{X}}_{\KV\omega}^{-1} - \tilde{\Pi}_{\KV\omega}} = \frac{\tilde{\mathcal{X}}_{\KV\omega}}{1 - \tilde{\Pi}_{\KV\omega}\tilde{\mathcal{X}}_{\KV\omega}},
\end{align}
we see that it represents a geometric series for the dual susceptibility $\tilde{X}$. 
In Appendix~\ref{app:dyson} we show that it diverges at exactly the same point as the physical susceptibility $X$ given by Eq.~\eqref{xdtox}. On the one hand, this is a manifestation of the fact that two-particle excitations are the same for dual and physical fermions. This is also the case in the DF approach~\cite{Hafermann09}. On the other hand, it means that when performing a dual expansion around a solution within the ordered phase, the above series will be summed beyond its convergence radius and the summation will fail.

\section{Computational scheme}
\label{sec:compscheme}

The DB computational scheme is similar to the one of EDMFT. In both schemes, the hybridization function $\Delta_{\nu\sigma}$ and retarded interaction $\Lambda_{\omega}$ are adapted iteratively within a self-consistency loop. In the DB approach, there is an additional step in each iteration where diagrammatic corrections are computed.
The computational scheme is also analogous to the one used in DF calculations~\cite{Hafermannphd}. The only difference is that in addition to the equations for the DF Green's functions, analogous equations for the bosonic Green's functions have to be handled simultaneously.

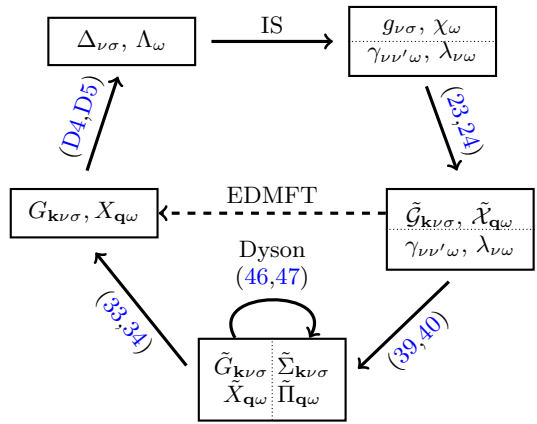
\begin{figure}
\begin{tikzpicture}
[
box/.style = {rectangle,draw=black,thick,inner sep=5pt,minimum size=0mm,align=center,text width=1.6cm},
abovebox/.style = {above}
]
 \coordinate (hybridization) at (0,0) ;
 \coordinate (impurity) at ($(hybridization)+(4,0)$) ;
 \coordinate (dualpt) at ($(impurity)+(0.5,-2.5)$) ;
 \coordinate (dualsigma) at ($(dualpt)+(-2.5,-2.)$) ;
 \coordinate (lattice) at ($(hybridization)+(-0.5,-2.5)$) ;
 
 \node (hybridization_box) at (hybridization) [box] {$\Delta_{\nu\sigma}$, $\Lambda_\omega$} ;
 \node (impurity_box) at (impurity) [box] {$g_{\nu\sigma}$, $\chi_\omega$ $\gamma_{\nu\nu'\omega}$, $\lambda_{\nu\omega}$} ;
 \node (dualpt_box) at (dualpt) [box] {$\tilde{\mathcal{G}}_{\kv\nu\sigma}$, $\tilde{\mathcal{X}}_{\KV\omega}$ $\gamma_{\nu\nu'\omega}$, $\lambda_{\nu\omega}$} ; 
 \node (dualsigma_box) at (dualsigma) [box] {$\tilde{G}_{\kv\nu\sigma}$ $\tilde{\Sigma}_{\kv\nu\sigma}$
 $\tilde{X}_{\KV\omega}$ $\tilde{\Pi}_{\KV\omega}$} ; 
 \node (lattice_box) at ($(dualpt_box.north)+(-5,0)$) [box,anchor=north] {$G_{\kv\nu\sigma}$,\! $X_{\KV\omega}$} ; 

 \node (hybridization_text) at (hybridization_box.north) [abovebox] {} ;
 \node (impurity_text) at (impurity_box.north) [abovebox] {} ;
 \node (dualpt_text) at (dualpt_box.north) [abovebox] {} ; 
 \node (dualsigma_text) at ($(dualsigma_box.south)$) [below] {} ; 
 \node (lattice_text) at (lattice_box.north) [abovebox] {} ; 
 
 \node[above] at ($(hybridization)!0.5!(impurity)$) {IS};
 
 \coordinate (helper1) at ($(lattice_box.north east)!0.5!(lattice_box.south east)$) ;
 \coordinate (helper) at (dualpt_box.west |- helper1) ;
 
 \draw[->,very thick] ($(impurity_box.west)!.9!(hybridization_box.east)$) to ($(impurity_box.west)!.1!(hybridization_box.east)$) ;
 \draw[->,very thick] ($(impurity_box.south)!0.1!(dualpt_text.north)$) to node[midway,sloped,above]{(\ref{gdual},\ref{chidual})} (dualpt_text);
 \draw[->,very thick] ($(dualsigma_box.east)!0.9!(dualpt_box.south)$) to node[midway,sloped,below]{(\ref{sigmaexample},\ref{piexample})} ($(dualsigma_box.east)!0.1!(dualpt_box.south)$) ;
 \draw[->,very thick] ($(lattice_box.south)!0.9!(dualsigma_box.west)$) to node[midway,sloped,below]{(\ref{gdtog},\ref{xdtox})} ($(lattice_box.south)!0.1!(dualsigma_box.west)$) ;
 \draw[->,very thick] (lattice_text) to node[midway,sloped,above]{(\ref{eq:updateG},\ref{eq:updateX})} ($(hybridization_box.south)!0.1!(lattice_text.north)$) ;
 \draw[dashed,->,very thick] (helper) to node [auto,swap]{EDMFT} ($(lattice_box.north east)!0.5!(lattice_box.south east)$) ;
 
 \draw[densely dotted,thin] (dualpt_box.west) to (dualpt_box.east) ;
 \draw[densely dotted,thin] (impurity_box.west) to (impurity_box.east) ;
 \draw[densely dotted,thin] (dualsigma_box.north) to (dualsigma_box.south) ;

 \draw[->,very thick] ($(dualsigma_box.north west)!0.25!(dualsigma_box.north east)$) to [bend left=120, looseness=2] node [auto,align=center]{Dyson\\ (\ref{gdyson},\ref{xdyson})} ($(dualsigma_box.north east)!0.25!(dualsigma_box.north west)$) ;
 
 \end{tikzpicture}
 \caption{
(Color online) Summary of the computational scheme. The loop is started with an initial guess for the hybridization function and retarded interaction. Local observables are calculated in the impurity solver (IS) step. Working with bare dual propagators and neglecting diagrammatic corrections is equivalent to EDMFT (dashed arrow). Corrections beyond EDMFT are taken into account by evaluating diagrams for the dual self-energies, which involve dual propagators and the vertex functions $\lambda$ and $\gamma$. The diagrams are renormalized self-consistently using Dyson equations.
From the dual Green's functions one obtains the physical lattice Green's functions and, in turn, an update for the hybridization and retarded interaction.}
 \label{fig:summary}
\end{figure}

The general computational scheme is depicted in Fig.~\ref{fig:summary} and can be summarized as follows:
\begin{enumerate}
 \item[(1)] Generate an initial guess for $\Delta_{\nu\sigma}$ and $\Lambda_\omega$. 
 \item[(2)] \label{enum:imp} Solve the impurity problem based on $\Delta_{\nu\sigma}$ and $\Lambda_\omega$ and compute $g_{\nu\sigma}$ and $\chi_\omega$ (sufficient for EDMFT) and additionally $\lambda_{\nu\omega}$ and $\gamma_{\nu\nu'\omega}$ for DB calculations.
 \item[(3)] Calculate $\tilde{\mathcal{G}}_{\kv\nu\sigma}$ and $\tilde{\mathcal{X}}_{\qv\omega}$ according to Eqs. \eqref{gdual} and \eqref{chidual}.
 \item[(4)] \label{enum:dual} Evaluate diagrams for $\tilde{\Sigma}_{\kv\nu\sigma}$ and $\tilde{\Pi}_{\qv\omega}$ using dual perturbation theory.
 \item[(5)] \label{enum:innersc} Compute renormalized dual propagators $\tilde{G}_{\kv\nu\sigma}$ and $\tilde{X}_{\qv\omega}$ using the dual Dyson equations (see below). Go back to step (4) and loop until convergence (inner self-consistency).
 \item[(6)] Once the inner loop is converged, calculate the physical propagators $G_{\kv\nu\sigma}$ and $X_{\qv\omega}$ according to Eqs. \eqref{gdtog} and \eqref{xdtox}.
 \item[(7)] Update the hybridization function $\Delta_{\nu\sigma}$ and retarded interaction $\Lambda_\omega$. Go back to step (2) and repeat until convergence is reached (outer self-consistency).
\end{enumerate}

If we skip steps (4) and (5), we work with the bare dual propagators and recover EDMFT as discussed in Sec.~\ref{sec:dualboson}. This is indicated by the dashed arrow in Fig.~\ref{fig:summary}. The vertices need not be calculated in this case.

When diagrammatic corrections are taken into account, we additionally compute the impurity vertex functions in the impurity solver step. This allows us to construct diagrammatic approximations to the dual self-energies $\tilde{\Sigma}_{\nu\kv\sigma}$ and $\tilde{\Pi}_{\omega\KV}$ in dual perturbation theory in step (4). From the self-energies, we then compute renormalized propagators using the Dyson equations
\begin{align}
\label{gdyson}
\tilde{G}_{\kv\nu\sigma}^{-1} &= \tilde{\mathcal{G}}_{\kv\nu\sigma}^{-1} - \tilde{\Sigma}_{\nu\kv\sigma},\\
\label{xdyson}
\tilde{X}_{\qv\omega}^{-1} &= \tilde{\mathcal{X}}_{\qv\omega}^{-1} - \tilde{\Pi}_{\omega\qv}.
\end{align}
The renormalized propagators are subsequently used in the diagrams in going back to step (4). This loop is repeated until convergence. We refer to this as the inner self-consistency loop. It is indicated at the bottom of Fig.~\ref{fig:summary}.

Variations of the full scheme are possible. 
For example, one may skip the outer self-consistency. After the converged EDMFT solution is found, vertex functions and diagrams are evaluated only once. This means that the bath is the same as in EDMFT and that such a scheme truly corresponds to a diagrammatic expansion around EDMFT. 

\section{Implementation}
\label{sec:impl}

The implementation of the DB approach is analogous to that of the DF method. For each equation involving DF propagators, there is a corresponding one for the bosonic propagator. We have therefore implemented the approach by integrating the additional equations into our existing DF implementation. We work with a fully parallelized code.
For the solution of the impurity problem~\eqref{simp} we employ the hybridization 
expansion quantum Monte Carlo method (CT-HYB)~\cite{Werner06}, which can treat a retarded interaction of density-density type without approximation (see, e.g., Ref.~\onlinecite{Ayral13} for details). Here we utilize a modified version of the open source implementation presented in Ref.~\onlinecite{Hafermann13}.  We use improved estimators for the impurity vertex functions and an efficient frequency measurement for the charge susceptibility $\chi_{\omega}$~\cite{Hafermann14}. These improvements reduce the Monte Carlo noise and the overall required computation time.
For the vertices, we use certain symmetry relations which improve the convergence with respect to the frequency cutoff (Appendixes \ref{app:impdetails:symmetries} and \ref{app:symmetries}). The diagrams are efficiently evaluated by exploiting lattice symmetries and employing fast Fourier transforms (FFTs) for the computation of the momentum convolutions on the discrete lattice with periodic boundary conditions. We use a standard size of $64\times 64$.
Individual components of the implementation are discussed in more detail in Appendix~\ref{app:impdetails}.

\section{EDMFT Results}
\label{sec:edmftresults}

Here and in the rest of the paper, we consider the two-dimensional, half-filled, extended Hubbard model described by the Hamiltonian \eqref{H}, with nearest-neighbor hopping $t$ and nearest-neighbor interaction parameter $V$. Hence, we have $t_{ij}=-t$ and $V_{ij}=V$ if $i$ and $j$ are nearest neighbors and $V_{ij}=0$ otherwise.
The dispersion and the Fourier transform of the nonlocal interaction thus read
\begin{align}
\epsilon_{\kv} &= -2t(\cos k_{x} + \cos k_{y}),\\
V_{\qv} &=\phantom{-}2V(\cos q_{x} + \cos q_{y})\label{eq:vq}.
\end{align}
The half-bandwidth $4t=1$ is taken as the energy unit.
We consider the paramagnetic phase only and, hence, omit spin labels in the following.

As we have seen in Sec.~\ref{sec:EDMTrel}, EDMFT is the starting point of the dual perturbation theory. As a basis for the discussion of the DB results, it is therefore instructive to discuss EDMFT results for the extended Hubbard model first.
The EDMFT phase diagram can be computed using the DB code, because it corresponds to a zero-order DB calculation (all diagrammatic corrections are neglected). This also serves as a first test of the implementation.

\subsection{EDMFT Phase diagram}

\begin{figure}[t]
\begin{center}
\includegraphics[scale=0.95,angle=0]{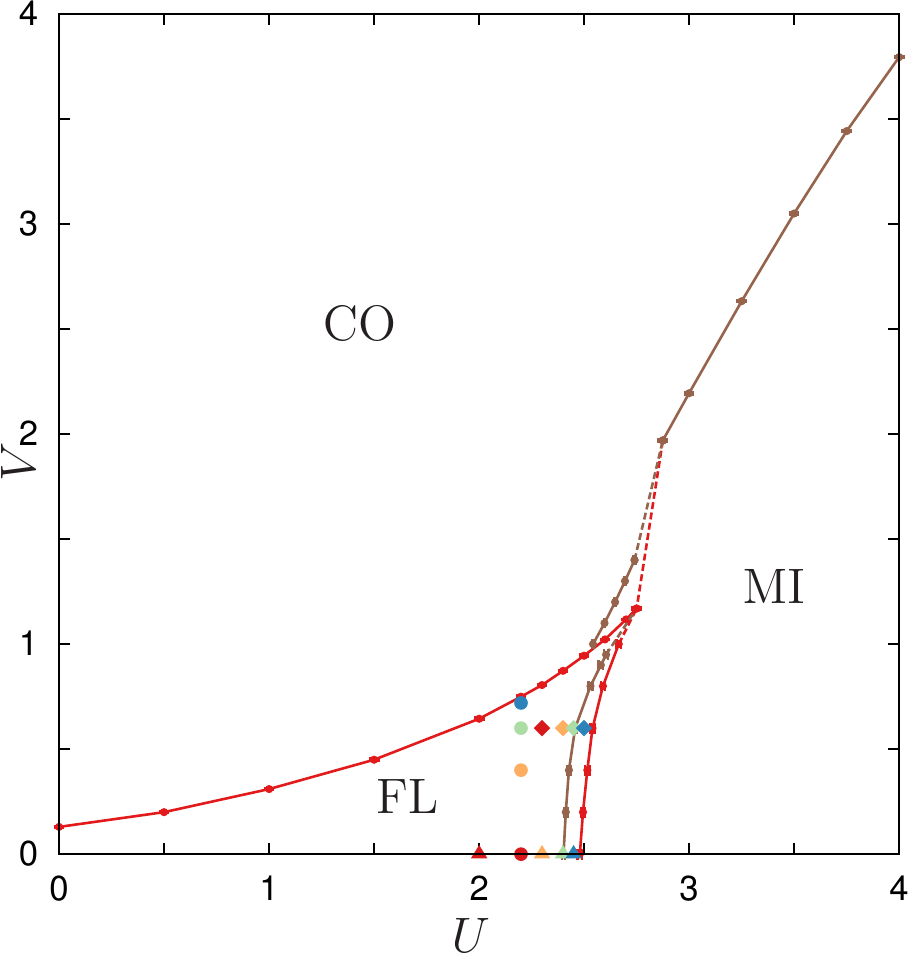} 
\end{center}
\caption{\label{fig:phasediagram} (Color online) EDMFT phase diagram for the extended Hubbard model in the plane of on-site interaction $U$ and nearest-neighbor interaction $V$ at temperature $T=0.01$. Phase boundaries marked in red have been obtained starting from a metallic seed, while the boundaries in brown were obtained starting from an insulating solution as the initial guess. CO, FL and MI denote charge-ordered, Fermi-liquid metallic, and Mott insulating phases, respectively. Colored points indicate positions at which quantities of interest, such as the self-energy, are evaluated.
}
\end{figure}

The EDMFT phase diagram in the $U$-$V$-plane at temperature $T=0.01$ is shown in Fig.~\ref{fig:phasediagram}.
It is compatible with the results of Refs.~\onlinecite{Huang14,Ayral13,Sun02}, showing a Fermi-liquid metal (FL) region for small to moderate values of $U$ and $V$, a charge-ordered (CO) phase with checkerboard order for sufficiently large nearest-neighbor interaction $V$ and small to moderate values of $U$, as well as a Mott insulating (MI) phase for sufficiently large values of the on-site interaction.

\begin{figure}[t]
\begin{center}
\includegraphics[scale=0.675,angle=0]{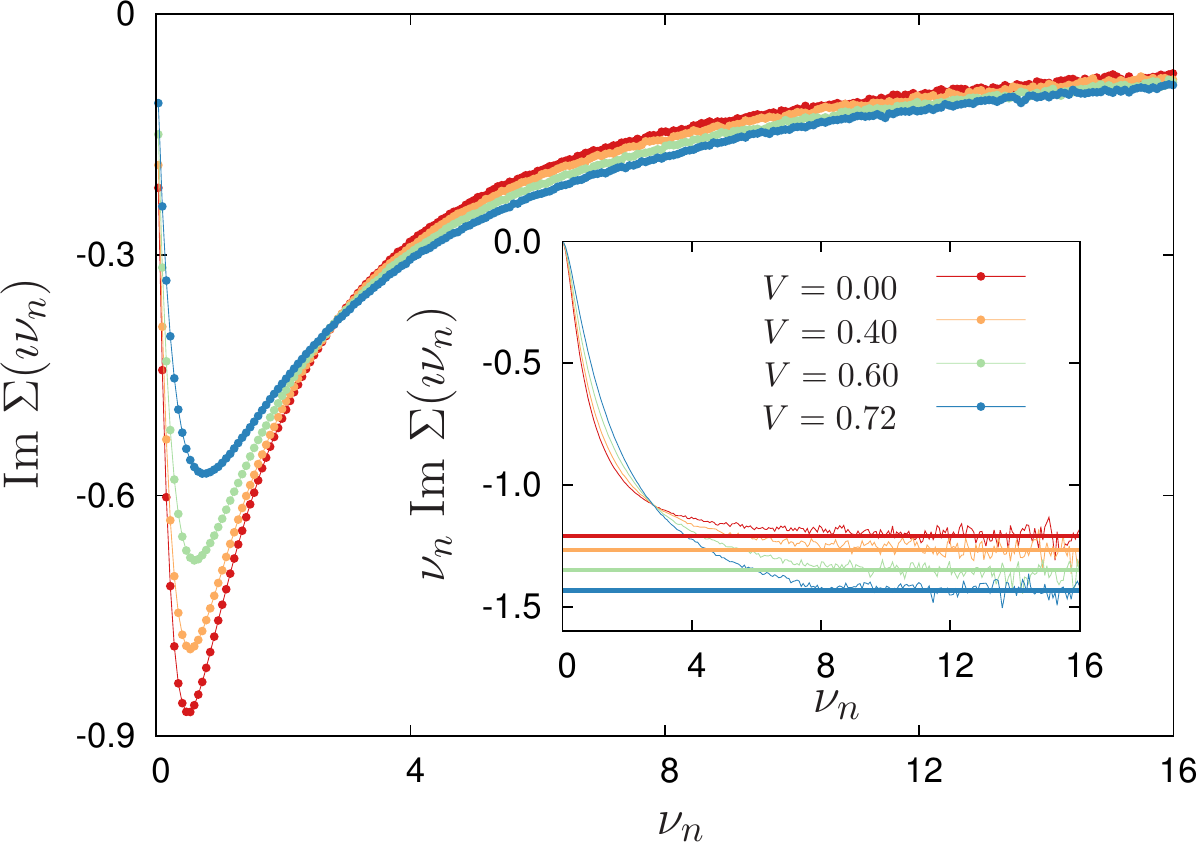} 
\end{center}
\caption{\label{fig:sigma_edmft_vertical} (Color online) EDMFT self-energy for fixed on-site interaction $U=2.2$ and different values of the nearest-neighbor interaction $V$. The different curves correspond to the positions marked by circles in the EDMFT phase diagram in Fig.~\ref{fig:phasediagram}. The inset shows the convergence of the self-energy to its high-frequency behavior. Horizontal lines mark the corresponding calculated values of the first moment.
}
\end{figure}

The checkerboard CO phase is characterized by a divergent charge susceptibility at the wave vector $\KV=(\pi,\pi)$. The phase boundary may therefore be located by looking for zeros of $X^{-1}_{\omega=0,\KV=(\pi,\pi)}$.\footnote{Here we have determined the critical value $V_{c}$ of the interaction from the equivalent condition~\cite{Ayral13} $1+(U-4V_{c})\Pi^{\text{imp}}_{\omega=0}=0$, where $\Pi^{\text{imp}}$ is the impurity polarization.} The phase boundary between the metallic and MI phases was determined from the quantity $(\beta/\pi) G(\beta/2)$, which undergoes a steep drop at the transition to the insulator (see Appendix \ref{sec:specweight}). Phase boundaries marked in red are obtained by approaching the respective phase boundary from the metallic side, while boundaries approached from the insulator are marked in brown.

Since for $V=0$ EDMFT reduces to DMFT, the first-order transition between the metallic and MI phases found in DMFT is reproduced here at $V=0$.
For finite values of $V$, the width of the coexistence region decreases, which is in agreement with the findings of Ref.~\onlinecite{Werner10}.

In the region around and directly above the top corner of the metallic region, fluctuations are strong and it is inherently difficult to obtain a converged solution. Starting with the usual metallic initial guess $\Sigma_{\nu}=0$, the impurity susceptibility is overestimated, which leads to an overestimation of the retarded interaction $\Lambda_\omega$ and to numerical instability. We circumvent this by obtaining a converged insulating solution for $V=0$ (or a smaller $V$) first and using this as an initial guess for calculations at finite $V$. If the initial guess is not sufficiently close to the actual solution, one may encounter the above-mentioned instabilities. The dashed lines indicate that we were not able to find a converged solution. The phase boundary might exhibit a jump here, as mentioned in Ref.~\cite{Huang14}. We find it difficult to make a definite statement though, because of the aforementioned convergence problems. Close to the lower part of the phase boundary separating the CO and MI phases, metastable metallic solutions are found close to the CO phase due to the screening effect of the nonlocal interaction, which, however, converge to insulating ones after a sufficiently large number of DMFT iterations. The top corner of the metallic region hence appears to be completely surrounded by the MI phase.

\subsection{Impurity quantities}

\begin{figure}[t]
\begin{center}
\includegraphics[scale=0.675,angle=0]{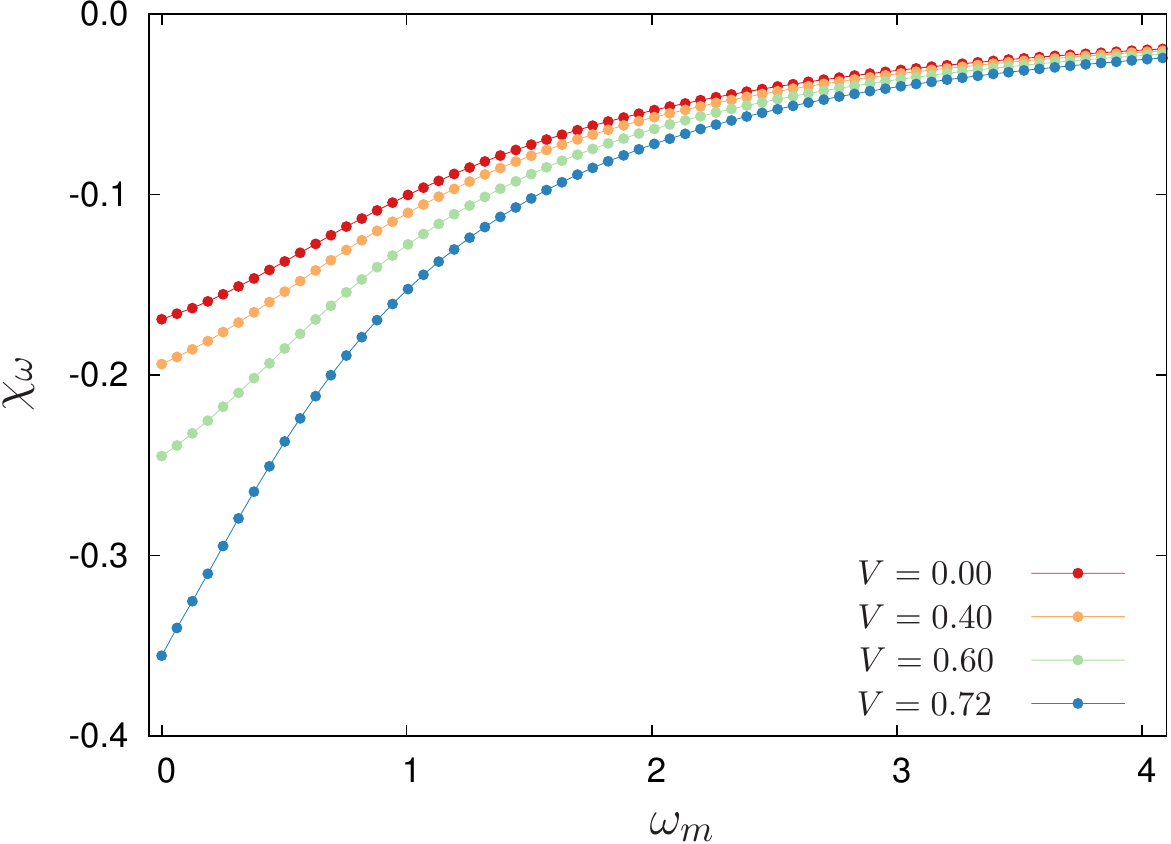} 
\end{center}
\caption{\label{fig:chi_edmft_vertical} (Color online) EDMFT charge susceptibility for the same parameters as in Fig.~\ref{fig:sigma_edmft_vertical}. 
}
\end{figure}

Let us consider some EDMFT impurity quantities, which enter the dual perturbation theory.
We start with the self-energy in Fig.~\ref{fig:sigma_edmft_vertical} for fixed Hubbard interaction $U$ and different values of $V$, corresponding to points marked by circles in the EDMFT phase diagram of Fig.~\ref{fig:phasediagram}. The self-energy exhibits the characteristics of a FL metal. With increasing nearest-neighbor interaction $V$, the solution clearly becomes less correlated. This is also reflected in an increase of the quasiparticle residue as $V$ increases (not shown). It may be an indication of the screening effect through the nearest-neighbor interaction.
As shown in the inset, the high-frequency behavior of the self-energy is also affected by the change in $V$. The first moment can be calculated from the charge susceptibility and retarded interaction (horizontal lines)~\cite{Hafermann14}. It is seen to be enhanced as $V$ increases, in line with an increase of the local charge susceptibility shown in Fig.~\ref{fig:chi_edmft_vertical}. Since in EDMFT, the susceptibility is related to the polarization through $\Pi_{\omega}^{\text{imp}} = -\chi_{\omega}/(1+\Lambda_{\omega}\chi_{\omega})$, this enhancement indicates the increased effect of screening as $V$ increases~\cite{Ayral13}.

\begin{figure}[t]
\begin{center}
\includegraphics[scale=0.675,angle=0]{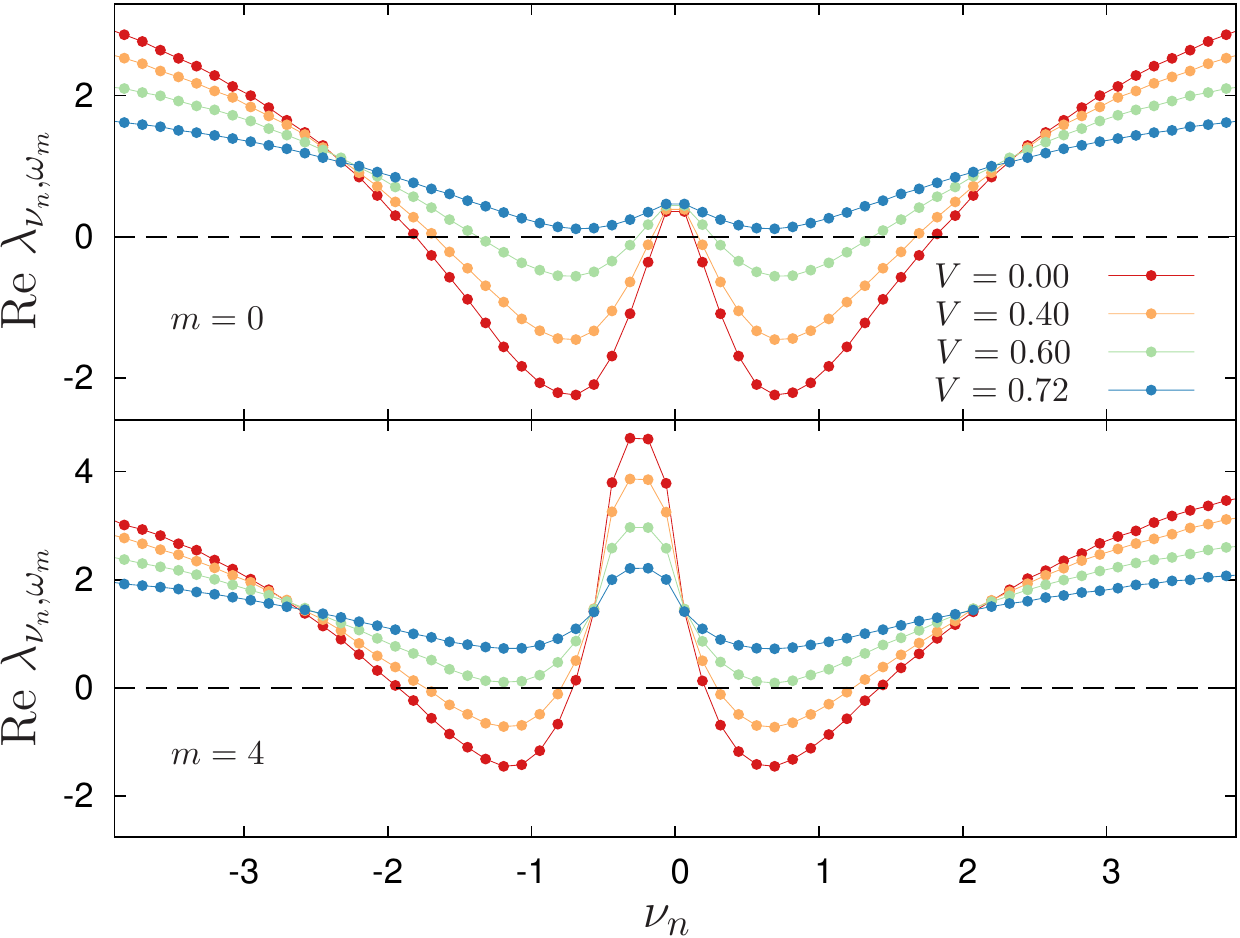} 
\end{center}
\caption{\label{fig:lambda_edmft_vertical} (Color online) Three-leg vertex in EDMFT for two bosonic frequencies $\omega_{m}=2m\pi/\beta$ with $m=0$ and $m=4$ and different values of the nearest-neighbor interaction $V$. The parameters are the same as in Fig.~\ref{fig:sigma_edmft_vertical}. With increasing $V$, the transition to the CO phase is approached and the vertex becomes flatter. Note the change of sign.
}
\end{figure}

\begin{figure}[t]
\begin{center}
\includegraphics[scale=0.675,angle=0]{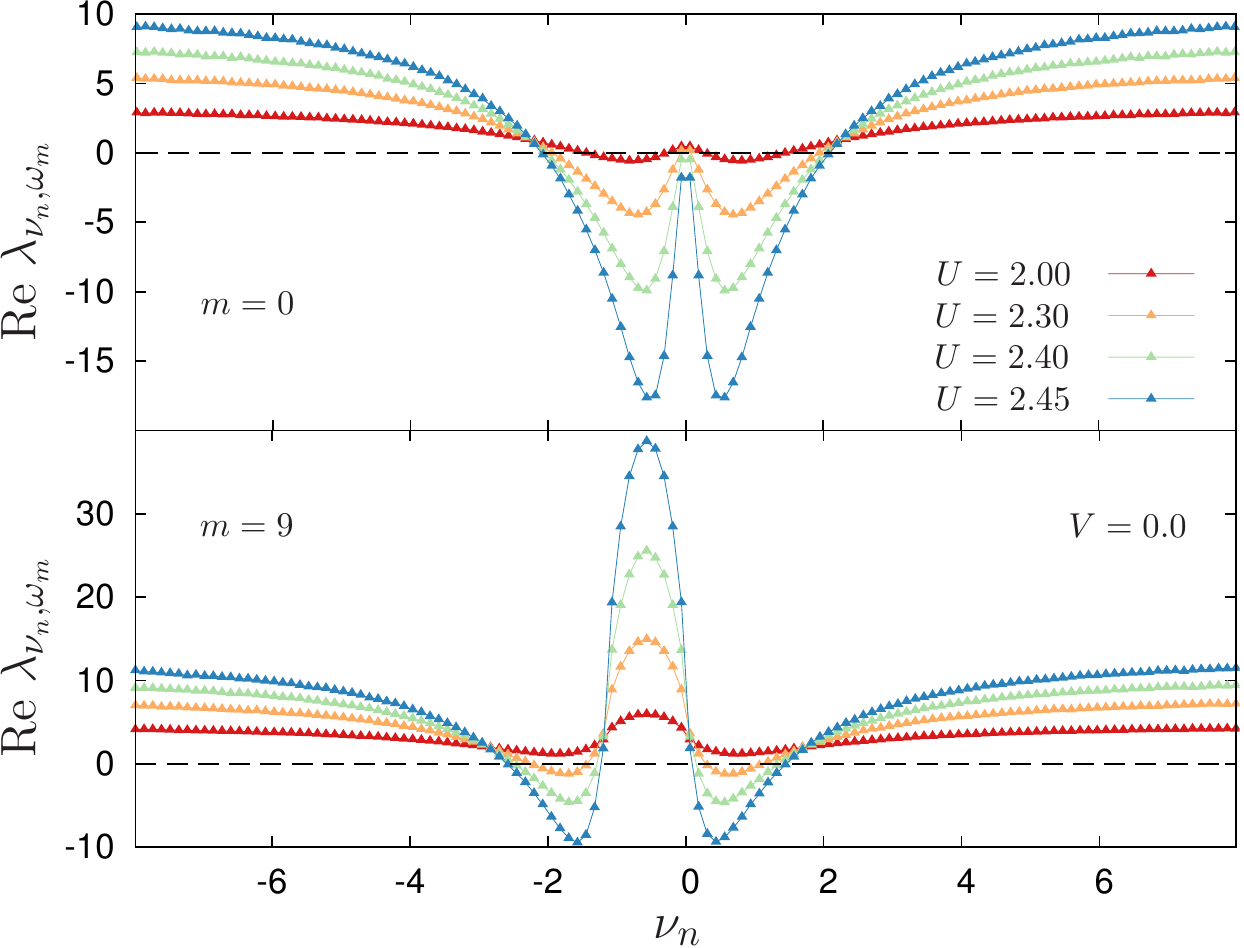} 
\end{center}
\caption{\label{fig:lambda_edmft_v00} (Color online) Three-leg vertex for two different bosonic frequencies as a function of fermionic frequency. The nearest-neighbor interaction is kept fixed at $V=0$ corresponding to a DMFT calculation. 
The corresponding points are marked by triangles in the phase diagram of Fig.~\ref{fig:phasediagram}. With increasing values of $U$, the Mott transition is approached from the metallic side.
}
\end{figure}

\begin{figure}[t]
\begin{center}
\includegraphics[scale=0.675,angle=0]{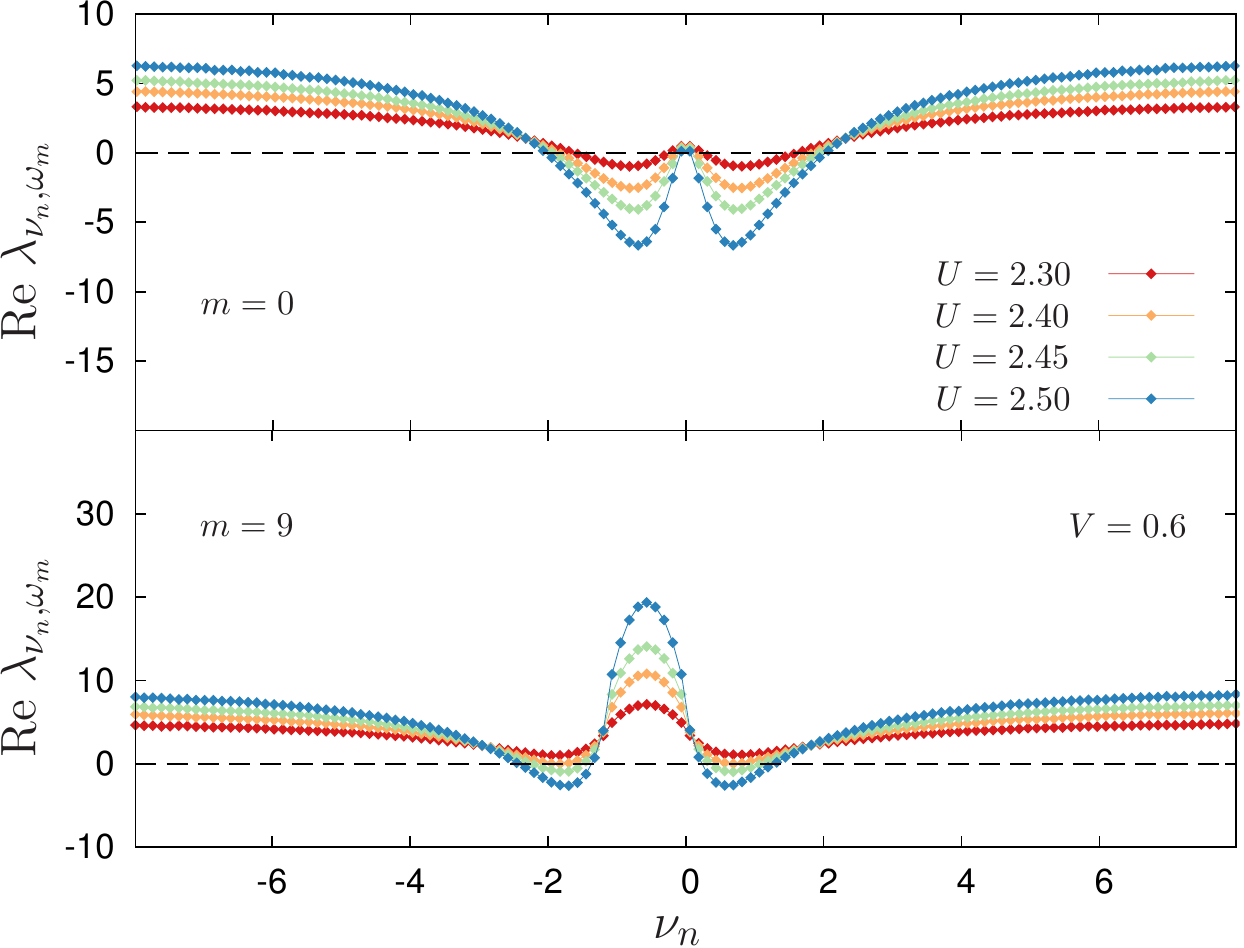} 
\end{center}
\caption{\label{fig:lambda_edmft_v06} (Color online) Three-leg vertex for different bosonic frequencies as a function of the fermionic frequency. The nearest-neighbor interaction is fixed at $V=0.6$. Parameters are otherwise the same as in Fig.~\ref{fig:lambda_edmft_v00}. The corresponding points are marked by diamonds in Fig.~\ref{fig:phasediagram}.
}
\end{figure}

In Fig.~\ref{fig:lambda_edmft_vertical} we plot the three-leg vertex at the same parameters as in the previous figures. It mediates the electron-boson interaction in the DB approach. Because of the particle-hole symmetry, it is purely real. We see that it exhibits less structure as the metallicity of the system is increased and becomes mostly flat as the phase boundary to the CO state is approached. Note that it changes sign, except very close to the phase boundary. We see later that this structure has an important effect on the DB results.
Interestingly, the vertex appears to be unaffected by the interaction at specific points, so that the curves cross. We have no explanation for this observation at the moment. The symmetry of the vertex under the transformation $\nu\rightarrow -\nu-\omega$ is clearly visible. We exploit this symmetry in the DB calculations (see Appendixes \ref{app:impdetails:symmetries} and \ref{app:symmetries}).
In Figs.~\ref{fig:lambda_edmft_v00} and~\ref{fig:lambda_edmft_v06} we show the behavior of the three-leg vertex when the Mott transition is approached. We plot it for different values of $U$ marked by triangles in the phase diagram in Fig.~\ref{fig:phasediagram}. Figure~\ref{fig:lambda_edmft_v00} is for fixed $V=0$ and corresponds to DMFT. As $U$ increases, the vertex develops structure and grows significantly in magnitude. This behavior reflects the enhancement of the fermion-fermion vertex when the transition is approached, according to Eq.~\eqref{lambdafromgamma}. The vertices diverge at the transition in the zero-temperature limit.
Figure~\ref{fig:lambda_edmft_v06} shows the vertex on the same scale as in the previous figure, albeit for $V=0.6$. 
As the Mott transition is approached, the magnitude increases, but the vertex shows less structure and is smaller in magnitude because the system is less correlated compared to the case $V=0$.

\section{Dual boson results}
\label{dbresults}

The purpose of this paper is to obtain a first understanding of the effects caused by different types of diagrams in the DB approach. We also aim to get a better understanding of EDMFT  and EDMFT~+~$GW$.
We therefore employ different diagrammatic approximations and analyze their physical content.

We mainly restrict ourselves to the following type of calculations: We start from a converged EDMFT solution, compute the vertices once and take into account diagrammatic corrections. In other words, the hybridization and retarded interaction have the same values as in EDMFT. The results can be interpreted in terms of a diagrammatic extension of EDMFT. This corresponds to DB calculations \emph{without} the outer self-consistency loop and is computationally significantly less expensive than the full scheme. We  nevertheless discuss the effect of the full self-consistency scheme at some selected points of the phase diagram.
The inner self-consistency loop is always iterated until convergence, corresponding to a self-consistent renormalization of self-energy diagrams and Green's functions.

In a first step, we examine the effect of polarization corrections only: We neglect diagrams to the fermionic self-energy and only include corrections to the EDMFT polarization via bosonic self-energy diagrams. In a second step, we additionally  consider the effect of fermionic self-energy diagrams.
Fermionic diagrams which explicitly contain the fermion-fermion vertex will not be considered. These diagrams also appear in the framework of the DF approach and their effect has been studied previously for the Hubbard model ($V=0$). For example, it is known that the second-order approximation includes dynamical short-range correlations which lead to a reduction of the critical $U$ of the Mott transition~\cite{Hafermannphd}. Similar effects can be expected for finite $V$. This renders a comparison with EDMFT or EDMFT~+~$GW$ more difficult. We therefore leave the study of more complete approximations for future work.

\subsection{Polarization corrections}

\subsubsection{Diagrams}

In terms of the charge susceptibility, the physical polarization $\Pi$ is defined through
\begin{align}
\label{Pidef}
X_{\qv\omega}=\frac{1}{-\Pi_{\qv\omega}^{-1}-(U+V_{\qv})}.
\end{align}
It should not be confused with the bosonic self-energy $\tilde{\Pi}$, which has different dimension.
Comparing with Eq.~\eqref{xedmft}, we see that the EDMFT polarization is independent of momentum:
\begin{align}
\label{eq:piedmft}
\Pi_{\omega}^{-1}=-\chi_{\omega}^{-1}-\Lambda_{\omega}.
\end{align}
On the other hand, the polarization in the DB approach, expressed in terms of the bosonic self-energy, becomes [cf. Eq.~\eqref{xdtox}]:
\begin{align}
\Pi_{\qv\omega}^{-1} = - (\chi_\omega+\chi_{\omega}\tilde{\Pi}_{\KV\omega}\chi_{\omega})^{-1}-\Lambda_{\omega}.\label{eq:pifromdual}
\end{align}
In these equations, $\Lambda_{\omega}$ contains the static $U$ (cf. Sec.~\ref{sec:invariance}).
We see that the momentum dependence introduced through diagrammatic corrections to the bosonic self-energy directly translates to a momentum dependence of the polarization. If the dual polarization is neglected, Eq.~\eqref{eq:pifromdual} evidently reduces to the EDMFT polarization.

\begin{figure}[t]
\subfloat[Second order]{
  \begin{tikzpicture}
    \coordinate (diagram1) at (0,0) ;

    \coordinate (leftend) at ($(diagram1)$ ) ;
    \coordinate (leftboson) at ($(leftend) + (-0.3,0)$) ;
    \coordinate (toplefttriangle) at ($(leftend) + (0.4,0.3)$) ;
    \coordinate (bottomlefttriangle) at ($(leftend) + (0.4,-0.3)$) ;

    \coordinate (toprighttriangle) at ($(toplefttriangle) + (0.7,0)$) ;
    \coordinate (bottomrighttriangle) at ($(toprighttriangle) + (0,-0.6)$) ;
    \coordinate (rightend) at ($(toprighttriangle) + (0.4,-0.3)$) ;
    \coordinate (rightboson) at ($(rightend) + (0.3,0)$) ;

    \draw[very thick] (leftend) -- (toplefttriangle) -- (bottomlefttriangle) -- cycle;

    \draw[very thick] (rightend) -- (toprighttriangle) -- (bottomrighttriangle) -- cycle;

    \draw[thick,decorate,decoration=snake] (leftboson) -- (leftend) ;
    \draw[thick,decorate,decoration=snake] (rightboson) -- (rightend) ;

    \draw[thick,-<-=0.5] (bottomlefttriangle) -- (bottomrighttriangle);
    \draw[thick,-<-=0.5] (toprighttriangle) -- (toplefttriangle);

    \node[below] at ($(bottomlefttriangle)!0.5!(bottomrighttriangle)$) {$\begin{array}{c}
    \kv+\qv\\\nu+\omega^{\phantom{0}}
    \end{array}$} ;
    \node[above] at ($(toplefttriangle)!0.5!(toprighttriangle)$) {$\begin{array}{c}
    \kv, \nu^{\phantom{0}}
    \end{array}$} ;
  \end{tikzpicture}
}\qquad\qquad
\subfloat[Ladder]{
  \begin{tikzpicture}
    \coordinate (diagram1) at (0,0) ;

    \coordinate (leftend) at ($(diagram1)$ ) ;
    \coordinate (leftboson) at ($(leftend) + (-0.3,0)$) ;
    \coordinate (toplefttriangle) at ($(leftend) + (0.4,0.3)$) ;
    \coordinate (bottomlefttriangle) at ($(leftend) + (0.4,-0.3)$) ;

    \coordinate (toprighttriangle) at ($(toplefttriangle) + (0.7,0)$) ;
    \coordinate (bottomrighttriangle) at ($(toprighttriangle) + (0,-0.6)$) ;
    \coordinate (rightend) at ($(toprighttriangle) + (0.4,-0.3)$) ;
    \coordinate (rightboson) at ($(rightend) + (0.3,0)$) ;

    \draw[very thick] (leftend) -- (toplefttriangle) -- (bottomlefttriangle) -- cycle;

    \draw[very thick] (rightend) -- (toprighttriangle) -- (bottomrighttriangle) -- cycle;

    \draw[thick,decorate,decoration=snake] (leftboson) -- (leftend) ;
    \draw[thick,decorate,decoration=snake] (rightboson) -- (rightend) ;

    \draw[thick,-<-=0.5] (bottomlefttriangle) -- (bottomrighttriangle);
    \draw[thick,-<-=0.5] (toprighttriangle) -- (toplefttriangle);

    \draw[very thick,fill=gray!50] (rightend) -- (toprighttriangle) -- (bottomrighttriangle) -- cycle;

    \node[below] at ($(bottomlefttriangle)!0.5!(bottomrighttriangle)$) {$\begin{array}{c}\kv+\qv\\
    \nu+\omega^{\phantom{0}}
    \end{array}$} ;
    \node[above] at ($(toplefttriangle)!0.5!(toprighttriangle)$) {$\begin{array}{c}
    \kv, \nu^{\phantom{0}}
    \end{array}$} ;
  \end{tikzpicture}
}
 \caption{ \label{fig:pi2_ladder} Diagrammatic approximations to the bosonic self-energy $\tilde{\Pi}_{\KV\omega}$. }
\end{figure}
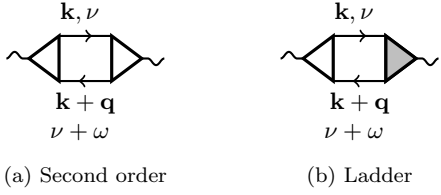

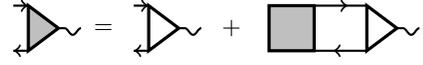
\begin{figure}[t]
  \begin{tikzpicture}

% Lambda  
    \coordinate (toprighttriangle) at ($(-3.2,0)$) ;
    \coordinate (bottomrighttriangle) at ($(toprighttriangle) + (0,-0.6)$) ;
    \coordinate (rightend) at ($(toprighttriangle) + (0.4,-0.3)$) ;
    \coordinate (rightboson) at ($(rightend) + (0.3,0)$) ;

    \draw[very thick,fill=gray!50] (rightend) -- (toprighttriangle) -- (bottomrighttriangle) -- cycle;

    \draw[thick,decorate,decoration=snake] (rightboson) -- (rightend) ;

    \draw[thick,-<-=0.5] ($(bottomrighttriangle)+(-0.2,0)$) -- (bottomrighttriangle);
    \draw[thick,-<-=0.5] (toprighttriangle) -- ($(toprighttriangle)+(-0.2,0)$);
    
% lambda  
    \coordinate (toprighttriangle) at ($(-1.6,0)$) ;
    \coordinate (bottomrighttriangle) at ($(toprighttriangle) + (0,-0.6)$) ;
    \coordinate (rightend) at ($(toprighttriangle) + (0.4,-0.3)$) ;
    \coordinate (rightboson) at ($(rightend) + (0.3,0)$) ;

    \draw[very thick] (rightend) -- (toprighttriangle) -- (bottomrighttriangle) -- cycle;

    \draw[thick,decorate,decoration=snake] (rightboson) -- (rightend) ;

    \draw[thick,-<-=0.5] ($(bottomrighttriangle)+(-0.2,0)$) -- (bottomrighttriangle);
    \draw[thick,-<-=0.5] (toprighttriangle) -- ($(toprighttriangle)+(-0.2,0)$);
  
% gamma lambda  
    \coordinate (squarenw) at ($(0,0)$) ;
    \coordinate (squaresw) at ($(squarenw) + (0,-0.6)$) ;
    \coordinate (squarene) at ($(squarenw) + (0.6,0)$) ;
    \coordinate (squarese) at ($(squaresw) + (0.6,0)$) ;

    \coordinate (toprighttriangle) at ($(squarene) + (0.7,0)$) ;
    \coordinate (bottomrighttriangle) at ($(toprighttriangle) + (0,-0.6)$) ;
    \coordinate (rightend) at ($(toprighttriangle) + (0.4,-0.3)$) ;
    \coordinate (rightboson) at ($(rightend) + (0.3,0)$) ;

    \draw[very thick] (rightend) -- (toprighttriangle) -- (bottomrighttriangle) -- cycle;

    \draw[very thick,fill=gray!50] (squarenw) -- (squarene) -- (squarese) -- (squaresw) -- cycle ;

    \draw[thick,decorate,decoration=snake] (rightboson) -- (rightend) ;

    \draw[thick,-<-=0.5] (squarese) -- (bottomrighttriangle);
    \draw[thick,-<-=0.5] (toprighttriangle) -- (squarene);

% Math symbols    
    \node at (-0.5,-0.3) {$+$};
    \node at (-2.2,-0.3) {$=$};
    
  \end{tikzpicture} 
  \caption{\label{fig:triangularladder}The renormalized triangular vertex $\Lambda_{\qv\nu\omega}$ in the ladder approximation. }
\end{figure}

\begin{figure}[b]
\begin{tikzpicture}
    \coordinate (diagram) at (0,0) ;

    \coordinate (Gammanw) at ($(diagram) + (-3.0,0)$) ;
    \coordinate (Gammasw) at ($(Gammanw) + (0,-0.6)$) ;
    \coordinate (Gammane) at ($(Gammanw) + (0.6,0)$) ;
    \coordinate (Gammase) at ($(Gammasw) + (0.6,0)$) ;    

    \coordinate (gammanw) at ($(diagram) + (-1.5,0)$) ;
    \coordinate (gammasw) at ($(gammanw) + (0,-0.6)$) ;
    \coordinate (gammane) at ($(gammanw) + (0.6,0)$) ;
    \coordinate (gammase) at ($(gammasw) + (0.6,0)$) ;

    \coordinate (square1nw) at ($(diagram) + (0.0,0)$) ;
    \coordinate (square1sw) at ($(square1nw) + (0,-0.6)$) ;
    \coordinate (square1ne) at ($(square1nw) + (0.6,0)$) ;
    \coordinate (square1se) at ($(square1sw) + (0.6,0)$) ;

    \coordinate (square2nw) at ($(square1ne) + (0.7,0)$) ;
    \coordinate (square2sw) at ($(square2nw) + (0,-0.6)$) ;
    \coordinate (square2ne) at ($(square2nw) + (0.6,0)$) ;
    \coordinate (square2se) at ($(square2sw) + (0.6,0)$) ;
    
    \draw[thick,-<-=0.5] (square1se) -- (square2sw);
    \draw[thick,-<-=0.5] (square2nw) -- (square1ne);

    \draw[very thick,fill=gray!50] (Gammane) -- (Gammanw) -- (Gammasw) -- (Gammase) -- cycle;
    \draw[very thick] (gammane) -- (gammanw) -- (gammasw) -- (gammase) -- cycle;

    \draw[very thick] (square1ne) -- (square1nw) -- (square1sw) -- (square1se) -- cycle;
    \draw[very thick,fill=gray!50] (square2ne) -- (square2nw) -- (square2sw) -- (square2se) -- cycle;
    
    \coordinate (plusleft) at ($(gammane)!0.5!(gammase)$) ;
    \coordinate (plusright) at ($(square1nw)!0.5!(square1sw)$) ;    
    \coordinate (plus) at ($(plusleft)!0.5!(plusright)$) ;    
    \node at (plus) {$+$} ;

    \coordinate (eqleft) at ($(Gammane)!0.5!(Gammase)$) ;
    \coordinate (eqright) at ($(gammanw)!0.5!(gammasw)$) ;    
    \coordinate (eq) at ($(eqleft)!0.5!(eqright)$) ;    
    \node at (eq) {$=$} ;

    \coordinate (square1w) at ($(square1nw)!0.5!(square1sw)$) ;
    \coordinate (square1e) at ($(square1ne)!0.5!(square1se)$) ;    
    \node at ($(square1w)!0.5!(square1e)$) {$\gamma$} ;

    \coordinate (square2w) at ($(square2nw)!0.5!(square2sw)$) ;
    \coordinate (square2e) at ($(square2ne)!0.5!(square2se)$) ;    
    \node at ($(square2w)!0.5!(square2e)$) {$\Gamma$} ;

    \coordinate (gammaw) at ($(gammanw)!0.5!(gammasw)$) ;
    \coordinate (gammae) at ($(gammane)!0.5!(gammase)$) ;    
    \node at ($(gammaw)!0.5!(gammae)$) {$\gamma$} ;

    \coordinate (Gammaw) at ($(Gammanw)!0.5!(Gammasw)$) ;
    \coordinate (Gammae) at ($(Gammane)!0.5!(Gammase)$) ;    
    \node at ($(Gammaw)!0.5!(Gammae)$) {$\Gamma$} ;
    
  \end{tikzpicture}
 \caption{Diagrammatic representation of the Bethe-Salpeter equation for the renormalized vertex function.}
 \label{fig:bse}
\end{figure}
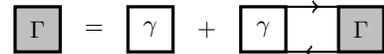

We consider the approximations to the bosonic self-energy diagrams depicted in Fig.~\ref{fig:pi2_ladder}. The first one shown in panel  (a) is a diagram which is second order in the electron-boson vertex $\lambda$. Using the diagrammatic rules, we obtain (cf. Appendix~\ref{app:feynmanrules})
\begin{align}
\label{pi2}
\tilde{\Pi}^{(2)}_{\KV\omega} &=  \frac{T}{N} \sum_{\kv\nu\sigma} \lambda_{\nu+\omega,-\omega} \tilde{G}_{\kv\nu\sigma} \tilde{G}_{\kv+\KV\nu+\omega\sigma}\lambda_{\nu\omega}.
\end{align}
In the approximation shown in Fig.~\ref{fig:pi2_ladder} (b), one of the triangular vertices has been replaced with the renormalized triangular vertex.
This diagram explicitly reads:
\begin{align}
\tilde{\Pi}^{(\text{ladder})}_{\KV\omega} &= \frac{T}{N}
\sum_{\kv\nu}\lambda_{\nu+\omega,-\omega}
\tilde{G}_{\kv\nu} \tilde{G}_{\kv+\KV\nu+\omega}
\Lambda_{\qv\nu\omega},\label{pib}
\end{align}
where the renormalized triangular vertex $\Lambda$ is momentum dependent. It is given in terms of the renormalized lattice vertex function $\Gamma$ in the charge channel in the form
\begin{align}
\Lambda_{\qv\nu\omega}=\lambda_{\nu\omega}-\frac{T}{N}\sum_{\kv'\nu'}\Gamma_{\qv\nu\nu'\omega}\tilde{G}_{\kv'\nu'} \tilde{G}_{\kv'+\KV\nu'+\omega}\lambda_{\nu'\omega},
\end{align}
which is depicted diagrammatically in Fig.~\ref{fig:triangularladder}. Equation~\eqref{pib} may be referred to as a ladder approximation: The renormalized lattice vertex $\Gamma$ contains a ladder-diagram series generated by the Bethe-Salpeter equation~\cite{Brener08},
\begin{align}
\label{eq:bse}
\Gamma_{\KV\nu\nu'\omega} = \gamma_{\nu\nu'\omega}\! -\! \frac{T}{N} \sum_{\kv''\nu''} \gamma_{\nu\nu''\omega}
 \tilde{G}_{\kv''\nu''}\tilde{G}_{\kv''+\KV\nu''+\omega} \Gamma_{\qv\nu''\nu'\omega},
\end{align}
shown in Fig.~\ref{fig:bse}. We solve it by matrix inversion according to
\begin{align}
[\Gamma_{\nu\nu'}]_{\qv\omega}^{-1} = [\gamma_{\nu\nu'}]^{-1}_{\omega} + T\tilde{\chi}_{\qv\nu\omega}^{0}\delta_{\nu\nu'},
\end{align}
with $\tilde{\chi}_{\qv\nu\omega}^{0}=(1/N)\sum_{\kv} \tilde{G}_{\kv\nu}\tilde{G}_{\kv+\KV\nu+\omega}$.
Note that only the charge channel $\Gamma^{\text{ch}}\Let\Gamma^{\uparrow\uparrow}+\Gamma^{\uparrow\downarrow}$ contributes to the polarization.
The ladder approximation is equivalent to the usual expression for the DMFT susceptibility~\cite{Georges96},
\begin{align}
X_{\qv\omega} = 2T\sum_{\nu}\chi^{0}_{\qv\nu\omega} - 2T^{2}\sum_{\nu\nu'} \chi^{0}_{\qv\nu\omega}\Gamma_{\qv\nu\nu'\omega} \chi^{0}_{\qv\nu'\omega},
\end{align}
as long as the (E)DMFT self-consistency condition \eqref{selfcG} is fulfilled~\cite{Hafermann14-2}. Here $\chi_{\qv\nu\omega}^{0}=(1/N)\sum_{\kv} G_{\kv\nu}G_{\kv+\KV\nu+\omega}$ denotes the usual particle-hole bubble.

\subsubsection{Results}

We now discuss the $U$-$V$ phase diagram computed by the DB method, and 
compare it to two established methods: the RPA and the EDMFT.
RPA is expected to be accurate in the low-$U$ regime and to fail at higher $U$.
On the contrary, EDMFT is expected to be accurate at large $U$, since it captures the atomic-like physics, and to fail at low $U$, because it lacks the momentum dependence of the polarization.
The strength of the DB method is precisely to interpolate between RPA at low $U$ and EDMFT at large $U$.
In the following, we first discuss these limits in detail and finally the intermediate coupling regime.

\begin{figure}[t]
\begin{center}
\includegraphics[scale=0.95,angle=0]{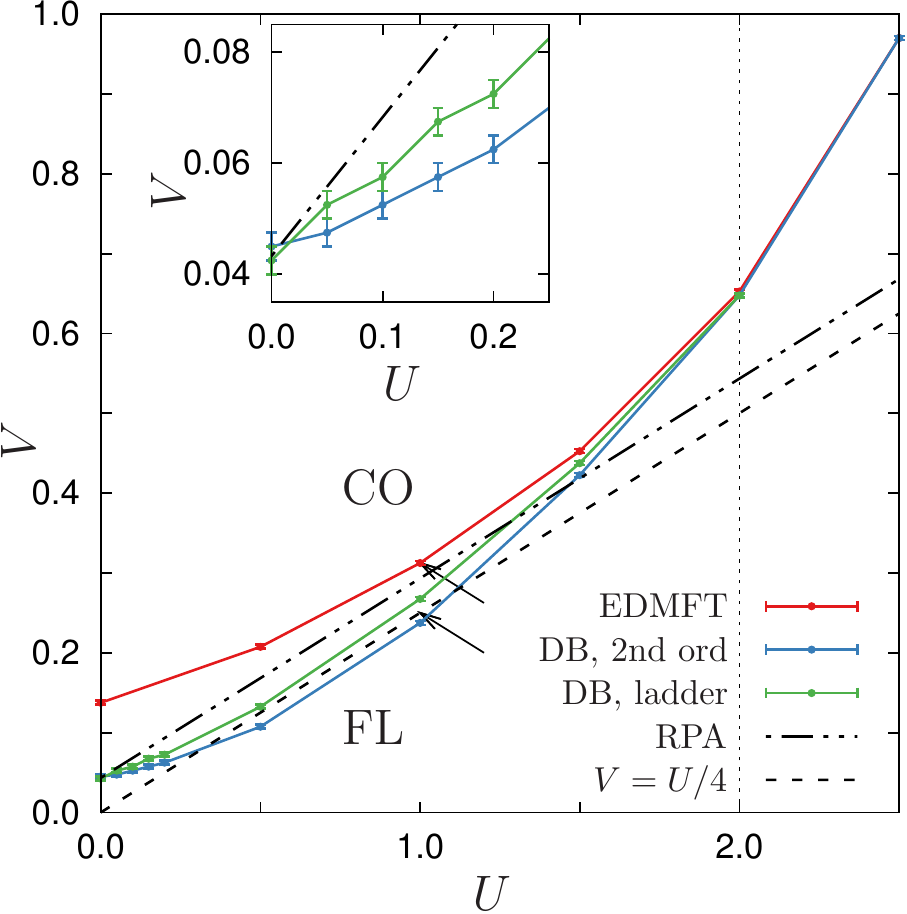} 
\end{center}
\caption{\label{fig:phasediagram_db_pionly} (Color online) $U$-$V$ phase diagram in EDMFT and dual boson (DB) with polarization corrections only, at $T=0.02$. The blue line is for the second-order diagrammatic correction [Fig.~\ref{fig:pi2_ladder} (a)] to the polarization and the green line includes the ladder diagrams [Fig.~\ref{fig:pi2_ladder} (b)]. Finite temperature RPA data (dash-dotted line) is shown for comparison. The dashed line corresponds to $V_{c}=U/z$, where $z=4$ is the coordination number. The value of the bandwidth is marked by the vertical dotted line.
Arrows bound the location of the transition according to lattice Monte Carlo results for $U=1$ and $T=0.125$ (see text for details).}
\end{figure}

The phase diagram of the Hamiltonian \eqref{H}, with nearest-neighbor hopping $t$ and nearest-neighbor interaction parameter $V$ is shown in Fig.~\ref{fig:phasediagram_db_pionly}, albeit at elevated temperature $T=0.02$ compared to Fig.~\ref{fig:phasediagram} (in units of the half bandwidth is $D=4t$). We focus on the region of $U$ values for which a metallic solution exists.
The phase boundaries have been determined from the zeros of $X_{\qv\omega}^{-1}$ at $\qv=(\pi,\pi)$ and $\omega=0$ and we have verified that the divergence happens at the $(\pi,\pi)$ point first. The EDMFT data in this figure is quantitatively very similar to the one in Fig.~\ref{fig:phasediagram}, revealing a very small overall temperature dependence of the phase boundary.

\begin{figure}[t]
\begin{center}
\includegraphics[scale=0.975]{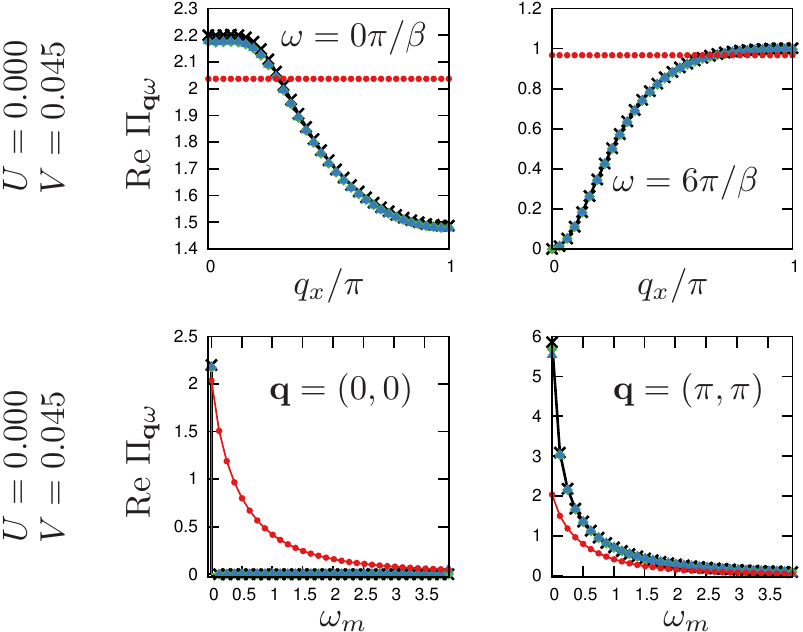} 
\end{center}
\caption{\label{fig:compRPA} (Color online) Comparison of the momentum dependence (top panels) and frequency dependence (bottom panels) of the polarization in EDMFT (red circles), second-order DB (blue triangles), ladder DB (green, diamonds) and RPA (black crosses) for $U=0$ and $V=0.045$ in the immediate vicinity of the charge-ordering transition.}
\end{figure}

 Let us first consider the low-$U$ regime and compare to RPA. In RPA, the polarization is given by the Lindhardt bubble, i.e. $\Pi^{\text{RPA}}_{\qv\omega} = -(T/N) \sum_{\kv\nu\sigma} G^{(0)}_{\kv\nu}G^{(0)}_{\kv+\qv\nu+\omega}$.
An explicit calculation of $\Pi^{\text{RPA}}_{\qv\omega}$ for $T=0.02$ shows $V^{\text{RPA}}_c \approx 0.043+U/4$, where the slope is determined by the number of nearest neighbors $z=4$. This phase boundary is shown by the dash-dotted line in Fig.~\ref{fig:phasediagram_db_pionly}.
The DB results approach the RPA result in the limit $U\to 0$. In Appendix \ref{app:sigmagw} we show that DB indeed reduces to RPA in the weak-coupling limit ($U,V\to 0$). For $U=0$ but finite $V$, the local retarded interaction is finite, but the deviations from RPA remain small. In Fig.~\ref{fig:compRPA} we compare the frequency- and momentum dependence of the ladder DB solution with RPA which confirms this picture. The second-order approximation forms the dominant part of the corrections by far as expected.
EDMFT fails in this low $U$ limit, because it completely neglects the momentum dependence of the polarization
(cf., e.g., Fig.~\ref{fig:compRPA} , top panel). The diagrammatic corrections restore this momentum dependence.

\begin{figure}[b]
\includegraphics[width=0.475\textwidth]{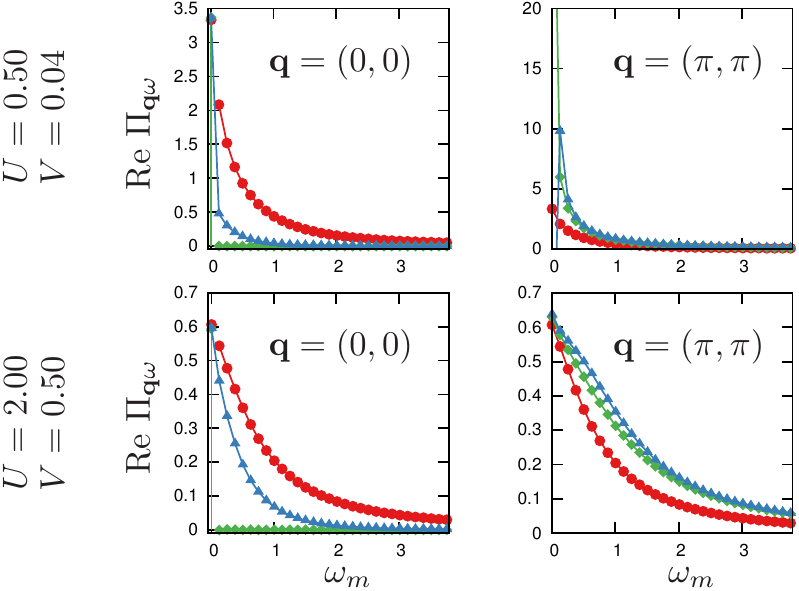} 
\caption{\label{fig:pi_lat_wdep}(Color online) Frequency dependence of the physical polarization $\Pi_{\qv\omega}$ for fixed momentum $\qv=(0,0)$ (left column) and $\qv=(\pi,\pi)$ (right column) in the ladder approximation. We show results for EDMFT (red  circles), second-order DB (blue triangles) and ladder DB (green  diamonds). The colors are the same as in the phase diagram Fig.~\ref{fig:phasediagram_db_pionly}. For $U=0.5$ and $\qv=(\pi,\pi)$, the polarization in second-order DB turns negative at $\omega=0$.
} 
\end{figure}

The DB corrections to EDMFT diminish in the opposite large-$U$ limit, where the MI phase is approached (the Mott transition roughly takes place at the right end of this figure; cf. Fig.~\ref{fig:phasediagram}). The physics is well described within EDMFT.
In this limit, the RPA clearly fails, because it is a weak coupling approach (the static self-energy and static irreducible vertex are not sufficient to describe the strong correlation physics).

\begin{figure}[t]
\includegraphics[width=0.475\textwidth]{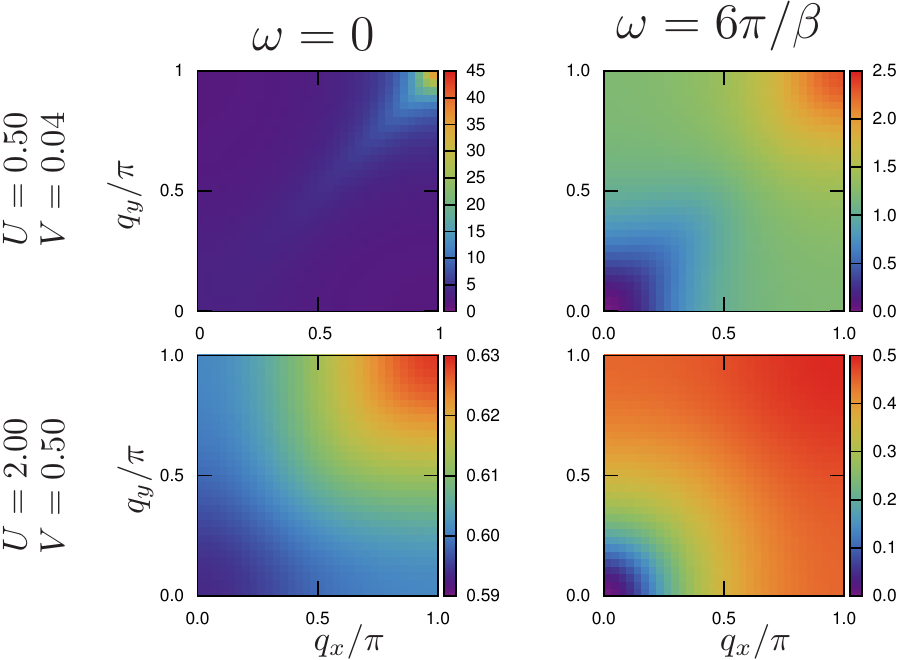} 
\caption{\label{fig:pi_lat_qdep}(Color online) Momentum dependence of the physical polarization $\Pi_{\qv\omega}$ for fixed Matsubara frequencies $\omega=0$ (left column) and $\omega=6\pi/\beta$ (right column) in the ladder approximation. In EDMFT, this quantity is a constant.} 
\end{figure}

In the intermediate interaction regime we obtain results which are compatible with currently available lattice Monte Carlo data for $U=1$ and $T=0.125$, which locate the transition roughly between $V=0.25$ and $V=0.3125$~\cite{Zhang89} (see arrows in Fig.~\ref{fig:phasediagram_db_pionly}).
The ladder and second-order DB approximation give close results. The four-leg vertex and the long-range vertex corrections built from it, which are included in the former but not in the latter, have a small effect in determining the phase boundary.
We further see that the DB phase boundaries run essentially parallel to the $V=U/4$-~line. This is in agreement with previous results (see, e.g., Ref.~\onlinecite{Davoudi07} and references therein). The lattice Monte Carlo results, however, suggest that the phase boundary is located \emph{above} this line.
While the ladder approximation agrees with this result, the second-order approximation lies below.
Related with this observation, the latter exhibits an artifact, which can be seen in the top right panel of Fig.~\ref{fig:pi_lat_wdep}. The polarization turns negative for $\qv=(\pi,\pi)$ and the lowest frequency $\omega=0$. This follows directly from the condition for charge ordering, $1+(U-4V)\Pi_{\qv=(\pi,\pi),\omega=0}=0$.
For larger $U$, the EDMFT and DB results are relatively close for all frequencies (see bottom right panel of Fig.~\ref{fig:pi_lat_wdep}).
The two DB approximations also differ qualitatively at $\qv=(0,0)$. The second-order approximation (and EDMFT) 
are finite at finite frequencies. This implies a violation of the Ward identity~\cite{Hafermann14-2}, which appears to be more severe for larger $U$. Ladder DB shows the required discontinuity at $\omega=0$ and appears to be conserving for all values of $U$~\cite{Rubtsov12}.
Long-range vertex corrections seem necessary to avoid such artifacts.

The phase diagram of Fig.~\ref{fig:phasediagram_db_pionly} can also be compared to cluster methods. It is qualitatively similar to cluster results on the two-dimensional triangular lattice~\cite{Hassan10}. 
For the square lattice, data for comparison are available only for $T=0$~\cite{Aichhorn07}. In these variational cluster approximation (VCA) calculations, the phase boundary is located close to the $V=U/4$ line.

In Fig.~\ref{fig:pi_lat_qdep} we show the momentum dependence of the polarization $\Pi_{\qv\omega}$. At moderate  $U=0.5$ it exhibits a rather strong momentum dependence and the correction is large compared to EDMFT ($\Pi_{\omega=0}\sim 1.25$ at these parameters). It is largest at the vector $\qv=(\pi,\pi)$ at which the transition occurs ($\abs{V_{\qv}}$ is maximal and $V_{\qv}<0$ at this point). 
At large $U$, the correction is much weaker and more isotropic (in EDMFT, $\Pi_{\omega=0}\sim 0.274$). At finite Matsubara frequencies, the polarization is rather flat, except in the vicinity of the $\qv=(0,0)$ point, where it decreases to zero. This is a further requirement imposed by charge conservation~\cite{Hafermann14-2}.

\begin{figure}[t]
\begin{center}
\includegraphics[scale=0.675]{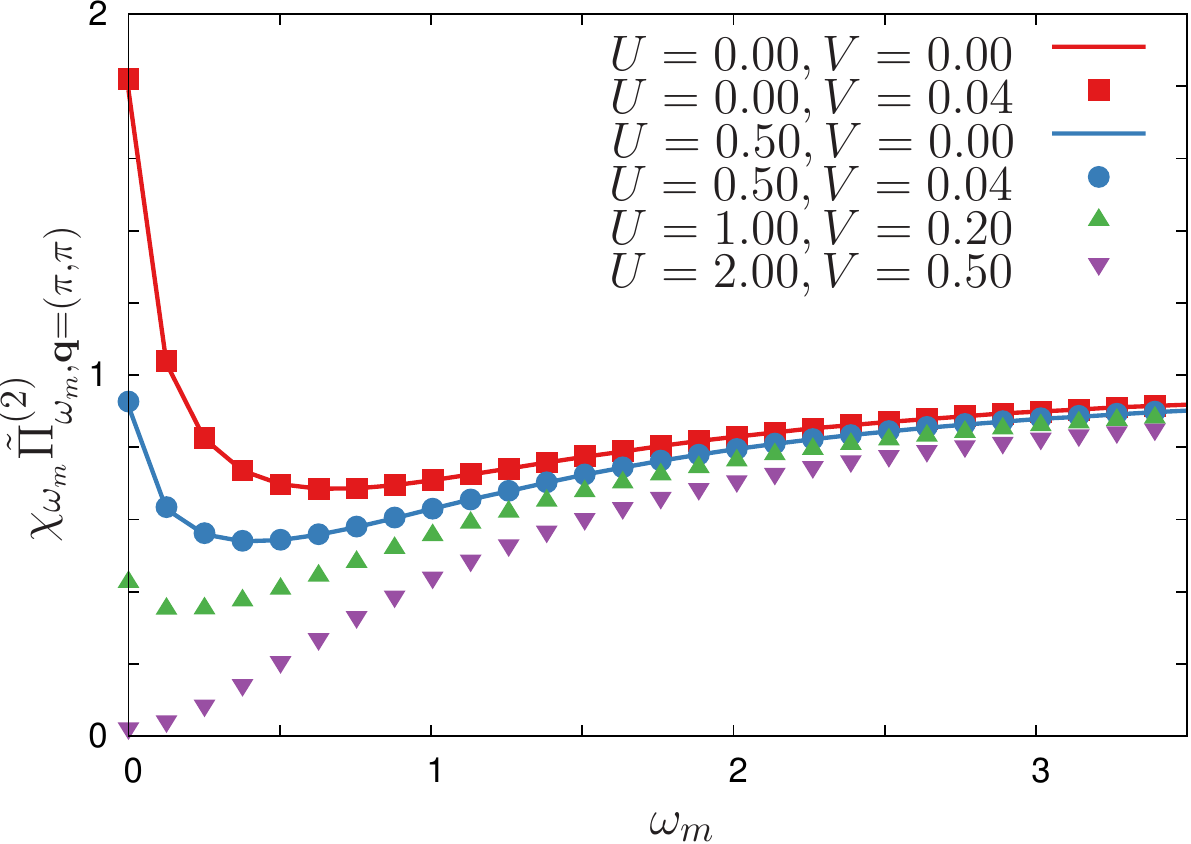} 
\end{center}
\caption{\label{fig:chipi} (Color online) The diagrammatic correction $\chi_\omega \tilde{\Pi}_{\omega,\KV=(\pi,\pi)}$ (see text) in second-order approximation, evaluated at the wave vector associated with charge order and shown as a function of Matsubara frequency. With increasing values of $U$, the diagrammatic correction is reduced. Its value is only weakly dependent on $V$.}
\end{figure}

In order to trace the interpolating behavior of the DB approach, it is instructive to rewrite Eq.~\eqref{xdtox} in the form
\begin{align}
X^{-1}_{\KV\omega} &\!\!=\! \chi_\omega^{-1}(1+\chi_{\omega}\tilde{\Pi}_{\KV\omega})^{-1}+\Lambda_{\omega}\!-\!V_{\KV}.\label{eq:xfrompi}
\end{align}
Comparing with the EDMFT susceptibility \eqref{xedmft}, one sees that the term in parentheses plays the role of a renormalization factor, which determines how much the solution is altered compared to EDMFT. The change of the phase boundary with respect to EDMFT is hence determined by the dimensionless quantity $\chi_\omega\tilde{\Pi}_{\omega,\KV}$ [at $\qv=(\pi,\pi)$ and $\omega=0$]. In Fig.~\ref{fig:chipi}, we show its frequency dependence in second-order approximation at various points in the phase diagram. For high frequencies it approaches a constant non-zero value because the asymptotic behavior of the constituents cancels. At low frequencies, diagrammatic corrections contribute significantly, even for small interaction. This is due to the fact that the fermion-boson vertex remains finite even for vanishing interaction [cf. Eq. \eqref{lambdafromcon}].
As $U$ increases, the magnitude of the vertex increases. At the same time, the dual Green's function and the local susceptibility $\chi_\omega$ become smaller in magnitude. As electrons become more and more localized, the net effect is that nonlocal corrections become less and less important: $\chi_{\omega}\tilde{\Pi}_{\qv\omega}$ decreases continuously at low frequencies as $U$ increases.  For example, at $U=2$, $V=0.5$, we have $\chi_{\omega=0}\tilde{\Pi}_{\qv=(\pi,\pi)\omega=0}\ll 1$ so that the DB phase boundary merges into the EDMFT one for large $U$.

Figure~\ref{fig:chipi}  further shows that diagrammatic corrections depend only weakly on $V$. For small $U$, results for $V=0$ and $V=0.04$ are virtually indistinguishable. The value $V=0.04$ is close to the critical value of the charge-ordering transition, which illustrates that there is no structural change in $\chi_{\omega}\tilde{\Pi}_{\qv\omega}$ near the transition. It is the interplay with the value of $V_{\qv}$ in \eqref{eq:xfrompi} that triggers the divergence in $X$.

\subsection{Fermionic self-energy diagrams}

In this section, we include fermionic self-energy diagrams in addition to the bosonic ones.
We see that these corrections have essentially no effect on the phase boundaries. They restore a weak momentum dependence of the self-energy missing in EDMFT.

\emph{A priori}, different approximations for the fermionic self-energy can be combined with the previously introduced polarization diagrams. In order to keep the discussion reasonably simple and reduce the number of possibilities, we impose the following guiding principle for constructing approximations: The fermionic and bosonic self-energies are chosen  such that they can be obtained from a common (dual) functional. In each case, the fermionic (bosonic) self-energy is given by a functional derivative with respect to the fermionic (bosonic) dual Green's function.

\subsubsection{Diagrams}

\begin{figure}[t]
\subfloat[]{
 \begin{tikzpicture}
    \coordinate (leftend) at ($(0,0)$ ) ;
    \coordinate (toplefttriangle) at ($(leftend) + (0.4,0.3)$) ;
    \coordinate (bottomlefttriangle) at ($(leftend) + (0.4,-0.3)$) ;

    \coordinate (toprighttriangle) at ($(toplefttriangle) + (-1.6,0)$) ;
    \coordinate (bottomrighttriangle) at ($(toprighttriangle) + (0,-0.6)$) ;
    \coordinate (rightend) at ($(toprighttriangle) + (0.4,-0.3)$) ;

    \draw[very thick] (leftend) -- (toplefttriangle) -- (bottomlefttriangle) -- cycle;

    \draw[very thick] (rightend) -- (toprighttriangle) -- (bottomrighttriangle) -- cycle;

    \draw[thick,decorate,decoration=snake] (leftend) -- (rightend) ;

    \draw[thick,-<-=0.5,bend left=90] (bottomlefttriangle) to (bottomrighttriangle);
    \draw[thick,-<-=0.5,bend left=90] (toprighttriangle) to (toplefttriangle);
 
 \end{tikzpicture} 
}
\subfloat[]{
\begin{tikzpicture}
  \coordinate (diagram1) at (0,0) ;

\coordinate (leftend) at ($(diagram1)$ ) ;
\coordinate (leftfermion) at ($(leftend) + (-0.3,0)$) ;
\coordinate (toplefttriangle) at ($(leftend) + (0.3,0.4)$) ;
\coordinate (rightlefttriangle) at ($(leftend) + (0.6,0.)$) ;

\coordinate (leftrighttriangle) at ($(rightlefttriangle) + (0.5,0)$) ;
\coordinate (toprighttriangle) at ($(leftrighttriangle) + (0.3,0.4)$) ;
\coordinate (rightend) at ($(leftrighttriangle) + (0.6,0.)$) ;
\coordinate (rightfermion) at ($(rightend) + (0.3,0)$) ;

\draw[very thick] (leftend) -- (toplefttriangle) -- (rightlefttriangle) -- cycle;

\draw[very thick] (rightend) -- (toprighttriangle) -- (leftrighttriangle) -- cycle;

\draw[thick,->-=0.5] (leftfermion) -- (leftend) ;
\draw[thick,-<-=0.5] (rightfermion) -- (rightend) ;

\draw[thick,->-=0.5] (rightlefttriangle) -- (leftrighttriangle);
\draw[thick,decorate,decoration=snake] (toprighttriangle) -- (toplefttriangle);
  \end{tikzpicture}
}
\subfloat[]{
  \begin{tikzpicture}
    \coordinate (diagram1) at (0,0) ;

    \coordinate (leftend) at ($(diagram1)$ ) ;
    \coordinate (leftboson) at ($(leftend) + (-0.3,0)$) ;
    \coordinate (toplefttriangle) at ($(leftend) + (0.4,0.3)$) ;
    \coordinate (bottomlefttriangle) at ($(leftend) + (0.4,-0.3)$) ;

    \coordinate (toprighttriangle) at ($(toplefttriangle) + (0.7,0)$) ;
    \coordinate (bottomrighttriangle) at ($(toprighttriangle) + (0,-0.6)$) ;
    \coordinate (rightend) at ($(toprighttriangle) + (0.4,-0.3)$) ;
    \coordinate (rightboson) at ($(rightend) + (0.3,0)$) ;

    \draw[very thick] (leftend) -- (toplefttriangle) -- (bottomlefttriangle) -- cycle;

    \draw[very thick] (rightend) -- (toprighttriangle) -- (bottomrighttriangle) -- cycle;

    \draw[thick,decorate,decoration=snake] (leftboson) -- (leftend) ;
    \draw[thick,decorate,decoration=snake] (rightboson) -- (rightend) ;

    \draw[thick,-<-=0.5] (bottomlefttriangle) -- (bottomrighttriangle);
    \draw[thick,-<-=0.5] (toprighttriangle) -- (toplefttriangle);
  \end{tikzpicture}
}
 \caption{\label{fig:functional_bubble} Second-order DB approximation for the fermionic (b) and bosonic (c) self-energies defined in terms of a common dual functional (a). The open triangles stand for the impurity triangular vertices, straight and wiggly lines denote fully dressed fermionic and bosonic propagators, respectively.}
\end{figure}
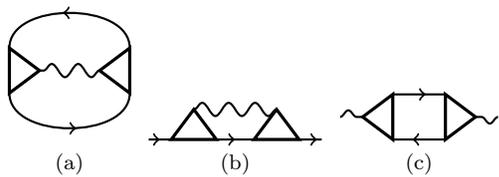

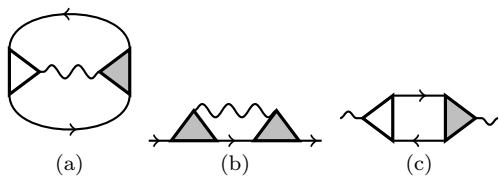
\begin{figure}[b]
\subfloat[]{
 \begin{tikzpicture}
    \coordinate (leftend) at ($(0,0)$ ) ;
    \coordinate (toplefttriangle) at ($(leftend) + (0.4,0.3)$) ;
    \coordinate (bottomlefttriangle) at ($(leftend) + (0.4,-0.3)$) ;

    \coordinate (toprighttriangle) at ($(toplefttriangle) + (-1.6,0)$) ;
    \coordinate (bottomrighttriangle) at ($(toprighttriangle) + (0,-0.6)$) ;
    \coordinate (rightend) at ($(toprighttriangle) + (0.4,-0.3)$) ;

    \draw[very thick,fill=gray!50] (leftend) -- (toplefttriangle) -- (bottomlefttriangle) -- cycle;

    \draw[very thick] (rightend) -- (toprighttriangle) -- (bottomrighttriangle) -- cycle;

    \draw[thick,decorate,decoration=snake] (leftend) -- (rightend) ;

    \draw[thick,-<-=0.5,bend left=90] (bottomlefttriangle) to (bottomrighttriangle);
    \draw[thick,-<-=0.5,bend left=90] (toprighttriangle) to (toplefttriangle);
 
 \end{tikzpicture} 
}
\subfloat[]{
\begin{tikzpicture}
  \coordinate (diagram1) at (0,0) ;

\coordinate (leftend) at ($(diagram1)$ ) ;
\coordinate (leftfermion) at ($(leftend) + (-0.3,0)$) ;
\coordinate (toplefttriangle) at ($(leftend) + (0.3,0.4)$) ;
\coordinate (rightlefttriangle) at ($(leftend) + (0.6,0.)$) ;

\coordinate (leftrighttriangle) at ($(rightlefttriangle) + (0.5,0)$) ;
\coordinate (toprighttriangle) at ($(leftrighttriangle) + (0.3,0.4)$) ;
\coordinate (rightend) at ($(leftrighttriangle) + (0.6,0.)$) ;
\coordinate (rightfermion) at ($(rightend) + (0.3,0)$) ;

\draw[very thick,fill=gray!50] (leftend) -- (toplefttriangle) -- (rightlefttriangle) -- cycle;

\draw[very thick,fill=gray!50] (rightend) -- (toprighttriangle) -- (leftrighttriangle) -- cycle;

\draw[thick,->-=0.5] (leftfermion) -- (leftend) ;
\draw[thick,-<-=0.5] (rightfermion) -- (rightend) ;

\draw[thick,->-=0.5] (rightlefttriangle) -- (leftrighttriangle);
\draw[thick,decorate,decoration=snake] (toprighttriangle) -- (toplefttriangle);
  \end{tikzpicture}
}
\subfloat[]{
  \begin{tikzpicture}
    \coordinate (diagram1) at (0,0) ;

    \coordinate (leftend) at ($(diagram1)$ ) ;
    \coordinate (leftboson) at ($(leftend) + (-0.3,0)$) ;
    \coordinate (toplefttriangle) at ($(leftend) + (0.4,0.3)$) ;
    \coordinate (bottomlefttriangle) at ($(leftend) + (0.4,-0.3)$) ;

    \coordinate (toprighttriangle) at ($(toplefttriangle) + (0.7,0)$) ;
    \coordinate (bottomrighttriangle) at ($(toprighttriangle) + (0,-0.6)$) ;
    \coordinate (rightend) at ($(toprighttriangle) + (0.4,-0.3)$) ;
    \coordinate (rightboson) at ($(rightend) + (0.3,0)$) ;

    \draw[very thick] (leftend) -- (toplefttriangle) -- (bottomlefttriangle) -- cycle;

    \draw[very thick,fill=gray!50] (rightend) -- (toprighttriangle) -- (bottomrighttriangle) -- cycle;

    \draw[thick,decorate,decoration=snake] (leftboson) -- (leftend) ;
    \draw[thick,decorate,decoration=snake] (rightboson) -- (rightend) ;

    \draw[thick,-<-=0.5] (bottomlefttriangle) -- (bottomrighttriangle);
    \draw[thick,-<-=0.5] (toprighttriangle) -- (toplefttriangle);
  \end{tikzpicture}
}
 \caption{\label{fig:functional_ladder} Dual boson ladder approximation constructed from a functional involving the renormalized triangular vertex shown in Fig.~\ref{fig:triangularladder} (shaded triangle).}
\end{figure}

\begin{figure}[t]
\begin{center}
\includegraphics[scale=0.95,angle=0]{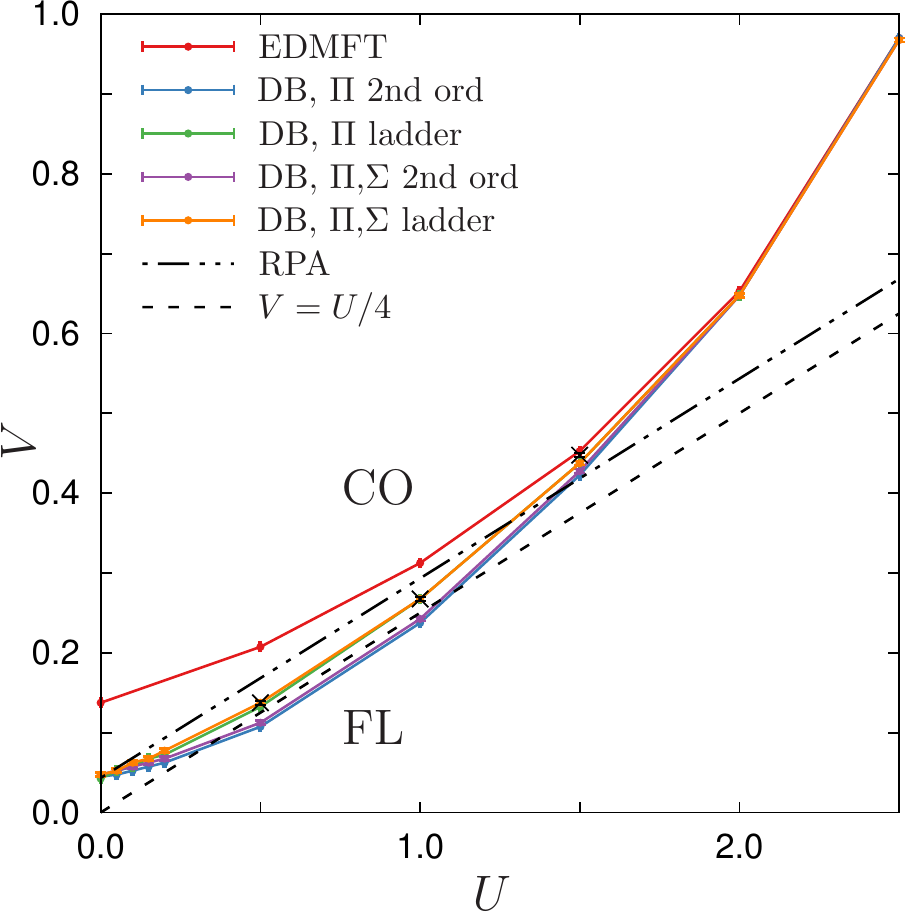} 
\end{center}
\caption{\label{fig:phasediagram_db} (Color online) Phase diagram for the extended Hubbard model in the plane of on-site interaction $U$ and nearest-neighbor interaction $V$ including fermionic self-energy corrections computed within the approximations depicted in Figs.~\ref{fig:functional_bubble} and \ref{fig:functional_ladder}.
The EDMFT and DB data with polarization corrections only (labeled ``DB, $\Pi$'') are the same as in Fig.~\ref{fig:phasediagram_db_pionly} and included for comparison. Black crosses mark ladder DB results with full outer self-consistency.
}
\end{figure}

The two approximations we consider are depicted in Figs.~\ref{fig:functional_bubble} and \ref{fig:functional_ladder}, which we refer to as ``second-order'' and ``ladder'' approximations, respectively, in accordance with the foregoing. The latter differs from the former in the use of the renormalized triangular vertex instead of the local impurity one. The approximations generate the same bosonic self-energies that we considered before. Note that in the ladder approximation, both vertices in the self-energy diagram are necessarily renormalized. This may be checked, for example, by symbolically inserting the graphical definition of the renormalized triangular vertex, Fig.~\ref{fig:triangularladder}, into Fig.~\ref{fig:functional_ladder} and taking the functional derivative (i.e., cutting lines) with respect to all internal Green's functions while obeying the product rule.

The self-energy in the second-order approximation reads
\begin{align}
 \tilde{\Sigma}^{(2)}_{\kv\nu\sigma} &= -\frac{T}{N} \sum_{\KV\omega} \lambda^\sigma_{\nu\omega}\tilde{G}_{\kv+\KV\nu+\omega} \tilde{X}_{\KV\omega} \lambda_{\nu+\omega,-\omega}\label{mixed2},
\end{align}
while in the ladder approximation we have
\begin{align}
 \tilde{\Sigma}^{(\text{ladder})}_{\kv\nu\sigma} &= -\frac{T}{N} \sum_{\KV\omega} \Lambda_{\qv\nu\omega}\tilde{G}_{\kv+\KV\nu+\omega} \tilde{X}_{\KV\omega} \Lambda_{\qv\nu+\omega,-\omega}.\label{app:mixedladder}
\end{align}
The bosonic self-energies are given by Eqs.~\eqref{pi2} and \eqref{pib}, respectively.

\subsubsection{Results}

Results for the phase diagram within this approximation are shown in Fig.~\ref{fig:phasediagram_db}. We include results from the previous phase diagram for comparison. The effect on the phase diagram is very small (the phase-boundaries are pushed to slightly higher values of $V$).

\begin{figure}[t]
\includegraphics[width=0.475\textwidth]{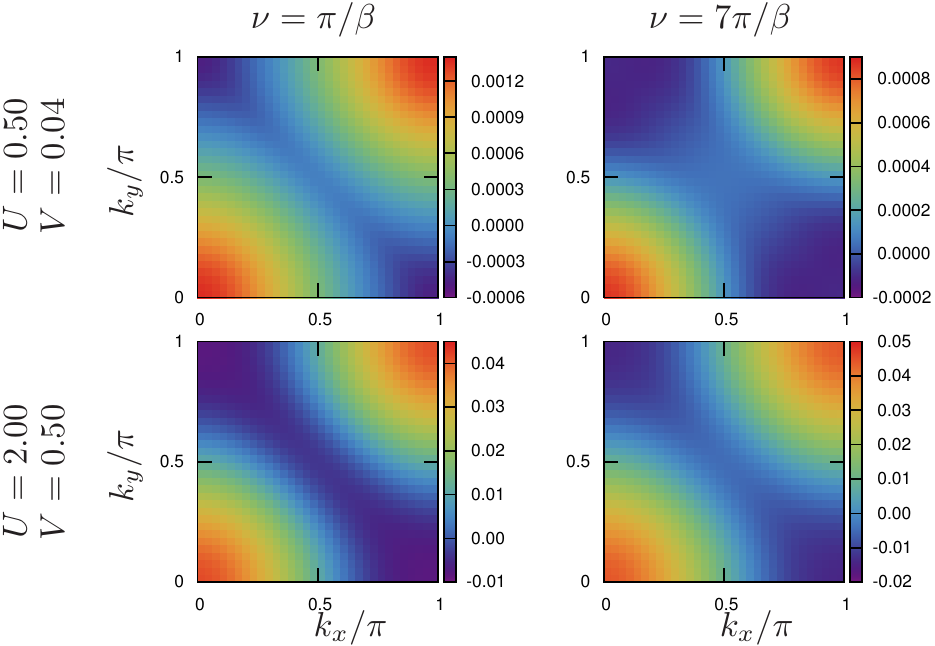} 
\caption{\label{fig:sigma_lat_qdep}(Color online) Momentum dependence of the \emph{nonlocal} part of physical self-energy $\Im \Sigma_{\kv\nu}-\Im \Sigma_{\nu}^{\text{EDMFT}}$ for fixed Matsubara frequencies $\nu=\pi/\beta$ (left column) and $\nu=7\pi/\beta$ (right column) in the ladder approximation. In EDMFT, this quantity is a constant (equal to zero). 
} 
\end{figure}

\begin{figure}[b]
\includegraphics[width=0.475\textwidth]{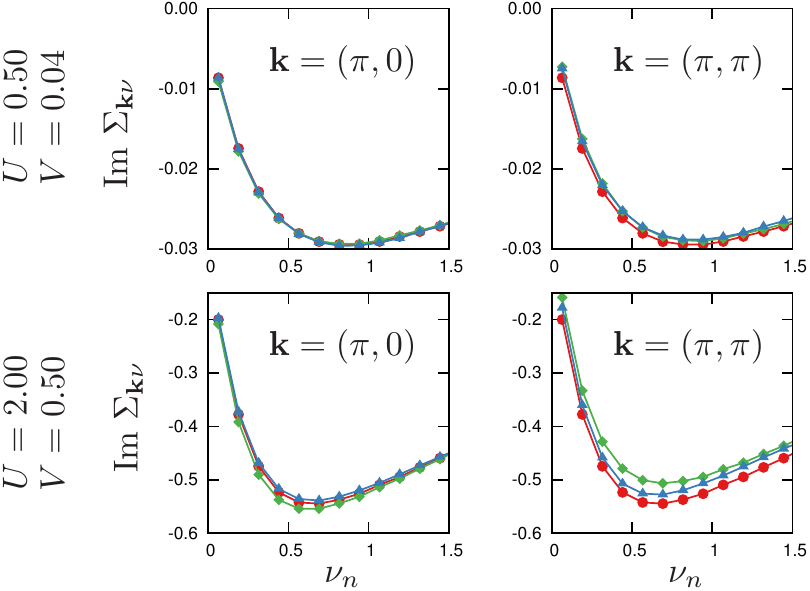} 
\caption{\label{fig:sigma_lat_wdep}(Color online) Frequency dependence of the physical self-energy $\Im \Sigma_{\kv\nu}$ for fixed momentum $\qv=(\pi,0)$ (left column) and $\qv=(\pi,\pi)$ (right column) in the ladder approximation. We show results for EDMFT (red circles), second-order DB (blue triangles), and ladder DB (green diamonds). The colors are the same as in the phase diagram of Fig.~\ref{fig:phasediagram_db_pionly}.
} 
\end{figure}

In Fig.~\ref{fig:sigma_lat_qdep} the momentum dependence of $\Im \Sigma_{\kv\nu}- \Im\Sigma_{\nu}^{\text{EDMFT}}$ is shown for the ladder approximation. The correction is smallest in the vicinity of the Fermi surface and increases away from it. It is generally negligible with respect to the EDMFT self-energy as we can see in Fig.~\ref{fig:sigma_lat_wdep}. The behavior is qualitatively similar compared to EDMFT~+~$GW$ as reported in Refs.~\onlinecite{Ayral13,Huang14}.
The polarization remains quantitatively similar as in the case without self-energy corrections (not shown).

\subsection{Effect of outer self-consistency}

We have performed ladder DB calculations with full outer self-consistency for  $U=0.5$, $U=1.0$, and $U=1.5$, to see the effect of converging the bath. The susceptibility is smaller in the fully self-consistent calculation, but the phase diagram is qualitatively unchanged. The results are marked by black crosses in Fig.~\ref{fig:phasediagram_db}. The corrections to the phase boundary increase for increasing $U$, but remain marginal.

\section{Simplified Approximation}
\label{sec:edmftgw}

We have seen that the nonlocal corrections to EDMFT lead to significant changes in the phase diagram and polarization. The DB approach, however, is computationally more expensive, because of the required computation of vertex functions. Simpler approximations which capture the essential features would be desirable. This leads to the question of how important vertex corrections are for an accurate description of the physics.

A simpler and well-known approximation that goes beyond EDMFT is the EDMFT~+~$GW$ approximation (for a recent discussion,  see Ref.~\onlinecite{Ayral13}). The basic idea of EDMFT~+~$GW$ is to treat the local self-energies within EDMFT and add nonlocal contributions from $GW$ diagrams. The two different decoupling schemes (see Sec.~\ref{sec:invariance}) give different results for the phase boundaries~\cite{Ayral13}, while the DB approach is invariant. Formally, the $V$-decoupling scheme is closer to the DB  approach: In both cases, we have an electron-electron and an electron-boson vertex ($U$ and $i$ in the EDMFT~+~$GW$) and the local interaction is taken into account on the level of the impurity model.
In this section, we formulate a simplified version of the DB equations that neglects corrections due to the fermion-fermion vertex. We then show that in the weak coupling regime, this approximation and EDMFT~+~$GW$ in the $V$-decoupling scheme yield similar results.\footnote{In EDMFT~+~$GW$ within the $UV$-decoupling scheme, the phase boundary to the CO state is essentially unaltered with respect to EDMFT~\cite{Ayral13}.}

We construct a simplified DB approximation (s-DB) which does not require the expensive calculation of the vertices $\gamma$ and $\lambda$ by setting $\gamma=0$ in the second-order diagrams (higher orders vanish in such an approach). With the relation (see Ref.~\onlinecite{Rubtsov12} and Appendix~\ref{app:lambda})
\begin{align}
\label{lambdafromgamma}
\lambda_{\nu\omega}^{\sigma} = \frac{1}{\chi_{\omega}} \left(T\sum_{\nu'\sigma'}\gamma_{\nu\nu'\omega}^{\sigma\sigma'}g_{\nu'\sigma'}g_{\nu'+\omega,\sigma'}-1\right),
\end{align}
we see that in this case $\lambda_{\nu\omega}=-\chi_{\omega}^{-1}$. This simplification hence neglects the fermionic frequency structure of the three-leg vertex function. The resulting expressions for the polarization and self-energy are given in Eqs.~\eqref{app:pi_our} and \eqref{app:sigmasdb}. The corresponding EDMFT~+~$GW$ expressions are provided in Eqs.~\eqref{app:pi_edmftgw} and \eqref{app:sigmaedmftgw}, respectively.

\begin{figure}[t]
\begin{center}
\includegraphics[scale=0.95,angle=0]{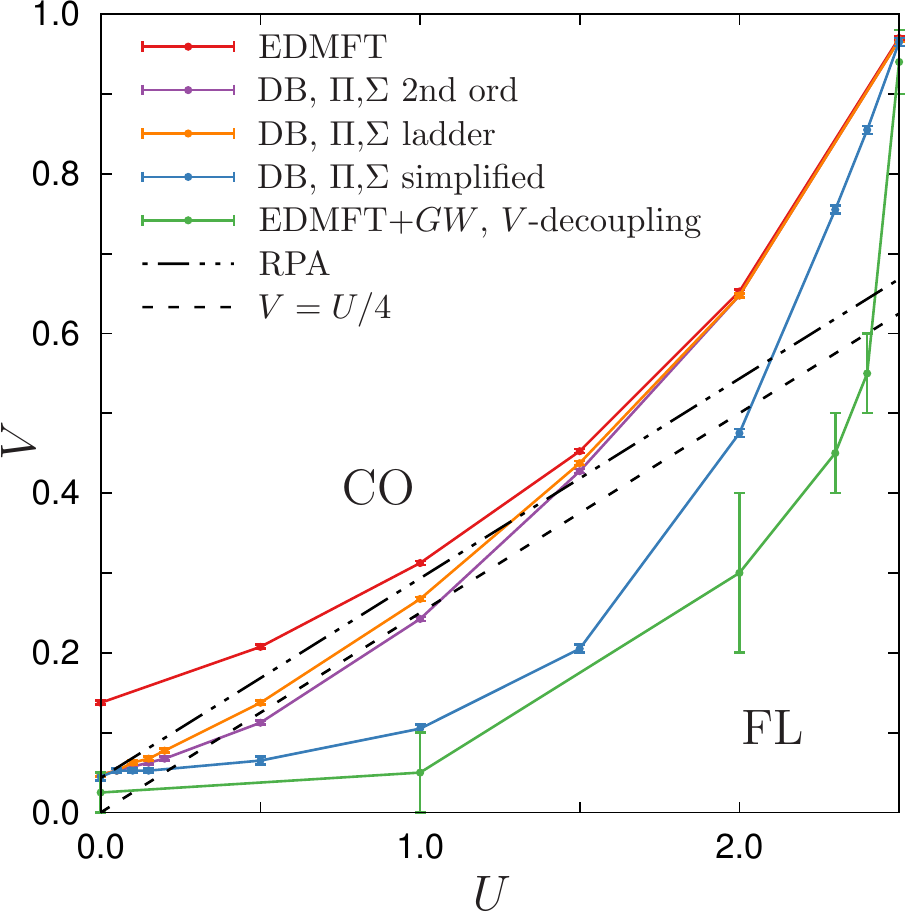} 
\end{center}
\caption{\label{fig:phasediagram_db_gw} (Color online) $U$-$V$ phase diagram in the DB approximation with and without vertex corrections and from EDMFT~+~$GW$ in the $V$-decoupling scheme. The EDMFT~+~$GW$ data obtained in the $V$-decoupling scheme are taken from Ref.~\onlinecite{Ayral13} (for $T=0.01$). Approximations that neglect the fermionic frequency structure of the three-leg vertex deviate strongly from those with vertex corrections and from the $V=U/4$ line, close to where the true phase boundary is expected. For details, see text.}
\end{figure}

In Fig.~\ref{fig:phasediagram_db_gw} we examine the difference between these approximations numerically. We include DB results from the phase diagram of Fig.~\ref{fig:phasediagram_db} for comparison.
Several observations can be made. First, s-DB agrees with the second-order approximation with vertex corrections for $U\to 0$. This is expected by construction. Note, however, that the two approximations are not exactly equivalent here because the former neglects the fermionic structure of the triangular vertex, which is present, but very weak at $U=0$ due to a small but finite retarded interaction.
EDMFT~+~$GW$ (for $T=0.01$) from Ref.~\onlinecite{Ayral13} agrees within error bars. We expect that a refined calculation at the same temperature would give closer (although not perfect, because of the finite $\Lambda_{\omega}$) agreement for $U\to 0$.

\begin{figure}[t]
\includegraphics[width=0.475\textwidth]{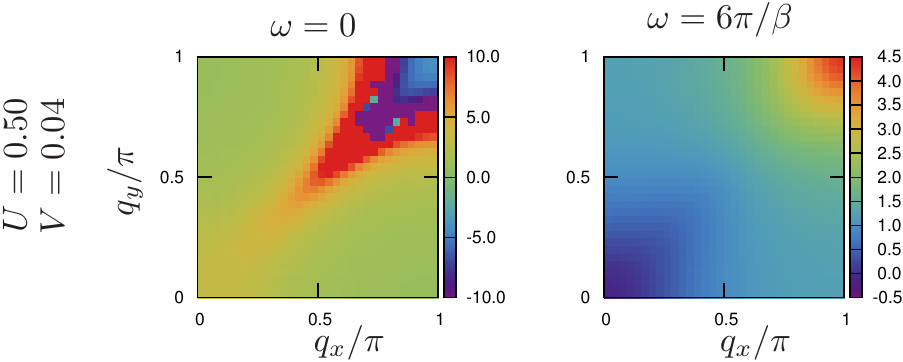} 
\caption{\label{fig:pi_lat_qdep_gw}(Color online) Momentum dependence of the physical polarization $\Pi_{\qv\omega}$ in s-DB, for fixed Matsubara frequencies $\omega=0$ (left) and $\omega=6\pi/\beta$ (right). In EDMFT, this quantity is a constant. The same quantity in DB is shown in Fig.~\ref{fig:pi_lat_qdep}. In the left panel, the plot range is restricted to improve contrast.} 
\end{figure}

EDMFT~+~$GW$ and s-DB, however, depart significantly from the DB result already for small interaction. In particular, the slope is different.
We can trace this back to the fact that the fermionic frequency structure of the fermion-boson vertex $\lambda_{\nu\omega}$ is neglected. 
From Figs.~\ref{fig:lambda_edmft_vertical}--\ref{fig:lambda_edmft_v06}, we see that $\lambda_{\nu\omega}$ changes sign as a function of $\nu$. Since s-DB ignores cancellations caused by this structure, we expect it to overestimate $\tilde{\Pi}_{\KV\omega=0}$ and, by Eq.~\eqref{eq:pifromdual}, the physical polarization $\Pi$. This leads to charge ordering at smaller  $V$ as observed [cf. Eq.~\eqref{Pidef}]. The deviations are largest in the intermediate coupling regime. EDMFT~+~$GW$ shows an overall similar trend as s-DB.\footnote{We note that the EDMFT~+~$GW$ result was computed including an outer-loop self-consistency (see Sec.~\ref{sec:compscheme}). We expect this to have a negligible effect in the weak-coupling region. For larger $U$, we are only interested in the qualitative behavior, which we do not expect to change.}

\begin{figure}[t]
\includegraphics[width=0.475\textwidth]{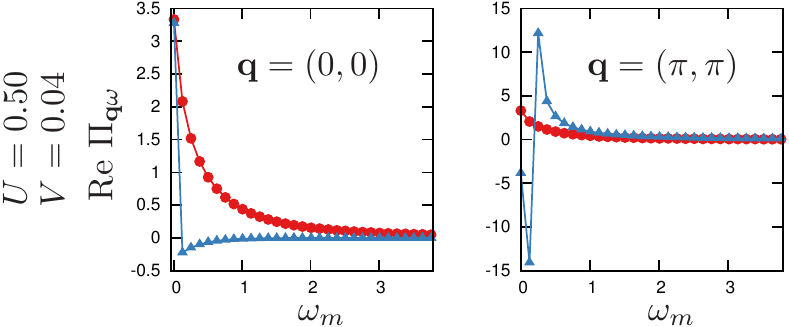} 
\caption{\label{fig:pi_lat_wdep_gw}(Color online) Frequency dependence of the physical polarization $\Pi_{\qv\omega}$ for fixed momentum $\qv=(0,0)$ (left) and $\qv=(\pi,\pi)$ (right) for s-DB (blue, triangles) and EDMFT (red, circles). The colors correspond to those in the phase diagram of Fig.~\ref{fig:phasediagram_db_gw}.
The same quantity in the DB scheme with vertex corrections is shown in Fig.~\ref{fig:pi_lat_wdep}.
} 
\end{figure}

Given that different methods (see Refs.~\onlinecite{Davoudi07,Vonsovsky79} and references therein), including the full DB calculation and lattice Monte Carlo results~\cite{Zhang89}, locate the phase boundary in the vicinity of the line $V=U/4$, we conclude from these results that the frequency-dependence of the three-leg vertex structure is crucial for an adequate determination of the phase boundary.

Neglecting the vertex corrections also comes at the cost of severe artifacts, even at relatively small local interaction $U$. This can be seen in Figs.~\ref{fig:pi_lat_qdep_gw} and \ref{fig:pi_lat_wdep_gw}. In the left panel of Fig.~\ref{fig:pi_lat_wdep_gw} we can see that the physical polarization in s-DB clearly violates the Ward identity. The right panel shows that it turns largely negative at small frequencies (see Fig.~\ref{fig:pi_lat_wdep} for the same quantity in the original DB approach). Nonlocal corrections are hence overestimated and the behavior is non-physical.

\begin{figure}[t]
\includegraphics[width=0.475\textwidth]{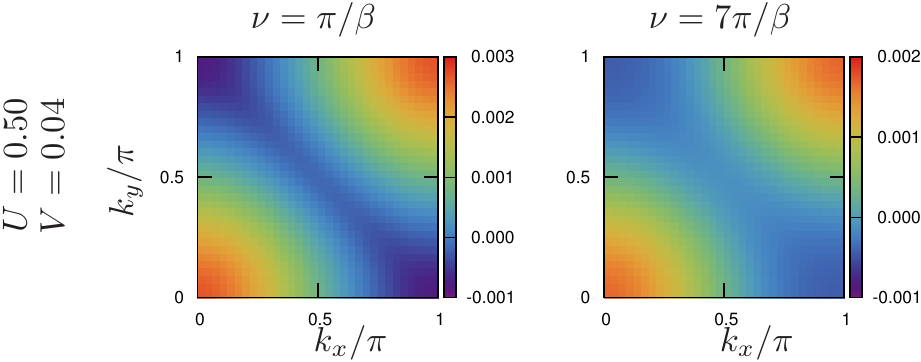} 
\caption{\label{fig:sigma_lat_qdep_gw}(Color online) Momentum dependence of the \emph{nonlocal} part of physical self-energy $\Im \Sigma_{\kv\nu}-\Im \Sigma_{\nu}^{\text{EDMFT}}$ for fixed Matsubara frequencies $\nu=\pi/\beta$ (left) and $\nu=7\pi/\beta$ (right) in s-DB. In EDMFT, this quantity is a constant. 
The corresponding plot with DB data is shown in Fig.~\ref{fig:sigma_lat_qdep}.
} 
\end{figure}

\begin{figure}[t]
\includegraphics[width=0.475\textwidth]{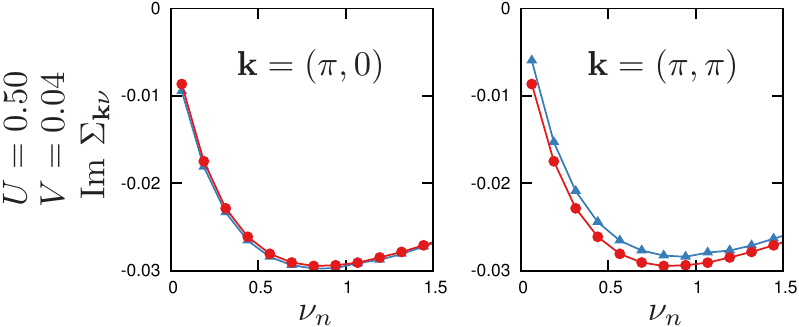} 
\caption{\label{fig:sigma_lat_wdep_gw}(Color online) Frequency dependence of the physical self-energy $\Sigma_{\kv\nu}$ for fixed momentum $\kv=(\pi,0)$ (left) and $\kv=(\pi,\pi)$ (right) in the DB approximation without vertex corrections (blue triangles). For comparison we show the same quantity in EDMFT (red circles).
The colors correspond to the ones used in the phase diagram of Fig.~\ref{fig:phasediagram_db_gw}.
The corresponding DB data are shown in Fig.~\ref{fig:sigma_lat_wdep}.
} 
\end{figure}

In Figs.~\ref{fig:sigma_lat_qdep_gw} and \ref{fig:sigma_lat_wdep_gw} we show the corresponding fermionic self-energy. Compared to the ladder approximation, the magnitude of the nonlocal corrections is larger. Nonlocal corrections hence also appear to be overestimated in the simplified approach.

\section{Conclusions}
\label{sec:summary}

We have presented the first implementation of the DB approach and applied it to the extended Hubbard model with nearest-neighbor interaction.

The DB method interpolates between RPA at low interaction and EDMFT at large interaction, as can be seen from the phase diagram in Fig.~\ref{fig:phasediagram_db_pionly}.
It captures the strong momentum dependence of the polarization, which is important at low to intermediate  interaction and which is absent in EDMFT.
Different choices of diagrammatic corrections have been shown to lead to very similar phase diagrams. They are compatible with the currently available lattice Monte Carlo data on this model.

Examining a simplified version of the DB method, we also concluded that the nontrivial frequency dependence of the local fermion-boson three-leg vertex is very important for the accuracy of the approximation and leads to large differences with EDMFT~+~$GW$ results.

\acknowledgments
The authors acknowledge very helpful discussions with Thomas Ayral, Silke Biermann, Junya Otsuki, Alexey Rubtsov  and Philipp Werner. We would further like to thank Thomas Ayral for providing the EDMFT~+~$GW$ data for comparison. E.G.C.P. v. L. and M.I.K. acknowledge support from ERC Advanced Grant No. 338957 FEMTO/NANO, A.I.L. acknowledges support from the DFG Research Unit FOR 1346 and H.H. and O.P. acknowledge support from the FP7/ERC, under Grant Agreement No. 278472-MottMetals.
Part of the calculations were performed using high-performance computing resources of GENCI-CCRT under Grant No. t2014056112.
The impurity solver and the DB implementation are based on the ALPS libraries~\cite{ALPS2}.

\appendix

\section{Bosonic Hubbard-Stratonovich transformation}
\label{app:hst}

In order for the integral in \eqref{realhst} to be convergent, the matrix $W=\Lambda-V$ must be positive definite, which is not necessarily the case. This may be resolved by adding a sufficiently large constant to $W$ and absorbing it into $\Lambda$. A similar approach was used in Ref.~\onlinecite{Sun02} and shown not to affect physical results. In practice, this constant appears not to be needed. The problem does not exist for a decoupling using a Hubbard-Stratonovich transformation based on complex fields. Convergence of the integral over the complex variables can be ensured by changing the integration path from the real to the imaginary axis for negative elements of the diagonalized matrix $W$ when it is not positive definite~\cite{Rubtsov12}.

\section{Feynman rules of the dual perturbation theory}
\label{app:feynmanrules}

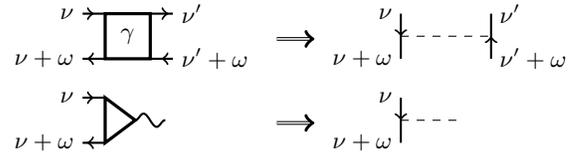
\begin{figure}[t]
  \begin{tikzpicture}
    \coordinate (fermionnw) at (0,0) ; 

    \coordinate (squarenw) at ($(fermionnw)+(0.3,0)$) ; 
    \coordinate (squarene) at ($(squarenw)+(0.6,0)$) ; 
    \coordinate (squarese) at ($(squarene)+(0,-0.6)$) ; 
    \coordinate (squaresw) at ($(squarenw)+(0,-0.6)$) ; 

    \coordinate (fermionne) at ($(squarene)+(0.3,0)$) ; 
    \coordinate (fermionse) at ($(squarese)+(0.3,0.)$) ; 
    \coordinate (fermionsw) at ($(squaresw)+(-0.3,0)$) ; 

    % gamma
    \draw[very thick] (squarenw) -- (squarene) -- (squarese) -- (squaresw) -- cycle ;
    % Fermion line  
    \draw[thick,-<-=0.5] (fermionsw) -- (squaresw);
    % Fermion line  
    \draw[thick,-<-=0.5] (fermionne) -- (squarene);
    % Fermion line  
    \draw[thick,->-=0.5]  (fermionnw) -- (squarenw);
    % Fermion line  
    \draw[thick,->-=0.5]  (fermionse) -- (squarese);
    \node [left] at ($(fermionsw) $) {$\nu+\omega$} ;
    \node [left] at ($(fermionnw) $) {$\nu$} ;
    \node [right] at ($(fermionse) $) {$\nu'+\omega$} ;
    \node [right] at ($(fermionne) $) {$\nu'$} ;
    
    \coordinate (squarew) at ($(squarenw)!0.5!(squaresw)$) ;
    \coordinate (squaree) at ($(squarene)!0.5!(squarese)$) ;    
    \node at ($(squarew)!0.5!(squaree)$) {$\gamma$} ;
    
  \end{tikzpicture}
  \begin{tikzpicture} % This is an arrow, drawn in tikz
   \coordinate (top) at (0,0) ; 
   \coordinate (bottom) at ($(top)+(0,-0.6)$) ;
   \coordinate (leftarr) at ($(top)!0.5!(bottom)$) ;
   \coordinate (rightarr) at ($(leftarr)+(0.5,0)$) ;
   \draw[->,double] (leftarr) -- (rightarr) ;
   \phantom{\node at ($(top) $) {$i$} ; ;} % These are for the vertical alignment of the arrow
   \phantom{\node at ($(bottom) $) {$i$} ; ;} % These are for the vertical alignment of the arrow
  \end{tikzpicture}
  \begin{tikzpicture}
    \coordinate (fermionnw) at (0,0) ; 

    \coordinate (squarenw) at ($(fermionnw)+(0.3,0)$) ; 
    \coordinate (squarene) at ($(squarenw)+(0.6,0)$) ; 
    \coordinate (squarese) at ($(squarene)+(0,-0.6)$) ; 
    \coordinate (squaresw) at ($(squarenw)+(0,-0.6)$) ; 

    \coordinate (fermionne) at ($(squarene)+(0.3,0)$) ; 
    \coordinate (fermionse) at ($(squarese)+(0.3,0.)$) ; 
    \coordinate (fermionsw) at ($(squaresw)+(-0.3,0)$) ; 

    % Fermion line  
    \draw[thick,->-=0.5] (fermionse) -- (fermionne);
    % Fermion line  
    \draw[thick,->-=0.5] (fermionnw) -- (fermionsw);
    \node [left] at ($(fermionsw) $) {$\nu+\omega$} ;
    \node [left] at ($(fermionnw) $) {$\nu$} ;
    \node [right] at ($(fermionse) $) {$\nu'+\omega$} ;
    \node [right] at ($(fermionne) $) {$\nu'$} ;
    
    \draw[dashed] ($ (fermionnw)!.5!(fermionsw)$) -- ($(fermionne)!.5!(fermionse)$) ;
  \end{tikzpicture}\\
  
  \begin{tikzpicture}
    \coordinate (diagram1) at (0,0) ;

    \coordinate (leftboson) at ($(diagram1)$ ) ;  
    \coordinate (rightboson) at ($(leftboson)+(0.4,0)$) ;  

    \coordinate (nw) at ($(leftboson) + (-0.4,0.3)$) ;
    \coordinate (sw) at ($(leftboson) + (-0.4,-0.3)$) ;
    \coordinate (ne) at ($(rightboson) + (0.1,0.3)$) ;
    \coordinate (se) at ($(rightboson) + (0.1,-0.3)$) ;

    \coordinate (fermionn) at ($(nw)+(-0.3,0.0)$) ;
    \coordinate (fermions) at ($(sw)+(-0.3,0.0)$) ;
    
    % Fermion line  
    \draw[thick,->-=0.5] (fermionn) -- (nw);
    % Fermion line  
    \draw[thick,-<-=0.5]  (fermions) -- (sw);

    \draw[very thick] (leftboson) -- (nw) -- (sw) -- cycle;
%    \draw[very thick] (rightboson) -- (ne) -- (se) -- cycle;

    \draw[thick,decorate,decoration=snake] (leftboson) -- (rightboson);  

    \node [left] at ($(fermions) $) {$\nu+\omega$} ;
    \node [left] at ($(fermionn) $) {$\nu$} ;
    \phantom{\node [right] at ($(se) $) {$\nu'+\omega$} ;}
    \phantom{\node [right] at ($(ne) $) {$\nu'$} ;}
    
  \end{tikzpicture}
  \begin{tikzpicture} % This is an arrow, drawn in tikz
   \coordinate (top) at (0,0) ; 
   \coordinate (bottom) at ($(top)+(0,-0.6)$) ;
   \coordinate (leftarr) at ($(top)!0.5!(bottom)$) ;
   \coordinate (rightarr) at ($(leftarr)+(0.5,0)$) ;
   \draw[->,double] (leftarr) -- (rightarr) ;
   \phantom{\node at ($(top) $) {$i$} ; ;} % These are for the vertical alignment of the arrow
   \phantom{\node at ($(bottom) $) {$i$} ; ;} % These are for the vertical alignment of the arrow
  \end{tikzpicture}  
  \begin{tikzpicture}
    \coordinate (diagram1) at (0,0) ;

    \coordinate (leftboson) at ($(diagram1)$ ) ;  
    \coordinate (rightboson) at ($(leftboson)+(0.4,0)$) ;  

    \coordinate (nw) at ($(leftboson) + (-0.4,0.3)$) ;
    \coordinate (sw) at ($(leftboson) + (-0.4,-0.3)$) ;
    \coordinate (ne) at ($(rightboson) + (0.4,0.3)$) ;
    \coordinate (se) at ($(rightboson) + (0.4,-0.3)$) ;

%    \draw[thick,decorate,decoration=snake] (leftboson) -- (rightboson);  

%    \node[shape=circle,draw,fill=black, minimum size=3, inner sep = 0pt] at (leftboson) {};
    \draw[dashed] ($ (nw)!.5!(sw)$) -- (rightboson) ;

    \draw[thick,->-=0.5] (nw) -- (sw) ;
    
    \node [left] at ($(sw) $) {$\nu+\omega$} ;
    \node [left] at ($(nw) $) {$\nu$} ;
    \phantom{\node [right] at ($(se) $) {$\nu'+\omega$} ;}
    \phantom{\node [right] at ($(ne) $) {$\nu'$} ;}
    
  \end{tikzpicture}

\caption{
Illustration of how to determine the sign of a diagram. The vertices (left) in a given diagram should symbolically be replaced by interaction lines as indicated on the right. Each closed fermion loop in the resulting diagram contributes a factor $-1$. See Appendix~\ref{app:feynmanrules} for details.}
  \label{fig:sign}
\end{figure}

\begin{figure}[b]
\subfloat[Second order]{
  \begin{tikzpicture}
  \coordinate (diagram1) at (0,0) ;

\coordinate (leftend) at ($(diagram1)$ ) ;
\coordinate (leftfermion) at ($(leftend) + (-0.3,0)$) ;
\coordinate (toplefttriangle) at ($(leftend) + (0.3,0.4)$) ;
\coordinate (rightlefttriangle) at ($(leftend) + (0.6,0.)$) ;

\coordinate (leftrighttriangle) at ($(rightlefttriangle) + (0.5,0)$) ;
\coordinate (toprighttriangle) at ($(leftrighttriangle) + (0.3,0.4)$) ;
\coordinate (rightend) at ($(leftrighttriangle) + (0.6,0.)$) ;
\coordinate (rightfermion) at ($(rightend) + (0.3,0)$) ;

\draw[very thick] (leftend) -- (toplefttriangle) -- (rightlefttriangle) -- cycle;

\draw[very thick] (rightend) -- (toprighttriangle) -- (leftrighttriangle) -- cycle;

\draw[thick,->-=0.5] (leftfermion) -- (leftend) ;
\draw[thick,-<-=0.5] (rightfermion) -- (rightend) ;

\draw[thick,->-=0.5] (rightlefttriangle) -- (leftrighttriangle);
\draw[thick,decorate,decoration=snake] (toprighttriangle) -- (toplefttriangle);

\node[below] at ($(rightlefttriangle)!0.5!(leftrighttriangle)$) {$\begin{array}{c}\kv+\qv,\\ \nu+\omega\end{array}$} ;
\node[above] at ($(toplefttriangle)!0.5!(toprighttriangle)$) {$\qv, _{\phantom{0}}\omega_{\phantom{0}}$} ;
\end{tikzpicture}
}
\subfloat[Second order]{
  \begin{tikzpicture}
    \coordinate (diagram1) at (0,0) ;

    \coordinate (leftend) at ($(diagram1)$ ) ;
    \coordinate (leftboson) at ($(leftend) + (-0.3,0)$) ;
    \coordinate (toplefttriangle) at ($(leftend) + (0.4,0.3)$) ;
    \coordinate (bottomlefttriangle) at ($(leftend) + (0.4,-0.3)$) ;

    \coordinate (toprighttriangle) at ($(toplefttriangle) + (0.7,0)$) ;
    \coordinate (bottomrighttriangle) at ($(toprighttriangle) + (0,-0.6)$) ;
    \coordinate (rightend) at ($(toprighttriangle) + (0.4,-0.3)$) ;
    \coordinate (rightboson) at ($(rightend) + (0.3,0)$) ;

    \draw[very thick] (leftend) -- (toplefttriangle) -- (bottomlefttriangle) -- cycle;

    \draw[very thick] (rightend) -- (toprighttriangle) -- (bottomrighttriangle) -- cycle;

    \draw[thick,decorate,decoration=snake] (leftboson) -- (leftend) ;
    \draw[thick,decorate,decoration=snake] (rightboson) -- (rightend) ;

    \draw[thick,-<-=0.5] (bottomlefttriangle) -- (bottomrighttriangle);
    \draw[thick,-<-=0.5] (toprighttriangle) -- (toplefttriangle);

    \node[below] at ($(bottomlefttriangle)!0.5!(bottomrighttriangle)$) {$\begin{array}{c}
    \kv+\qv\\
    \nu+\omega^{\phantom{0}}
    \end{array}$} ;
    \node[above] at ($(toplefttriangle)!0.5!(toprighttriangle)$) {$\kv, \nu^{\phantom{0}}$} ;
  \end{tikzpicture}
}
\subfloat[Third order]{
  \begin{tikzpicture}
    \coordinate (diagram2) at (3,0) ;

    \coordinate (leftend) at ($(diagram2)$ ) ;
    \coordinate (leftboson) at ($(leftend) + (-0.3,0)$) ;
    \coordinate (toplefttriangle) at ($(leftend) + (0.4,0.3)$) ;
    \coordinate (bottomlefttriangle) at ($(leftend) + (0.4,-0.3)$) ;

    \coordinate (squarenw) at ($(toplefttriangle) + (0.7,0)$) ;
    \coordinate (squaresw) at ($(squarenw) + (0,-0.6)$) ;
    \coordinate (squarene) at ($(squarenw) + (0.6,0)$) ;
    \coordinate (squarese) at ($(squaresw) + (0.6,0)$) ;

    \coordinate (toprighttriangle) at ($(squarene) + (0.7,0)$) ;
    \coordinate (bottomrighttriangle) at ($(toprighttriangle) + (0,-0.6)$) ;
    \coordinate (rightend) at ($(toprighttriangle) + (0.4,-0.3)$) ;
    \coordinate (rightboson) at ($(rightend) + (0.3,0)$) ;

    \draw[very thick] (leftend) -- (toplefttriangle) -- (bottomlefttriangle) -- cycle;

    \draw[very thick] (rightend) -- (toprighttriangle) -- (bottomrighttriangle) -- cycle;

    \draw[very thick] (squarenw) -- (squarene) -- (squarese) -- (squaresw) -- cycle ;

    \draw[thick,decorate,decoration=snake] (leftboson) -- (leftend) ;
    \draw[thick,decorate,decoration=snake] (rightboson) -- (rightend) ;

    \draw[thick,-<-=0.5] (bottomlefttriangle) -- (squaresw);
    \draw[thick,-<-=0.5] (squarese) -- (bottomrighttriangle);
    \draw[thick,-<-=0.5] (squarenw) -- (toplefttriangle);
    \draw[thick,-<-=0.5] (toprighttriangle) -- (squarene);

    \node[below] at ($(bottomlefttriangle)!0.5!(squaresw)$) {$\begin{array}{c}
    \kv+\qv\\
    \nu+\omega^{\phantom{0}}
    \end{array}$} ;
    \node[below] at ($(squarese)!0.5!(bottomrighttriangle)$) {$\begin{array}{c}
    \kv'+\qv\\
    \nu'+\omega^{\phantom{0}}
    \end{array}$} ;
    \node[above] at ($(toplefttriangle)!0.5!(squarenw)$) {$\kv, \nu$} ;
    \node[above] at ($(squarene)!0.5!(toprighttriangle)$) {$\kv^{\prime},\nu^\prime$} ;
    
    \coordinate (squarew) at ($(squarenw)!0.5!(squaresw)$) ;
    \coordinate (squaree) at ($(squarene)!0.5!(squarese)$) ;    
    \node at ($(squarew)!0.5!(squaree)$) {$\gamma$} ;
    
  \end{tikzpicture}
}
 \caption{\label{fig:app:diagrams}Second-order diagram for the dual fermionic self-energy $\tilde{\Sigma}_{\kv\nu\sigma}$ (a) and second-order (b)  and third-order (c) diagrams contributing to the DB self-energy $\tilde{\Pi}_{\KV\omega}$.}
\end{figure}
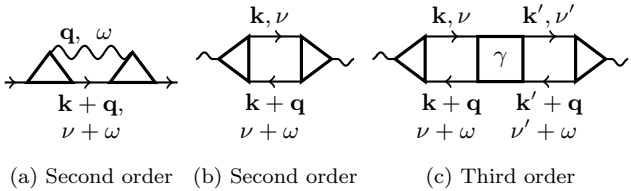

The Feynman rules for evaluating the expression corresponding to a given diagram may be stated as follows:

\begin{itemize}
 \item[(i)] Draw all topologically distinct, connected diagrams involving the elements of Fig.~\ref{fig:feynmanrules}.
 \item[(ii)] With each fermion line associate a dual Green's function $\tilde{G}$.
 \item[(iii)] With each boson line associate a dual Green's function $\tilde{X}$.
 \item[(iv)] With each triangular vertex associate a fermion-boson interaction $\lambda$.
 \item[(v)] With each square vertex associate a fermion-fermion interaction $\gamma$.
 \item[(vi)] Assign a frequency, momentum and spin label to each line, taking into account the conservation of these quantities at each vertex.
 \item[(vii)] Sum over all internal variables. 
 Include a factor $T$ for every frequency summation and $N^{-1}$ for every momentum summation, where $N$ is the number of lattice sites.
 \item[(viii)] Divide the end result by the symmetry factor. 
 Every set of $n$ topologically equivalent lines (i.e. connecting the same two vertices and pointing in the same direction) contributes a factor $n!$ to the symmetry factor.
 \item[(ix)] Determine the sign of the diagram as follows: A diagram for $\tilde{\Pi}$ has a global minus sign resulting from the definition $-\av{\phi\phi}$. Every internal boson line contributes an additional factor $(-1)$. An additional minus sign may arise from fermionic closed loops. To determine this sign, symbolically replace all vertices with interaction lines as illustrated in Fig.~\ref{fig:sign} and count the number of resulting closed fermion loops. Every closed loop contributes a factor $(-1)$.
\end{itemize}

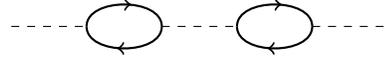
\begin{figure}[t]
   \begin{tikzpicture}
   \coordinate (loop0) at ($(0,0)$) ;
   \coordinate (loop1) at ($(loop0)+(1,0)$) ;
   \coordinate (loop2) at ($(loop1)+(1,0)$) ;
   \coordinate (loop3) at ($(loop2)+(1,0)$) ;
   \coordinate (loop4) at ($(loop3)+(1,0)$) ;
   \coordinate (loop5) at ($(loop4)+(1,0)$) ;
   
   \draw[dashed] (loop0) to (loop1) ;
   \draw[dashed] (loop2) to (loop3) ;
   \draw[dashed] (loop4) to (loop5) ;

   \draw[thick,-<-=0.5,bend right=90] (loop1) to (loop2);
   \draw[thick,-<-=0.5,bend right=90] (loop2) to (loop1);   

   \draw[thick,-<-=0.5,bend right=90] (loop3) to (loop4);
   \draw[thick,-<-=0.5,bend right=90] (loop4) to (loop3);   

  \end{tikzpicture}
  \caption{Illustration of how to count the number of closed fermion loops to determine the sign of the diagram in Fig.~\ref{fig:app:diagrams}(c). The symbolic rules of Fig.~\ref{fig:sign} yield the diagram depicted here. It contains two closed fermion loops, which contribute a factor $(-1)^2$.
  }
  \label{fig:sign3order}
\end{figure}

Let us exemplify these rules for the diagrams shown in the figures. 
The diagram in Fig.~\ref{fig:app:diagrams}(a) straightforwardly evaluates to
\begin{align}
 \tilde{\Sigma}^{(2)}_{\kv\nu\sigma} &= -\frac{T}{N} \sum_{\KV\omega} \lambda^\sigma_{\nu\omega}\tilde{G}^\sigma_{\kv+\KV\nu+\omega} \tilde{X}_{\KV\omega} \lambda^\sigma_{\nu+\omega,-\omega} .\label{app:mixed2}
\end{align}
There is no closed loop, but a single boson line. The overall sign is hence $-1$.

For the second-order diagram to the bosonic self-energy, Fig.~\ref{fig:app:diagrams}(b), we start with a minus sign and account for the one resulting from the closed fermion loop after replacing the vertices by interaction lines. The overall sign of the diagram is hence $+1$. In the third-order diagram [Fig.~\ref{fig:app:diagrams}(c)] we count two closed loops after replacing the vertices by interaction lines as shown in Fig.~\ref{fig:sign3order}. Its sign is hence $-1$. We therefore obtain
\begin{align}
\tilde{\Pi}^{(2)}_{\KV\omega} &=  \frac{T}{N} \sum_{\kv\nu\sigma} \lambda^\sigma_{\nu+\omega,-\omega} \tilde{G}_{\kv\nu\sigma} \tilde{G}_{\kv+\KV\nu+\omega\sigma}\lambda^\sigma_{\nu\omega} \label{app:pi2},\\
\tilde{\Pi}^{(3)}_{\KV\omega} &= -(\frac{T}{N})^2
\sum_{\kv\nu\sigma}\sum_{\kv'\nu'\sigma'}\lambda^{\sigma}_{\nu+\omega,-\omega}
\tilde{G}^\sigma_{\kv\nu} \tilde{G}^\sigma_{\kv+\KV\nu+\omega}  \notag \\
&\phantom{= } 
\qquad\qquad\qquad\times
\gamma^{\sigma\sigma'}_{\nu\nu'\omega}
\tilde{G}^{\sigma'}_{\kv'\nu'} \tilde{G}^{\sigma'}_{\kv'+\KV\nu'+\omega} 
\lambda^{\sigma'}_{\nu'\omega}.\label{pi3}
\end{align}

\section{Instabilities and the Dyson equation}
\label{app:dyson}

Here we show that the Dyson equation, which is a geometric series for the dual susceptibility $\tilde{X}$, diverges at the same point as the physical susceptibility $X$.

The Dyson equation for bosons reads
\begin{align}
\tilde{X}_{\KV\omega}^{-1} &= \tilde{\mathcal{X}}_{\KV\omega}^{-1} - \tilde{\Pi}_{\KV\omega} = \frac{1-\tilde{\Pi}_{\qv\omega}\tilde{\mathcal{X}}_{\qv\omega}}{\tilde{\mathcal{X}}_{\qv\omega}}.
\end{align}
The dual susceptibility $\tilde{X}$ diverges when the numerator in the above equation vanishes,
\begin{align}
1-\tilde{\Pi}_{\qv\omega}\tilde{\mathcal{X}}_{\qv\omega} &= 0.\label{eq:dysondiverges}
\end{align}
The bare dual Green's function \eqref{chidual} can be brought into an alternative form,
\begin{align}
\label{chidualalt}
\tilde{\mathcal{X}}_{\KV\omega}^{-1} = \frac{1}{(\chi_{\omega}^{-1} + \Lambda_{\omega}-V_{\qv})^{-1}-\chi_{\omega}} =\frac{\chi_{\omega}^{-1}+\Lambda_{\omega}-V_{\qv}}{-\chi_{\omega}(\Lambda_{\omega}-V_{\qv})}.
\end{align}
Multiplying Eq.~\eqref{eq:dysondiverges} by an appropriate (non zero) factor and using the form \eqref{chidualalt} eventually leads to
\begin{align}
 \frac{1+\chi_{\omega}(\Lambda_{\omega}-V_{\qv})}{\chi_{\omega}(1+\chi_{\omega}\tilde{\Pi}_{\qv\omega})}\left(1-\tilde{\Pi}_{\qv\omega}\tilde{\mathcal{X}}_{\qv\omega}\right) &= 0\notag \\
 \frac{1+\chi_{\omega}(\Lambda_{\omega}-V_{\qv})}{\chi_{\omega}(1+\chi_{\omega}\tilde{\Pi}_{\qv\omega})}\left(1-\tilde{\Pi}_{\qv\omega}\frac{-\chi_{\omega}(\Lambda_{\omega}-V_{\qv})\chi_{\omega}}{1+(\Lambda_{\omega}-V_{\qv})\chi_{\omega}}\right) &= 0 \notag\\
 \frac{1+\chi_{\omega}(\Lambda_{\omega}-V_{\qv})}{\chi_{\omega}(1+\chi_{\omega}\tilde{\Pi}_{\qv\omega})}+ \tilde{\Pi}_{\qv\omega}\frac{\chi_{\omega}(\Lambda_{\omega}-V_{\qv})\chi_{\omega}}{\chi_{\omega}(1+\chi_{\omega}\tilde{\Pi}_{\qv\omega})} &= 0 \notag\\
 \frac{1}{\chi_{\omega}(1+\chi_{\omega}\tilde{\Pi}_{\qv\omega})} + \Lambda_{\omega} - V_{\qv} &= 0 \notag\\
 X^{-1}_{\qv\omega} &= 0,
\end{align}
where \eqref{xdtox} was used to obtain the last line.
This shows that the physical susceptibility $X$ diverges at the same point as the  dual susceptibility.

\section{Implementation details}
\label{app:impdetails}

\subsection{Impurity solver}

We use a hybridization expansion quantum Monte Carlo impurity solver (CT-HYB)~\cite{Hafermann13}. The solver takes the retarded interaction kernel $K(\tau)$ and its derivative $K'(\tau)$ as input.\footnote{The derivative is required for the improved estimator; see Ref.~\onlinecite{Hafermann14}.} The kernel is defined such that $K''(\tau)=\Lambda(\tau)$, with boundary conditions $K(\tau=0)=0$ and $K(\tau=\beta)=0$.
In order to interface the DB program with the impurity solver, we compute $K(\tau)$ directly from the retarded interaction known on Matsubara frequencies $\Lambda_{\omega}$. We assume that the infinite frequency limit $\lim_{\omega\to\infty}\Lambda_{\omega}\teL\Lambda_{\infty}$ of the latter vanishes. Otherwise, we subtract the tail contribution $\Lambda_{\infty}$ and add it to the instantaneous part of the interaction inside the solver, which needs to be treated separately. With this $\Lambda$, we found the following expressions to give maximal numerical accuracy:
\begin{align}
\label{kfromlambda}
K(\tau)&=\frac{\Lambda_{0}}{2}\left(\frac{\tau}{\beta}-1\right)\tau\!+\!\frac{1}{\beta}\mathop{\sum_{m=-\infty}^{\infty}}_{m\neq 0} \frac{\Lambda_{\iom_{m}}}{(\iom_{m})^{2}} \left(e^{-\iom_{m}\tau}\!-\!1\right),\\
K'(\tau)&=\frac{\Lambda_{0}}{2}\left(\frac{2\tau}{\beta}-1\right)\!-\!\frac{1}{\beta}\mathop{\sum_{m=-\infty}^{\infty}}_{m\neq 0} \frac{\Lambda_{\iom_{m}}}{\iom_{m}} e^{-\iom_{m}\tau}.
\label{kprimefromlambda}
\end{align}
These can be obtained by Fourier transform while treating the static contribution explicitly. In practice we compute the sums with a finite frequency cutoff. The formula \eqref{kfromlambda} has been obtained independently in Ref.~\onlinecite{Otsuki13}. It is valid for $\tau\in [0,\beta]$. This result slightly differs from the formula used in Ref.~\onlinecite{Ayral13} in that the static component $\Lambda_{0}$ is treated separately. This has the advantage that the summand decays faster than $1/(\iom)^{2}$ so that the error due to the finite frequency cutoff of the sum is negligible. 

\subsection{Self-consistency loops; Initial guess}
\label{app:impdetails:scloop}

In practice, the propagators in subsequent inner self-consistency loop iterations are mixed linearly according to
\begin{align}
\tilde{G}_{\kv\nu\sigma}^{\text{new},(n)} &= \xi_{\tilde{G}}\tilde{G}_{\kv\nu\sigma}^{(n)}+(1-\xi_{\tilde{G}})\tilde{G}_{\kv\nu\sigma}^{(n-1)},\notag\\
\tilde{X}_{\qv\omega}^{\text{new},(n)} &= \xi_{\tilde{X}}\tilde{\mathcal{X}}_{\qv\omega}^{(n)} +
(1-\xi_{\tilde{X}})\tilde{\mathcal{X}}_{\qv\omega}^{(n-1)}.
\end{align}
where $\xi_{\tilde{G}},\xi_{\tilde{X}}\in(0,1]$ are mixing parameters in order to avoid oscillations. We typically take $\xi_{\tilde{G}}=\xi_{\tilde{X}}=0.9$. The iterations are stopped once the difference between two successive iterations according to a suitable difference measure\footnote{We use the measure $d=(T/N)\sum_{\kv\nu\sigma}\abs{\tilde{G}_{\kv\nu\sigma}^{(n)} - \tilde{G}_{\kv\nu\sigma}^{(n-1)}}$ and an analogous expression for the bosonic Green's function.}
becomes smaller than some small number $\varepsilon$. Apart from the first few iterations, the convergence of this inner loop is found to be exponential. The number of required iterations increases in the vicinity of phase transitions.

In the outer self-consistency loop, we update the hybridization function and retarded interaction according to the rules
\begin{align}
\label{eq:updateG}
 \Delta_{\nu\sigma}^{\text{new}} &= \Delta_{\nu\sigma}^{\text{old}} + \xi_{\Delta}[g^{-1}_{\nu\sigma} \tilde{G}^{\text{local}}_{\nu\sigma} (G_{\nu\sigma}^{\text{local}})^{-1}],\\
\label{eq:updateX}
 \Lambda_\omega^{\text{new}} &= \Lambda_\omega^{\text{old}} + \xi_{\Lambda}[\chi^{-1}_\omega \tilde{X}^{\text{local}}_{\omega} (X_{\omega}^{\text{local}})^{-1} ].
\end{align}
Here $G^{\text{local}}_{\nu\sigma}$ and $X^{\text{local}}_\omega$ denote the local parts of the lattice propagators and $\tilde{G}^{\text{local}}_{\nu\sigma}$ and $\tilde{X}^{\text{local}}_\omega$ the local parts of the dual propagators.\footnote{Note that $G^{\text{local}}_{\nu\sigma}$ and $X^{\text{local}}_\omega$ as well as $g_{\nu\sigma}$ and $\chi_{\omega}$ merely serve as scaling factors in these equations: Inserting the EDMFT relations $\tilde{\mathcal{G}}=G^{\text{EDMFT}}-g$ and $\tilde{\mathcal{X}}=X^{\text{EDMFT}}-\chi$ in place of $\tilde{G}$ and $\tilde{X}$ one sees  that the terms in angular brackets reduce to $g^{-1}-G^{-1}$ and $\chi^{-1}-X^{-1}$, respectively.}
These equations have the same fixed point as the self-consistency conditions \eqref{eq:gsc} and \eqref{eq:xsc}. 
Typical values for the mixing parameters of the update are $\xi_{\Delta}=\xi_{\Lambda}=0.9$. The convergence is well behaved and similar as one is used to from DMFT.

As an initial guess for EDMFT or DB calculations, one may start from  $\Lambda_{\omega}=0$ and a hybridization function corresponding to the noninteracting case, i.e., $\Delta_{\nu\sigma}$ is given by Eq.~\eqref{eq:updateG} with $\Delta_{\nu\sigma}^{\text{old}}=0$, $\Sigma_{\nu\sigma}=\Sigma_{\nu\sigma}^{\text{imp}}=0$ and $\xi_{\Delta}=1$.

\subsection{Symmetry relations for vertices; Frequency cutoffs}
\label{app:impdetails:symmetries}

In order to reduce computation time, we exploit the symmetries of the vertex functions.
The vertices need to be computed for positive bosonic  frequencies $\omega_m\geq 0$ only, i.e. for $m = 0,\ldots, N_\omega -1$, where $N_{\omega}$ is the number of frequencies. For negative $\omega$, they are related to their values at positive frequencies by $(\lambda^{\sigma}_{\nu,-\omega})^{*}=\lambda^{\sigma}_{-\nu,\omega}$ and $(\gamma^{\sigma\sigma'}_{\nu,\nu',-\omega})^{*}=\gamma^{\sigma\sigma'}_{-\nu,-\nu',\omega}$, which follow from the definition of the Fourier transform. 
In our implementation, we further use the following symmetry relations for the vertex functions $\lambda$ and $\gamma$:
\begin{align}
\label{lambdasym}
(\lambda^{\sigma}_{\nu,\omega})^{*} &= \lambda^{\sigma}_{-\nu-\omega,\omega},\\
\label{gammasym}
(\gamma^{\sigma\sigma'}_{\nu\nu'\omega})^{*} &= \gamma^{\sigma'\sigma}_{-\nu'-\omega,-\nu-\omega,\omega}.
\end{align}
They are derived in Appendix \ref{app:symmetries} below. These relations do not assume particle-hole symmetry. Note the exchange of spin labels in the second line.

One sees that for a given bosonic frequency $\omega_{m}$ the vertex $\lambda$ as a function of \emph{fermionic} frequency is symmetric around $-\omega_{m}/2$. This can directly be observed in Figs.~\ref{fig:lambda_edmft_vertical}--\ref{fig:lambda_edmft_v06}. As a result, for a given bosonic frequency $\omega_{m}$ the vertex only needs to be computed for fermionic frequency $\nu_{n} \geq -\omega_{m}/2$. In terms of indices, $n\geq -(m-1)/2$ for $m$ odd and $n\geq-m/2$ for $m$ even, respectively. 

The diagrams, such as Eqs.~\eqref{pi2} and \eqref{pi3} contain convolutions of Green's functions which are of the form $\tilde{G}_{\nu}\tilde{G}_{\nu+\omega}$. Important contributions to these diagrams stem from those combinations of frequencies for which the product is large. This is the case for $\nu\approx 0$ and $\nu+\omega\approx 0$, because $\tilde{G}$ is largest for small frequencies [it decays as $1/(\inu)^{2}$]. 
The dominant contributions $\nu\approx0$ and $\nu\approx -\omega$ lie symmetrically around $-\omega/2$.
If one computes the vertex functions on an interval for the fermionic frequencies which is symmetric around 0, i.e., $n=-N_{\nu},\ldots N_{\nu}-1$, where $N_{\nu}$ is the cutoff frequency, the symmetry relations ensure that the dominant contribution at $\nu\approx -\omega$ is captured even for bosonic frequencies $m>N_{\nu}$. 

For larger $\omega$, the dominant contributions are less strongly peaked at $\nu\approx 0$ and $\nu+\omega\approx 0$. It is important to take a sufficiently large number of fermionic frequencies into account to obtain the correct large $\omega$ behavior of the diagrams.

\subsection{Optimized evaluation of diagrams}

The diagrammatic expressions can be decomposed into a sequence of convolutions in momentum space: Consider, for example, the diagram \eqref{mixed2}. Notwithstanding the frequency summations, it contains the convolution $\sum_{\qv} \tilde{G}_{\kv+\qv} \tilde{X}_{\qv}$. $\tilde{X}_{\qv}$ in turn involves diagrammatic corrections such as the one in \eqref{pi2}, which also contains a convolution $\sum_{\kv} \tilde{G}_{\kv}\tilde{G}_{\kv+\qv}$. We compute these using fast convolution. This amounts to, e.g., computing $\tilde{G}_{\rv}$ in real space using an FFT and then computing the FFT of $\tilde{G}_{\rv}\tilde{G}_{-\rv}$. With FFTs, we can compute diagrams at a cost that scales as $N\log N$ with the number of \kv~points $N$ instead of at least $N^2$ for straightforward momentum summations.

We gain additional speedup by using the lattice symmetries. The Bethe-Salpeter equation, which is introduced below, has to be evaluated for all points in the Brillouin zone and the evaluation is computationally costly. Using lattice symmetries, we can reduce the computational effort by roughly a factor of $\sim 8$ for the two-dimensional square lattice and a factor $\sim 48$ in the three-dimensional cubic lattice.

\subsection{Parallelization}

We use the message passing interface (MPI) for the parallelization.
The Monte Carlo impurity solver is parallelized in the standard way by running different Markov chains on different threads. The evaluation of the DB equations is most straightforwardly parallelized over the bosonic frequencies. The Bethe-Salpeter equation \eqref{eq:bse}, the diagrammatic expressions for $\tilde{\Pi}_\omega$ \eqref{pi2}, \eqref{pi3} and contributions to $\tilde{\Sigma}$ from individual bosonic frequencies may be computed independently for different $\omega$. Communication between threads is only needed at the end of each inner renormalization iteration when the results and the contributions to the self-energies from different bosonic frequencies are collected on the master.

\section{Vertex functions}
\label{app:lambda}

\begin{figure}[t]
\begin{center}
\includegraphics[scale=0.6,angle=0]{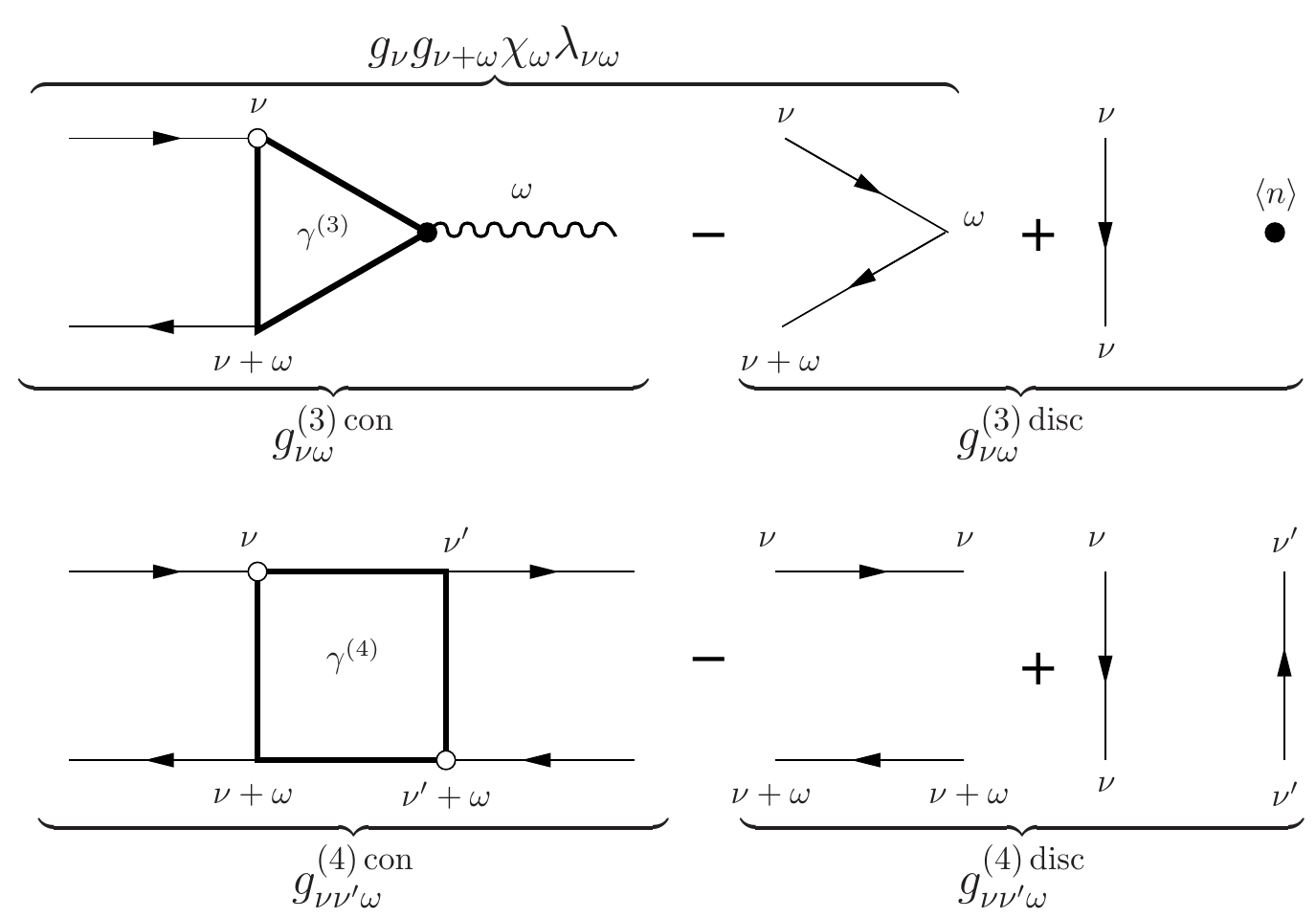} 
\end{center}
\caption{\label{vertices} Diagrammatic representation of the three-leg $g^{(3)}$ (top panel) and four-leg correlation function $g^{(4)}$ (bottom panel) and the definition of the vertex functions in terms of them.
The figure illustrates the relation between the three-leg and four-leg vertices $\gamma^{(4)}$ and $\lambda$. Wavy lines represent the charge susceptibility and straight lines with arrows denote fully dressed single-particle Green's functions.
}
\end{figure}

In the DB approach, we have two types of interaction vertices: a fermion-fermion (four-leg) and a fermion-boson (three-leg) vertex. They are local and computed from the QIM.
Their relation and definition in terms of the impurity correlation functions $g^{(3)}$ and $g^{(4)}$ are illustrated diagrammatically in Fig. \ref{vertices}. The four-leg vertex $\gamma^{(4)}$ is obtained from the connected part $g^{\text{(4)con}}$ of the two-particle Green's function by amputating the legs. $g^{\text{(4)con}}$ [the numerator in \eqref{gammadef}], in turn, is obtained by subtracting the disconnected part $g^{\text{(4)disc}}$ from the two-particle Green's function $g^{\text{(4)}}$.

The structure of the three- and four-leg correlation functions is very similar. The three-leg vertex $\lambda$, however, is not obtained by amputating the legs from the connected part (this would give $\gamma^{(3)}$). The reason is essentially that in the decoupling of the interaction, Eq.~\eqref{realhst}, the field $\phi$ couples to the density $n$ as an entity and not to $c^{*}$ and $c$ individually. As a consequence, the three-leg vertex is non zero even in absence of interaction. In terms of the connected correlation function, the three-leg vertex is given by
\begin{align}
\label{lambdafromcon}
\lambda_{\nu\omega}^{\sigma} &= \frac{g_{\nu\omega}^{\sigma(3)\text{con}} }{g_{\nu\sigma}g_{\nu+\omega,\sigma}\chi_{\omega}} -\frac{1}{\chi_{\omega}}.
\end{align}
In the noninteracting case, $\gamma^{(3)}$ and the connected part $g^{(3)\text{con}}$ vanish and it follows that $\lambda_{\nu\omega}^{\sigma}=-\chi_{\omega}^{-1}$.

A relation between the fermion-boson vertex and $\gamma$ may be derived by replacing $n_\omega = \beta^{-1}\sum_\omega c^{*}_{\nu'} c_{\nu'+\omega}$ in the expectation value of $g^{(3)}$~\cite{Rubtsov12},
\begin{align}
\label{app:ggchilambda}
g_{\nu+\omega\sigma}g_{\nu\sigma}\chi_\omega \lambda^{\sigma}_{\nu\omega} =& -\av{c^{\phantom{*}}_{\nu\sigma}c^{*}_{\nu+\omega\sigma}n_\omega} - \beta g_{\nu\sigma}\av{n}\delta_{\omega}\notag\\
=& -\frac{1}{\beta}\sum_{\sigma'\nu'}\av{c^{\phantom{*}}_{\nu\sigma}c^{*}_{\nu+\omega\sigma}c^{*}_{\nu'\sigma'}c^{\phantom{*}}_{\nu'+\omega\sigma'}} \notag \\
\phantom{=}&-g_{\nu\sigma}\sum_{\sigma'\nu'}\av{c^{*}_{\nu'\sigma'}c^{\phantom{*}}_{\nu'+\omega\sigma'}}\delta_{\omega}
\notag\\
=& +\frac{1}{\beta}\sum_{\sigma'\nu'}g^{(4)\sigma\sigma'}_{\nu\nu'\omega} -g_{\nu\sigma}\sum_{\sigma'\nu'}g_{\nu'\sigma'}\delta_{\omega},
\end{align}
where we have used the definition \eqref{chi4def} of the two-particle correlation function, $g^{(4)\sigma\sigma'}_{\nu\nu'\omega}\Let\av{c^{\phantom{*}}_{\nu\sigma}c^{*}_{\nu+\omega\sigma}c^{\phantom{*}}_{\nu'+\omega\sigma'}c^{*}_{\nu'\sigma'}}$
Taking the definition \eqref{gammadef} of the vertex $\gamma$ and summing over $\sigma',\nu'$,  we can write
\begin{align}
&\frac{1}{\beta}\sum_{\sigma'\nu'}g^{(4)\sigma\sigma'}_{\nu\nu'\omega}
-g_{\nu\sigma}\sum_{\sigma'\nu'}g_{\nu'\sigma'}\delta_{\omega}+g_{\nu+\omega\sigma}g_{\nu\sigma}\notag\\
&=g_{\nu+\omega\sigma}g_{\nu\sigma}\frac{1}{\beta}\sum_{\sigma'\nu'}\gamma^{\sigma\sigma'}_{\nu\nu'\omega}g_{\nu'\sigma'}g_{\nu'+\omega\sigma'} 
\label{app:gggggamma}.
\end{align}
Combining \eqref{app:ggchilambda} and \eqref{app:gggggamma} yields the desired relation:
\begin{align}
\lambda_{\nu\omega}^{\sigma} = \chi^{-1}_{\omega} \left(\frac{1}{\beta}\sum_{\sigma'\nu'}\gamma^{\sigma\sigma'}_{\nu\nu'\omega}g_{\nu'\sigma'}g_{\nu'+\omega\sigma'} -1\right).
\label{app:lambdafromgamma}
\end{align}

\section{Symmetry relations}
\label{app:symmetries}

The symmetry relation \eqref{gammasym} for the fermion-fermion vertex can be proven starting from the definition of the Fourier transform of the two-particle correlation function,
\begin{align}
\label{app:ftg4def}
g^{(4)\sigma\sigma'}_{\nu,\nu',\omega} \Let &
\frac{1}{\beta}\int d\tau_i \av{c^{\phantom{*}}_{\sigma}(\tau_1)c^*_{\sigma}(\tau_2)c^{\phantom{*}}_{\sigma'}(\tau_3)c^*_{\sigma'}(\tau_4)}\notag\\
&\times e^{i\left[ \nu\tau_1 - (\nu+\omega)\tau_2+(\nu'+\omega)\tau_3-\nu'\tau_4\right]}.
\end{align}
Taking the conjugate, commuting the Grassmann variables and relabeling $\tau_1 \leftrightarrow \tau_3$ and $\tau_2 \leftrightarrow \tau_4$, we obtain
\begin{align}
\left(g^{(4)\sigma\sigma'}_{\nu,\nu',\omega}\right)^{*} = &
\frac{1}{\beta}\int d\tau_i \av{c^{\phantom{*}}_{\sigma}(\tau_1)c^*_{\sigma}(\tau_2)c^{\phantom{*}}_{\sigma'}(\tau_3)c^*_{\sigma'}(\tau_4)}\notag\\
&\times e^{-i\left\{ \nu\tau_1 - (\nu+\omega)\tau_2+(\nu'+\omega)\tau_3-\nu'\tau_4\right\}}\notag\\
=&\frac{1}{\beta}\int d\tau_i \av{c^{\phantom{*}}_{\sigma'}(\tau_3)c^*_{\sigma'}(\tau_4)c^{\phantom{*}}_{\sigma}(\tau_1)c^*_{\sigma}(\tau_2)}\notag\\
&\times e^{-i\left\{(\nu'+\omega)\tau_3-\nu'\tau_4+ \nu\tau_1 - (\nu+\omega)\tau_2\right\}}\notag\\
=&\frac{1}{\beta}\int d\tau_i \av{c^{\phantom{*}}_{\sigma'}(\tau_1)c^*_{\sigma'}(\tau_2)c^{\phantom{*}}_{\sigma}(\tau_3)c^*_{\sigma}(\tau_4)}\notag\\
&\times e^{-i\left\{(\nu'+\omega)\tau_1-\nu'\tau_2+ \nu\tau_3 - (\nu+\omega)\tau_4\right\}}\notag\\
=&\frac{1}{\beta}\int d\tau_i \av{c^{\phantom{*}}_{\sigma'}(\tau_1)c^*_{\sigma'}(\tau_2)c^{\phantom{*}}_{\sigma}(\tau_3)c^*_{\sigma}(\tau_4)}\notag\\
&\times e^{i\left\{[-(\nu'+\omega)]\tau_1-(-\nu')\tau_2 + (-\nu)\tau_3 - [-(\nu+\omega)]\tau_4\right\}}\notag\\
\teL & g^{(4)\sigma'\sigma}_{-(\nu'+\omega),-(\nu+\omega),\omega},
\end{align}
where the last equality follows from comparison with the definition of the Fourier transform \eqref{app:ftg4def}.
It remains to show the analogous relations for the disconnected part and the denominator in the definition \eqref{gammadef} of the vertex function.
We treat the two terms in the disconnected part separately:
\begin{align}
G^{(4)\,\sigma\sigma',\text{disc}}_{\nu\nu'\omega}&\Let  \beta g_{\nu\sigma}g_{\nu+\omega\sigma}\delta_{\nu\nu'}\delta_{\sigma\sigma'} - \beta g_{\nu\sigma}g_{\nu'\sigma'}\delta_{\omega}\notag\\
&\teL G^{(a)\,\sigma\sigma',\text{disc}}_{\nu\nu'\omega} - G^{(b)\,\sigma\sigma',\text{disc}}_{\nu\nu'\omega}.
\end{align}
Using that $(g_{\nu,\sigma})^{*} = g_{-\nu,\sigma}$, we obtain
\begin{align}
(G^{(a)\,\sigma\sigma',\text{disc}}_{\nu\nu'\omega})^{*} &=  \beta g_{-\nu,\sigma}g_{-\nu-\omega,\sigma}\delta_{\nu\nu'}\delta_{\sigma\sigma'}\notag\\
&= \beta g_{-\nu',\sigma'}g_{-\nu'-\omega,\sigma'}\delta_{\nu\nu'}\delta_{\sigma\sigma'}\notag\\
&= \beta g_{-\nu'-\omega,\sigma'}g_{-\nu',\sigma'}\delta_{\nu'\nu}\delta_{\sigma'\sigma}\notag\\
&= \beta g_{-\nu'-\omega,\sigma'}g_{-\nu',\sigma'}\delta_{-\nu'-\omega,-\nu-\omega}\delta_{\sigma'\sigma} \notag\\
&\teL G^{(a)\,\sigma'\sigma,\text{disc}}_{-\nu'-\omega,-\nu-\omega,\omega},\notag\\
(G^{(b)\,\sigma\sigma',\text{disc}}_{\nu\nu'\omega})^{*}  &= \beta g_{-\nu\sigma}g_{-\nu'\sigma'}\delta_{\omega}\notag\\
&= \beta g_{-\nu'\sigma'}g_{-\nu\sigma}\delta_{\omega}\notag\\
&= \beta g_{-\nu'-\omega\sigma'}g_{-\nu-\omega\sigma}\delta_{\omega}\notag\\
&\teL G^{(b)\,\sigma'\sigma,\text{disc}}_{-\nu'-\omega,-\nu-\omega,\omega}.
\end{align}
Let us abbreviate the product in the denominator of the vertex function \eqref{gammadef} by $\pi^{\sigma\sigma'}_{\nu\nu'\omega}$:
\begin{align}
\pi^{\sigma\sigma'}_{\nu\nu'\omega} \Let g_{\nu,\sigma}g_{\nu+\omega,\sigma}g_{\nu',\sigma'}g_{\nu'+\omega,\sigma'}.
\end{align}
Under conjugation we find
\begin{align}
(\pi^{\sigma\sigma'}_{\nu\nu'\omega})^{*} &\Let g_{-\nu,\sigma}g_{-\nu-\omega,\sigma}g_{-\nu',\sigma'}g_{-\nu'-\omega,\sigma'}\notag\\
&= g_{-\nu'-\omega,\sigma'}g_{-\nu',\sigma'}g_{-\nu,\sigma}g_{-\nu-\omega,\sigma}\notag\\
&\teL \pi^{\sigma'\sigma}_{-\nu'-\omega,-\nu-\omega,\omega}.
\end{align}
Because the same symmetry relations hold for all constituents of the vertex function, we can write
\begin{align}
(\gamma^{\sigma\sigma'}_{\nu,\nu',\omega})^{*}
&=\gamma^{\sigma'\sigma}_{-\nu'-\omega,-\nu-\omega,\omega} \label{eq:gammasym:app}.
\end{align}

The corresponding symmetry relation for the electron-boson vertex $\lambda$ can be derived by similar means. Alternatively, we can use the above symmetry relation for the electron-electron vertex $\gamma$ by employing the relation \eqref{app:lambdafromgamma} between $\lambda$ and  $\gamma$ . With $\chi_{\omega}^{*}=\chi_{\omega}$ we have
\begin{align}
 (\lambda^\sigma_{\nu,\omega})^\ast
 &=  \frac{1}{\chi_{\omega}} \left[\frac{1}{\beta}\sum_{\nu'\sigma'}(\gamma^{\sigma\sigma'}_{\nu\nu'\omega})^{*}(g_{\nu'\sigma'})^{*}(g_{\nu'+\omega,\sigma'})^{*} -1\right] \notag\\
 &=  \frac{1}{\chi_{\omega}} \left[\frac{1}{\beta}\sum_{\nu'\sigma'}\gamma^{\sigma'\sigma}_{-\nu'-\omega,-\nu-\omega,\omega}g_{-\nu'\sigma'}g_{-\nu'-\omega,\sigma'}\! -\!1\right]\!.
\end{align}
Shifting the summation variable according to $\nu' \to -\nu'-\omega$, we obtain
\begin{align}
(\lambda^\sigma_{\nu,\omega})^\ast
&=  \frac{1}{\chi_{\omega}} \left[\frac{1}{\beta}\sum_{\nu'\sigma'}\gamma^{\sigma'\sigma}_{\nu',-\nu-\omega,\omega}g_{\nu'+\omega,\sigma'}g_{\nu'\sigma'} -1\right].
\end{align}
Time-reversal symmetry finally implies~\cite{Rohringer12,Brown94} 
$\gamma^{\sigma\sigma'}_{\nu\nu'\omega} = \gamma^{\sigma'\sigma}_{\nu'\nu\omega}$, which gives the desired relation
\begin{align}
(\lambda^\sigma_{\nu,\omega})^\ast
&=  \frac{1}{\chi_{\omega}} \left[\frac{1}{\beta}\sum_{\nu'\sigma'}\gamma^{\sigma\sigma'}_{-\nu-\omega,\nu',\omega}g_{\nu'\sigma'}g_{\nu'+\omega,\sigma'} -1\right]
\notag\\
&\teL \lambda^\sigma_{-\nu-\omega,\omega}.
\end{align}

\section{Approximation without vertex corrections}
\label{app:sigmagw}

\subsection{Polarization}

We construct a simplified approximation by letting $\gamma\equiv 0$. In this case, $\lambda_{\nu\omega}=-\chi_{\omega}^{-1}$,  according to  \eqref{app:lambdafromgamma}.
Inserting this into the second-order approximation for the dual bosonic self-energy, Eq. \eqref{pi2} and Fig.~\ref{fig:functional_bubble}, one obtains
\begin{align}
\label{app:pi2approx}
\tilde{\Pi}_{\qv\omega} = \frac{T}{N}\chi_{\omega}^{-1}\sum_{\kv\nu\sigma}\tilde{\mathcal{G}}_{\kv+\qv\nu+\omega}\tilde{\mathcal{G}}_{\kv\nu}\chi_{\omega}^{-1}.
\end{align}
Using $\tilde{\mathcal{G}}=G^{\text{EDMFT}}-g$, which follows from \eqref{gdual} and \eqref{gedmft}, we can rewrite the convolution as (the label EDMFT is omitted in the following)
\begin{align}
\label{app:ggnonloc}
\frac{T}{N}\sum_{\kv\nu\sigma}\tilde{\mathcal{G}}_{\kv\nu}\tilde{\mathcal{G}}_{\kv+\qv\nu+\omega} = &\frac{T}{N}\sum_{\kv\nu\sigma}G_{\kv\nu}G_{\kv+\qv\nu+\omega}\notag\\
& - T\sum_{\nu\sigma}g_{\nu}g_{\nu+\omega} \teL \{GG\}_{\qv\omega}^{\text{nonloc}}.
\end{align}
The result can be interpreted as the nonlocal part of the bubble because for a self-consistent EDMFT solution, the local part of the lattice Green's function $G$ equals the impurity Green's function $g$, i.e. $(1/N)\sum_{\kv}G_{\kv\nu}=g_{\nu}$.
In order to obtain an expression for the physical polarization $\Pi$ in terms of the bosonic self-energy $\tilde{\Pi}$, we use \eqref{eq:pifromdual}
\begin{align}
\Pi_{\qv\omega}^{-1} = - (\chi_\omega+\chi_{\omega}\tilde{\Pi}_{\KV\omega}\chi_{\omega})^{-1}-\Lambda_{\omega}.
\end{align}
Inserting the above result for $\tilde{\Pi}$ yields
\begin{align}
\label{app:pi_our}
\Pi_{\qv\omega}^{-1} =-(\chi_{\omega}+\{GG\}_{\qv\omega}^{\text{nonloc}})^{-1}-\Lambda_{\omega}.
\end{align}
The susceptibility can be decomposed into a bubble contribution and a part containing local vertex corrections $\Delta\chi_{\omega}$,
\begin{align}
\label{app:chidecomp1}
\chi_{\omega}=& T\sum_{\nu\sigma}g_{\nu}g_{\nu+\omega} + \Delta\chi_{\omega},
\end{align}
where
\begin{align}
\label{app:chidecomp2}
\Delta\chi_{\omega} \Let 
T^{2}\sum_{\nu\nu'\sigma}g_{\nu}g_{\nu+\omega}\gamma_{\nu\nu'\omega}g_{\nu'}g_{\nu'+\omega}.
\end{align}
Inserting \eqref{app:chidecomp1} into \eqref{app:pi_our}, we can write the polarization in the form
\begin{align}
\label{app:pivertexcorr}
\Pi_{\qv\omega}^{-1} =-(\{GG\}_{\qv\omega}+\Delta\chi_{\omega})^{-1}-\Lambda_{\omega},
\end{align}
where the term in brackets is a lattice bubble with local vertex corrections added.

In EDMFT~+~$GW$, the polarization is the sum of the local EDMFT part and the nonlocal part of the lattice bubble~\cite{Ayral13}. In our notation, it reads:
\begin{align}
\Pi_{\qv\omega}^{-1} = \left[\Big(-\chi_{\omega}^{-1} - \Lambda_{\omega}\Big)^{-1} - \{GG\}_{\qv\omega}^{\text{nonloc}}\right]^{-1},
\label{app:pi_edmftgw}
\end{align}
where the first term is the impurity polarization
\begin{align}
\label{app:pimp}
\Pi_{\omega}^{\text{imp}} \Let (-\chi_{\omega}^{-1} - \Lambda_{\omega})^{-1}=\frac{-\chi_{\omega}}{1+\Lambda_{\omega}\chi_{\omega}}. 
\end{align}
The two expressions \eqref{app:pi_our} and \eqref{app:pi_edmftgw} obviously give different results in general. In the weak-coupling limit  ($U$ and $V$ small), when $\Lambda$ is small, they give the same result. In this limit, $\Delta\chi_{\omega}$ is negligible and $\chi_{\omega}$ can be approximated by the bare local bubble according to \eqref{app:chidecomp1}. The polarization in both approximations correspondingly reduces to $\Pi_{\qv\omega}\approx-\{GG\}_{\qv\omega}$, i.e., to the (bare) lattice bubble. This corresponds to RPA.

\subsection{Fermionic self-energy}

In EDMFT~+~$GW$, the fermionic self-energy is decomposed into the local impurity part and a nonlocal correction from $GW$ diagrams. In DB, there is an analogous decomposition~\cite{Rubtsov12}:
\begin{align}
\Sigma_{\kv\nu} &= \Sigma^{\text{imp}}_{\nu} + \frac{\tilde{\Sigma}_{\kv\nu}}{1+g_{\nu}\tilde{\Sigma}_{\kv\nu}}. \label{eq:sigmaprime}
\end{align}
For weak coupling, the dual self-energy and the denominator are small, i.e., $\Sigma_{\kv\nu} \approx \Sigma^{\text{imp}}_{\nu} + \tilde{\Sigma}_{\kv\nu}$. The nonlocal part is then given by the dual self-energy itself. 
Letting $\lambda_{\nu\omega}\approx -\chi_{\omega}^{-1}$ as for the polarization, we obtain for the second-order diagram of Fig.~\ref{fig:functional_bubble} and Eq.~\eqref{mixed2}:
\begin{align}
\label{app:sigma2}
 \tilde{\Sigma}_{\kv\nu} &= -\frac{T}{N} \sum_{\KV\omega} \chi_{\omega}^{-1}\tilde{G}_{\kv+\KV\nu+\omega} \tilde{X}_{\KV\omega} \chi_{\omega}^{-1}.
\end{align}
In analogy to the $GW$ approximation, which is constructed using the renormalized interaction, we evaluate it with the renormalized bosonic propagator $\tilde{X}$. Using Dyson's equation,
\begin{align}
\tilde{X}_{\KV\omega}^{-1} &= \tilde{\mathcal{X}}_{\KV\omega}^{-1} - \tilde{\Pi}_{\KV\omega},
\end{align}
as well as \eqref{chidualalt} and the result \eqref{app:pi2approx} and \eqref{app:ggnonloc} for the polarization, we obtain for $\tilde{X}_{\KV\omega}^{-1}$:
\begin{align}
\tilde{\mathcal{X}}_{\KV\omega}^{-1} - \tilde{\Pi}_{\KV\omega} &= \frac{\chi_{\omega}^{-1}-(V_{\qv}-\Lambda_{\omega})}{\chi_{\omega}(V_{\qv}-\Lambda_{\omega})}-\frac{\{GG\}_{\qv\omega}^{\text{nonloc}}}{\chi_{\omega}\chi_{\omega}}\notag\\
&=\frac{1-(V_{\qv}-\Lambda_{\omega})\chi_{\omega}-(V_{\qv}-\Lambda_{\omega})\{GG\}_{\qv\omega}^{\text{nonloc}}}{\chi_{\omega}(V_{\qv}-\Lambda_{\omega})\chi_{\omega}}.
\end{align}
Inserting this into \eqref{app:sigma2} gives the result
\begin{align}
\label{app:sigmasdb}
 \tilde{\Sigma}_{\kv\nu} =& -\frac{T}{N} \sum_{\KV\omega}\tilde{\mathcal{G}}_{\kv+\KV\nu+\omega}\notag\\
 &\qquad \times\frac{(V_{\qv}-\Lambda_{\omega})}{1+(V_{\qv}-\Lambda_{\omega})(-\chi_{\omega}-\{GG\}_{\qv\omega}^{\text{nonloc}})}.
\end{align}
The fraction in the second line has the form of a screened interaction $V_{\qv}-\Lambda_{\omega}$ with a polarization given by $-\chi_{\omega}-\{GG\}_{\qv\omega}^{\text{nonloc}}$. As noted above, it corresponds to a bubble of lattice Green's functions with local vertex corrections added.
The corresponding EDMFT~+~$GW$ result of Ref.~\onlinecite{Ayral13}, in our notation and with our conventions, reads
\begin{align}
\label{app:sigmaedmftgw}
\tilde{\Sigma}_{\kv\nu} =& -\frac{T}{N} \sum_{\KV\omega}G_{\kv+\KV\nu+\omega} \frac{V_{\qv}}{1+V_{\qv}(\Pi^{\text{imp}}_{\omega}-\{GG\}_{\qv\omega}^{\text{nonloc}})}.
\end{align}
Here we do not consider an additional term of order $U^{2}$, which is irrelevant in the low-$U$ region of the phase diagram on which we want to focus. A similar contribution arises in the DB approach from the local fermion-fermion vertex $\gamma$.
There are three main differences. (i) The impurity polarization $\Pi_{\omega}^{\text{imp}}$ in the denominator in EDMFT~+~$GW$ is replaced by $-\chi_{\omega}$ in DB.
According to \eqref{app:pimp}, $\Pi^{\text{imp}}$ contains $\chi_{\omega}$ to leading order, and they will be similar to $\chi_{\omega}$ when $\Lambda_{\omega}$ is small.
(ii) The full lattice Green's function enters the EDMFT~+~$GW$ self-energy, while in the DB self-energy, it is only its nonlocal part. (iii) The interaction in DB is given by $V_{\qv}-\Lambda_\omega$, while it is simply $V_{\qv}$ in EDMFT~+~$GW$.
The latter two differences are not fundamental: To leading order in the interaction, the two expressions are the same. In this case \eqref{app:sigmasdb} becomes
\begin{align}
 \tilde{\Sigma}_{\kv\nu} &= -(T/N)\sum_{\qv\omega} \tilde{\mathcal{G}}_{\kv+\qv\nu+\omega}(V_{\qv}-\Lambda_{\omega})\notag\\
 &= -(T/N)\sum_{\qv\omega} \tilde{\mathcal{G}}_{\kv+\qv\nu+\omega}V_{\qv}\notag\\
 &= -(T/N)\sum_{\qv\omega} G_{\kv+\qv\nu+\omega}V_{\qv},
\end{align}
where we have used $\sum_{\kv}\tilde{\mathcal{G}}_{\kv\nu}=0$ to obtain the second line and $\tilde{\mathcal{G}}_{\kv\nu}=G_{\kv\nu}-g_{\nu}$ and $\sum_{\qv}V_{\qv}=0$ for the third line. For higher orders, the two expression differ, however, because the above cancellations do not occur for the mixed terms in the expansion in powers of $(V_{\qv}-\Lambda_{\omega})$.

\section{Estimating the spectral weight at the Fermi level}
\label{sec:specweight}

The local density of states (DOS) can be determined from the local imaginary time Green's function, e.g., by maximum entropy (MaxEnt) methods~\cite{Jarrell1996133}.
The DOS at the Fermi level indicates if the system is metallic or insulating.  
In the limit of low temperatures, it can be obtained directly from the imaginary time Green's function at $\tau=\beta/2$ without analytical continuation.
To prove this, use that $\beta/\cosh(E'\beta/2)\rightarrow 2\pi\delta(E')$ for $\beta\rightarrow \infty$, so
\begin{align}
G\bigl(\frac{\beta}{2}\bigr) &= -\int_{-\infty}^{\infty} dE' A(E')  \frac{1}{2\cosh(\frac{E'\beta}{2})} \\
&\approx -\int_{-\infty}^{\infty} dE' A(E')  \frac{\pi}{\beta} \delta(E') \\
&\approx -A(0)\frac{\pi}{\beta}, \\
A(0) &\approx -\frac{\beta}{\pi} G\bigl(\frac{\beta}{2}\bigr).
\end{align}
The minus sign with respect to Ref.~\onlinecite{Jarrell1996133} is due to our definition of the Green's function as $-\av{cc^{*}}$.

\bibliography{main}

%merlin.mbs apsrev4-1.bst 2010-07-25 4.21a (PWD, AO, DPC) hacked
%Control: key (0)
%Control: author (8) initials jnrlst
%Control: editor formatted (1) identically to author
%Control: production of article title (-1) disabled
%Control: page (0) single
%Control: year (1) truncated
%Control: production of eprint (0) enabled
\begin{thebibliography}{54}%
\makeatletter
\providecommand \@ifxundefined [1]{%
 \@ifx{#1\undefined}
}%
\providecommand \@ifnum [1]{%
 \ifnum #1\expandafter \@firstoftwo
 \else \expandafter \@secondoftwo
 \fi
}%
\providecommand \@ifx [1]{%
 \ifx #1\expandafter \@firstoftwo
 \else \expandafter \@secondoftwo
 \fi
}%
\providecommand \natexlab [1]{#1}%
\providecommand \enquote  [1]{``#1''}%
\providecommand \bibnamefont  [1]{#1}%
\providecommand \bibfnamefont [1]{#1}%
\providecommand \citenamefont [1]{#1}%
\providecommand \href@noop [0]{\@secondoftwo}%
\providecommand \href [0]{\begingroup \@sanitize@url \@href}%
\providecommand \@href[1]{\@@startlink{#1}\@@href}%
\providecommand \@@href[1]{\endgroup#1\@@endlink}%
\providecommand \@sanitize@url [0]{\catcode `\\12\catcode `\$12\catcode
  `\&12\catcode `\#12\catcode `\^12\catcode `\_12\catcode `\%12\relax}%
\providecommand \@@startlink[1]{}%
\providecommand \@@endlink[0]{}%
\providecommand \url  [0]{\begingroup\@sanitize@url \@url }%
\providecommand \@url [1]{\endgroup\@href {#1}{\urlprefix }}%
\providecommand \urlprefix  [0]{URL }%
\providecommand \Eprint [0]{\href }%
\providecommand \doibase [0]{http://dx.doi.org/}%
\providecommand \selectlanguage [0]{\@gobble}%
\providecommand \bibinfo  [0]{\@secondoftwo}%
\providecommand \bibfield  [0]{\@secondoftwo}%
\providecommand \translation [1]{[#1]}%
\providecommand \BibitemOpen [0]{}%
\providecommand \bibitemStop [0]{}%
\providecommand \bibitemNoStop [0]{.\EOS\space}%
\providecommand \EOS [0]{\spacefactor3000\relax}%
\providecommand \BibitemShut  [1]{\csname bibitem#1\endcsname}%
\let\auto@bib@innerbib\@empty
%</preamble>
\bibitem [{\citenamefont {Hubbard}(1963)}]{Hubbard63}%
  \BibitemOpen
  \bibfield  {author} {\bibinfo {author} {\bibfnamefont {J.}~\bibnamefont
  {Hubbard}},\ }\href {\doibase 10.1098/rspa.1963.0204} {\bibfield  {journal}
  {\bibinfo  {journal} {Proc. R. Soc. London}\ }\textbf {\bibinfo {volume}
  {276}},\ \bibinfo {pages} {238} (\bibinfo {year} {1963})}\BibitemShut
  {NoStop}%
\bibitem [{\citenamefont {Anderson}(1987)}]{Anderson87}%
  \BibitemOpen
  \bibfield  {author} {\bibinfo {author} {\bibfnamefont {P.~W.}\ \bibnamefont
  {Anderson}},\ }\href {\doibase 10.1126/science.235.4793.1196} {\bibfield
  {journal} {\bibinfo  {journal} {Science}\ }\textbf {\bibinfo {volume}
  {235}},\ \bibinfo {pages} {1196} (\bibinfo {year} {1987})}\BibitemShut
  {NoStop}%
\bibitem [{\citenamefont {Hansmann}\ \emph {et~al.}(2013)\citenamefont
  {Hansmann}, \citenamefont {Ayral}, \citenamefont {Vaugier}, \citenamefont
  {Werner},\ and\ \citenamefont {Biermann}}]{Hansmann13}%
  \BibitemOpen
  \bibfield  {author} {\bibinfo {author} {\bibfnamefont {P.}~\bibnamefont
  {Hansmann}}, \bibinfo {author} {\bibfnamefont {T.}~\bibnamefont {Ayral}},
  \bibinfo {author} {\bibfnamefont {L.}~\bibnamefont {Vaugier}}, \bibinfo
  {author} {\bibfnamefont {P.}~\bibnamefont {Werner}}, \ and\ \bibinfo {author}
  {\bibfnamefont {S.}~\bibnamefont {Biermann}},\ }\href {\doibase
  10.1103/PhysRevLett.110.166401} {\bibfield  {journal} {\bibinfo  {journal}
  {Phys. Rev. Lett.}\ }\textbf {\bibinfo {volume} {110}},\ \bibinfo {pages}
  {166401} (\bibinfo {year} {2013})}\BibitemShut {NoStop}%
\bibitem [{\citenamefont {Wehling}\ \emph {et~al.}(2011)\citenamefont
  {Wehling}, \citenamefont {\ifmmode \mbox{\c{S}}\else \c{S}\fi{}a\ifmmode
  \mbox{\c{s}}\else \c{s}\fi{}\ifmmode \imath \else \i
  \fi{}o\ifmmode~\breve{g}\else \u{g}\fi{}lu}, \citenamefont {Friedrich},
  \citenamefont {Lichtenstein}, \citenamefont {Katsnelson},\ and\ \citenamefont
  {Bl\"ugel}}]{Wehling11}%
  \BibitemOpen
  \bibfield  {author} {\bibinfo {author} {\bibfnamefont {T.~O.}\ \bibnamefont
  {Wehling}}, \bibinfo {author} {\bibfnamefont {E.}~\bibnamefont {\ifmmode
  \mbox{\c{S}}\else \c{S}\fi{}a\ifmmode \mbox{\c{s}}\else \c{s}\fi{}\ifmmode
  \imath \else \i \fi{}o\ifmmode~\breve{g}\else \u{g}\fi{}lu}}, \bibinfo
  {author} {\bibfnamefont {C.}~\bibnamefont {Friedrich}}, \bibinfo {author}
  {\bibfnamefont {A.~I.}\ \bibnamefont {Lichtenstein}}, \bibinfo {author}
  {\bibfnamefont {M.~I.}\ \bibnamefont {Katsnelson}}, \ and\ \bibinfo {author}
  {\bibfnamefont {S.}~\bibnamefont {Bl\"ugel}},\ }\href {\doibase
  10.1103/PhysRevLett.106.236805} {\bibfield  {journal} {\bibinfo  {journal}
  {Phys. Rev. Lett.}\ }\textbf {\bibinfo {volume} {106}},\ \bibinfo {pages}
  {236805} (\bibinfo {year} {2011})}\BibitemShut {NoStop}%
\bibitem [{\citenamefont {Sch\"uler}\ \emph {et~al.}(2013)\citenamefont
  {Sch\"uler}, \citenamefont {R\"osner}, \citenamefont {Wehling}, \citenamefont
  {Lichtenstein},\ and\ \citenamefont {Katsnelson}}]{Schuler13}%
  \BibitemOpen
  \bibfield  {author} {\bibinfo {author} {\bibfnamefont {M.}~\bibnamefont
  {Sch\"uler}}, \bibinfo {author} {\bibfnamefont {M.}~\bibnamefont {R\"osner}},
  \bibinfo {author} {\bibfnamefont {T.~O.}\ \bibnamefont {Wehling}}, \bibinfo
  {author} {\bibfnamefont {A.~I.}\ \bibnamefont {Lichtenstein}}, \ and\
  \bibinfo {author} {\bibfnamefont {M.~I.}\ \bibnamefont {Katsnelson}},\ }\href
  {\doibase 10.1103/PhysRevLett.111.036601} {\bibfield  {journal} {\bibinfo
  {journal} {Phys. Rev. Lett.}\ }\textbf {\bibinfo {volume} {111}},\ \bibinfo
  {pages} {036601} (\bibinfo {year} {2013})}\BibitemShut {NoStop}%
\bibitem [{\citenamefont {Schubin}\ and\ \citenamefont
  {Wonsowsky}(1934)}]{schubin1934}%
  \BibitemOpen
  \bibfield  {author} {\bibinfo {author} {\bibfnamefont {S.}~\bibnamefont
  {Schubin}}\ and\ \bibinfo {author} {\bibfnamefont {S.}~\bibnamefont
  {Wonsowsky}},\ }\href@noop {} {\bibfield  {journal} {\bibinfo  {journal}
  {Proc. R. Soc. London}\ }\textbf {\bibinfo {volume} {145}},\ \bibinfo {pages}
  {159} (\bibinfo {year} {1934})}\BibitemShut {NoStop}%
\bibitem [{\citenamefont {Georges}\ \emph {et~al.}(1996)\citenamefont
  {Georges}, \citenamefont {Kotliar}, \citenamefont {Krauth},\ and\
  \citenamefont {Rozenberg}}]{Georges96}%
  \BibitemOpen
  \bibfield  {author} {\bibinfo {author} {\bibfnamefont {A.}~\bibnamefont
  {Georges}}, \bibinfo {author} {\bibfnamefont {G.}~\bibnamefont {Kotliar}},
  \bibinfo {author} {\bibfnamefont {W.}~\bibnamefont {Krauth}}, \ and\ \bibinfo
  {author} {\bibfnamefont {M.~J.}\ \bibnamefont {Rozenberg}},\ }\href {\doibase
  10.1103/RevModPhys.68.13} {\bibfield  {journal} {\bibinfo  {journal} {Rev.
  Mod. Phys.}\ }\textbf {\bibinfo {volume} {68}},\ \bibinfo {pages} {13}
  (\bibinfo {year} {1996})}\BibitemShut {NoStop}%
\bibitem [{\citenamefont {Kotliar}\ \emph {et~al.}(2006)\citenamefont
  {Kotliar}, \citenamefont {Savrasov}, \citenamefont {Haule}, \citenamefont
  {Oudovenko}, \citenamefont {Parcollet},\ and\ \citenamefont
  {Marianetti}}]{Kotliar06}%
  \BibitemOpen
  \bibfield  {author} {\bibinfo {author} {\bibfnamefont {G.}~\bibnamefont
  {Kotliar}}, \bibinfo {author} {\bibfnamefont {S.~Y.}\ \bibnamefont
  {Savrasov}}, \bibinfo {author} {\bibfnamefont {K.}~\bibnamefont {Haule}},
  \bibinfo {author} {\bibfnamefont {V.~S.}\ \bibnamefont {Oudovenko}}, \bibinfo
  {author} {\bibfnamefont {O.}~\bibnamefont {Parcollet}}, \ and\ \bibinfo
  {author} {\bibfnamefont {C.~A.}\ \bibnamefont {Marianetti}},\ }\href
  {\doibase 10.1103/RevModPhys.78.865} {\bibfield  {journal} {\bibinfo
  {journal} {Rev. Mod. Phys.}\ }\textbf {\bibinfo {volume} {78}},\ \bibinfo
  {pages} {865} (\bibinfo {year} {2006})}\BibitemShut {NoStop}%
\bibitem [{\citenamefont {Maier}\ \emph
  {et~al.}(2005{\natexlab{a}})\citenamefont {Maier}, \citenamefont {Jarrell},
  \citenamefont {Pruschke},\ and\ \citenamefont {Hettler}}]{Maier05}%
  \BibitemOpen
  \bibfield  {author} {\bibinfo {author} {\bibfnamefont {T.}~\bibnamefont
  {Maier}}, \bibinfo {author} {\bibfnamefont {M.}~\bibnamefont {Jarrell}},
  \bibinfo {author} {\bibfnamefont {T.}~\bibnamefont {Pruschke}}, \ and\
  \bibinfo {author} {\bibfnamefont {M.~H.}\ \bibnamefont {Hettler}},\ }\href
  {\doibase 10.1103/RevModPhys.77.1027} {\bibfield  {journal} {\bibinfo
  {journal} {Rev. Mod. Phys.}\ }\textbf {\bibinfo {volume} {77}},\ \bibinfo
  {pages} {1027} (\bibinfo {year} {2005}{\natexlab{a}})}\BibitemShut {NoStop}%
\bibitem [{\citenamefont {Werner}\ \emph {et~al.}(2009)\citenamefont {Werner},
  \citenamefont {Gull}, \citenamefont {Parcollet},\ and\ \citenamefont
  {Millis}}]{Werner09}%
  \BibitemOpen
  \bibfield  {author} {\bibinfo {author} {\bibfnamefont {P.}~\bibnamefont
  {Werner}}, \bibinfo {author} {\bibfnamefont {E.}~\bibnamefont {Gull}},
  \bibinfo {author} {\bibfnamefont {O.}~\bibnamefont {Parcollet}}, \ and\
  \bibinfo {author} {\bibfnamefont {A.~J.}\ \bibnamefont {Millis}},\ }\href
  {\doibase 10.1103/PhysRevB.80.045120} {\bibfield  {journal} {\bibinfo
  {journal} {Phys. Rev. B}\ }\textbf {\bibinfo {volume} {80}},\ \bibinfo
  {pages} {045120} (\bibinfo {year} {2009})}\BibitemShut {NoStop}%
\bibitem [{\citenamefont {Ferrero}\ \emph {et~al.}(2009)\citenamefont
  {Ferrero}, \citenamefont {Cornaglia}, \citenamefont {De~Leo}, \citenamefont
  {Parcollet}, \citenamefont {Kotliar},\ and\ \citenamefont
  {Georges}}]{Ferrero09}%
  \BibitemOpen
  \bibfield  {author} {\bibinfo {author} {\bibfnamefont {M.}~\bibnamefont
  {Ferrero}}, \bibinfo {author} {\bibfnamefont {P.~S.}\ \bibnamefont
  {Cornaglia}}, \bibinfo {author} {\bibfnamefont {L.}~\bibnamefont {De~Leo}},
  \bibinfo {author} {\bibfnamefont {O.}~\bibnamefont {Parcollet}}, \bibinfo
  {author} {\bibfnamefont {G.}~\bibnamefont {Kotliar}}, \ and\ \bibinfo
  {author} {\bibfnamefont {A.}~\bibnamefont {Georges}},\ }\href {\doibase
  10.1103/PhysRevB.80.064501} {\bibfield  {journal} {\bibinfo  {journal} {Phys.
  Rev. B}\ }\textbf {\bibinfo {volume} {80}},\ \bibinfo {pages} {064501}
  (\bibinfo {year} {2009})}\BibitemShut {NoStop}%
\bibitem [{\citenamefont {Gull}\ \emph {et~al.}(2010)\citenamefont {Gull},
  \citenamefont {Ferrero}, \citenamefont {Parcollet}, \citenamefont {Georges},\
  and\ \citenamefont {Millis}}]{Gull10}%
  \BibitemOpen
  \bibfield  {author} {\bibinfo {author} {\bibfnamefont {E.}~\bibnamefont
  {Gull}}, \bibinfo {author} {\bibfnamefont {M.}~\bibnamefont {Ferrero}},
  \bibinfo {author} {\bibfnamefont {O.}~\bibnamefont {Parcollet}}, \bibinfo
  {author} {\bibfnamefont {A.}~\bibnamefont {Georges}}, \ and\ \bibinfo
  {author} {\bibfnamefont {A.~J.}\ \bibnamefont {Millis}},\ }\href {\doibase
  10.1103/PhysRevB.82.155101} {\bibfield  {journal} {\bibinfo  {journal} {Phys.
  Rev. B}\ }\textbf {\bibinfo {volume} {82}},\ \bibinfo {pages} {155101}
  (\bibinfo {year} {2010})}\BibitemShut {NoStop}%
\bibitem [{\citenamefont {Maier}\ \emph
  {et~al.}(2005{\natexlab{b}})\citenamefont {Maier}, \citenamefont {Jarrell},
  \citenamefont {Schulthess}, \citenamefont {Kent},\ and\ \citenamefont
  {White}}]{Maier05-2}%
  \BibitemOpen
  \bibfield  {author} {\bibinfo {author} {\bibfnamefont {T.~A.}\ \bibnamefont
  {Maier}}, \bibinfo {author} {\bibfnamefont {M.}~\bibnamefont {Jarrell}},
  \bibinfo {author} {\bibfnamefont {T.~C.}\ \bibnamefont {Schulthess}},
  \bibinfo {author} {\bibfnamefont {P.~R.~C.}\ \bibnamefont {Kent}}, \ and\
  \bibinfo {author} {\bibfnamefont {J.~B.}\ \bibnamefont {White}},\ }\href
  {\doibase 10.1103/PhysRevLett.95.237001} {\bibfield  {journal} {\bibinfo
  {journal} {Phys. Rev. Lett.}\ }\textbf {\bibinfo {volume} {95}},\ \bibinfo
  {pages} {237001} (\bibinfo {year} {2005}{\natexlab{b}})}\BibitemShut
  {NoStop}%
\bibitem [{\citenamefont {Gull}\ \emph {et~al.}(2013)\citenamefont {Gull},
  \citenamefont {Parcollet},\ and\ \citenamefont {Millis}}]{Gull13}%
  \BibitemOpen
  \bibfield  {author} {\bibinfo {author} {\bibfnamefont {E.}~\bibnamefont
  {Gull}}, \bibinfo {author} {\bibfnamefont {O.}~\bibnamefont {Parcollet}}, \
  and\ \bibinfo {author} {\bibfnamefont {A.~J.}\ \bibnamefont {Millis}},\
  }\href {\doibase 10.1103/PhysRevLett.110.216405} {\bibfield  {journal}
  {\bibinfo  {journal} {Phys. Rev. Lett.}\ }\textbf {\bibinfo {volume} {110}},\
  \bibinfo {pages} {216405} (\bibinfo {year} {2013})}\BibitemShut {NoStop}%
\bibitem [{\citenamefont {Kusunose}(2006)}]{Kusunose06}%
  \BibitemOpen
  \bibfield  {author} {\bibinfo {author} {\bibfnamefont {H.}~\bibnamefont
  {Kusunose}},\ }\href {\doibase 10.1143/JPSJ.75.054713} {\bibfield  {journal}
  {\bibinfo  {journal} {J. Phys. Soc. Jpn}\ }\textbf {\bibinfo {volume} {75}},\
  \bibinfo {pages} {054713} (\bibinfo {year} {2006})}\BibitemShut {NoStop}%
\bibitem [{\citenamefont {Toschi}\ \emph {et~al.}(2007)\citenamefont {Toschi},
  \citenamefont {Katanin},\ and\ \citenamefont {Held}}]{Toschi07}%
  \BibitemOpen
  \bibfield  {author} {\bibinfo {author} {\bibfnamefont {A.}~\bibnamefont
  {Toschi}}, \bibinfo {author} {\bibfnamefont {A.~A.}\ \bibnamefont {Katanin}},
  \ and\ \bibinfo {author} {\bibfnamefont {K.}~\bibnamefont {Held}},\ }\href
  {\doibase 10.1103/PhysRevB.75.045118} {\bibfield  {journal} {\bibinfo
  {journal} {Phys. Rev. B}\ }\textbf {\bibinfo {volume} {75}},\ \bibinfo {eid}
  {045118} (\bibinfo {year} {2007})}\BibitemShut {NoStop}%
\bibitem [{\citenamefont {Rohringer}\ \emph {et~al.}(2013)\citenamefont
  {Rohringer}, \citenamefont {Toschi}, \citenamefont {Hafermann}, \citenamefont
  {Held}, \citenamefont {Anisimov},\ and\ \citenamefont
  {Katanin}}]{Rohringer13}%
  \BibitemOpen
  \bibfield  {author} {\bibinfo {author} {\bibfnamefont {G.}~\bibnamefont
  {Rohringer}}, \bibinfo {author} {\bibfnamefont {A.}~\bibnamefont {Toschi}},
  \bibinfo {author} {\bibfnamefont {H.}~\bibnamefont {Hafermann}}, \bibinfo
  {author} {\bibfnamefont {K.}~\bibnamefont {Held}}, \bibinfo {author}
  {\bibfnamefont {V.~I.}\ \bibnamefont {Anisimov}}, \ and\ \bibinfo {author}
  {\bibfnamefont {A.~A.}\ \bibnamefont {Katanin}},\ }\href {\doibase
  10.1103/PhysRevB.88.115112} {\bibfield  {journal} {\bibinfo  {journal} {Phys.
  Rev. B}\ }\textbf {\bibinfo {volume} {88}},\ \bibinfo {pages} {115112}
  (\bibinfo {year} {2013})}\BibitemShut {NoStop}%
\bibitem [{\citenamefont {Rubtsov}\ \emph {et~al.}(2008)\citenamefont
  {Rubtsov}, \citenamefont {Katsnelson},\ and\ \citenamefont
  {Lichtenstein}}]{Rubtsov08}%
  \BibitemOpen
  \bibfield  {author} {\bibinfo {author} {\bibfnamefont {A.~N.}\ \bibnamefont
  {Rubtsov}}, \bibinfo {author} {\bibfnamefont {M.~I.}\ \bibnamefont
  {Katsnelson}}, \ and\ \bibinfo {author} {\bibfnamefont {A.~I.}\ \bibnamefont
  {Lichtenstein}},\ }\href {\doibase 10.1103/PhysRevB.77.033101} {\bibfield
  {journal} {\bibinfo  {journal} {Phys. Rev. B}\ }\textbf {\bibinfo {volume}
  {77}},\ \bibinfo {pages} {033101} (\bibinfo {year} {2008})}\BibitemShut
  {NoStop}%
\bibitem [{\citenamefont {Rubtsov}\ \emph {et~al.}(2009)\citenamefont
  {Rubtsov}, \citenamefont {Katsnelson}, \citenamefont {Lichtenstein},\ and\
  \citenamefont {Georges}}]{Rubtsov09}%
  \BibitemOpen
  \bibfield  {author} {\bibinfo {author} {\bibfnamefont {A.~N.}\ \bibnamefont
  {Rubtsov}}, \bibinfo {author} {\bibfnamefont {M.~I.}\ \bibnamefont
  {Katsnelson}}, \bibinfo {author} {\bibfnamefont {A.~I.}\ \bibnamefont
  {Lichtenstein}}, \ and\ \bibinfo {author} {\bibfnamefont {A.}~\bibnamefont
  {Georges}},\ }\href {\doibase 10.1103/PhysRevB.79.045133} {\bibfield
  {journal} {\bibinfo  {journal} {Phys. Rev. B}\ }\textbf {\bibinfo {volume}
  {79}},\ \bibinfo {pages} {045133} (\bibinfo {year} {2009})}\BibitemShut
  {NoStop}%
\bibitem [{\citenamefont {Bolech}\ \emph {et~al.}(2003)\citenamefont {Bolech},
  \citenamefont {Kancharla},\ and\ \citenamefont {Kotliar}}]{Bolech03}%
  \BibitemOpen
  \bibfield  {author} {\bibinfo {author} {\bibfnamefont {C.~J.}\ \bibnamefont
  {Bolech}}, \bibinfo {author} {\bibfnamefont {S.~S.}\ \bibnamefont
  {Kancharla}}, \ and\ \bibinfo {author} {\bibfnamefont {G.}~\bibnamefont
  {Kotliar}},\ }\href {\doibase 10.1103/PhysRevB.67.075110} {\bibfield
  {journal} {\bibinfo  {journal} {Phys. Rev. B}\ }\textbf {\bibinfo {volume}
  {67}},\ \bibinfo {pages} {075110} (\bibinfo {year} {2003})}\BibitemShut
  {NoStop}%
\bibitem [{\citenamefont {Poteryaev}\ \emph {et~al.}(2004)\citenamefont
  {Poteryaev}, \citenamefont {Lichtenstein},\ and\ \citenamefont
  {Kotliar}}]{Poteryaev04}%
  \BibitemOpen
  \bibfield  {author} {\bibinfo {author} {\bibfnamefont {A.~I.}\ \bibnamefont
  {Poteryaev}}, \bibinfo {author} {\bibfnamefont {A.~I.}\ \bibnamefont
  {Lichtenstein}}, \ and\ \bibinfo {author} {\bibfnamefont {G.}~\bibnamefont
  {Kotliar}},\ }\href {\doibase 10.1103/PhysRevLett.93.086401} {\bibfield
  {journal} {\bibinfo  {journal} {Phys. Rev. Lett.}\ }\textbf {\bibinfo
  {volume} {93}},\ \bibinfo {pages} {086401} (\bibinfo {year}
  {2004})}\BibitemShut {NoStop}%
\bibitem [{\citenamefont {Tong}(2005)}]{Tong05}%
  \BibitemOpen
  \bibfield  {author} {\bibinfo {author} {\bibfnamefont {N.-H.}\ \bibnamefont
  {Tong}},\ }\href {\doibase 10.1103/PhysRevB.72.115104} {\bibfield  {journal}
  {\bibinfo  {journal} {Phys. Rev. B}\ }\textbf {\bibinfo {volume} {72}},\
  \bibinfo {pages} {115104} (\bibinfo {year} {2005})}\BibitemShut {NoStop}%
\bibitem [{\citenamefont {Merino}(2007)}]{Merino07}%
  \BibitemOpen
  \bibfield  {author} {\bibinfo {author} {\bibfnamefont {J.}~\bibnamefont
  {Merino}},\ }\href {\doibase 10.1103/PhysRevLett.99.036404} {\bibfield
  {journal} {\bibinfo  {journal} {Phys. Rev. Lett.}\ }\textbf {\bibinfo
  {volume} {99}},\ \bibinfo {pages} {036404} (\bibinfo {year}
  {2007})}\BibitemShut {NoStop}%
\bibitem [{\citenamefont {Aichhorn}\ \emph {et~al.}(2004)\citenamefont
  {Aichhorn}, \citenamefont {Evertz}, \citenamefont {von~der Linden},\ and\
  \citenamefont {Potthoff}}]{Aichhorn07}%
  \BibitemOpen
  \bibfield  {author} {\bibinfo {author} {\bibfnamefont {M.}~\bibnamefont
  {Aichhorn}}, \bibinfo {author} {\bibfnamefont {H.~G.}\ \bibnamefont
  {Evertz}}, \bibinfo {author} {\bibfnamefont {W.}~\bibnamefont {von~der
  Linden}}, \ and\ \bibinfo {author} {\bibfnamefont {M.}~\bibnamefont
  {Potthoff}},\ }\href {\doibase 10.1103/PhysRevB.70.235107} {\bibfield
  {journal} {\bibinfo  {journal} {Phys. Rev. B}\ }\textbf {\bibinfo {volume}
  {70}},\ \bibinfo {pages} {235107} (\bibinfo {year} {2004})}\BibitemShut
  {NoStop}%
\bibitem [{\citenamefont {Hassan}\ and\ \citenamefont {de'
  Medici}(2010)}]{Hassan10}%
  \BibitemOpen
  \bibfield  {author} {\bibinfo {author} {\bibfnamefont {S.~R.}\ \bibnamefont
  {Hassan}}\ and\ \bibinfo {author} {\bibfnamefont {L.}~\bibnamefont {de'
  Medici}},\ }\href {\doibase 10.1103/PhysRevB.81.035106} {\bibfield  {journal}
  {\bibinfo  {journal} {Phys. Rev. B}\ }\textbf {\bibinfo {volume} {81}},\
  \bibinfo {pages} {035106} (\bibinfo {year} {2010})}\BibitemShut {NoStop}%
\bibitem [{\citenamefont {Aryasetiawan}\ and\ \citenamefont
  {Gunnarsson}(1998)}]{Aryasetiawan98}%
  \BibitemOpen
  \bibfield  {author} {\bibinfo {author} {\bibfnamefont {F.}~\bibnamefont
  {Aryasetiawan}}\ and\ \bibinfo {author} {\bibfnamefont {O.}~\bibnamefont
  {Gunnarsson}},\ }\href {http://stacks.iop.org/0034-4885/61/i=3/a=002}
  {\bibfield  {journal} {\bibinfo  {journal} {Rep Prog. Phys.}\ }\textbf
  {\bibinfo {volume} {61}},\ \bibinfo {pages} {237} (\bibinfo {year}
  {1998})}\BibitemShut {NoStop}%
\bibitem [{\citenamefont {Si}\ and\ \citenamefont {Smith}(1996)}]{Si96}%
  \BibitemOpen
  \bibfield  {author} {\bibinfo {author} {\bibfnamefont {Q.}~\bibnamefont
  {Si}}\ and\ \bibinfo {author} {\bibfnamefont {J.~L.}\ \bibnamefont {Smith}},\
  }\href {\doibase 10.1103/PhysRevLett.77.3391} {\bibfield  {journal} {\bibinfo
   {journal} {Phys. Rev. Lett.}\ }\textbf {\bibinfo {volume} {77}},\ \bibinfo
  {pages} {3391} (\bibinfo {year} {1996})}\BibitemShut {NoStop}%
\bibitem [{\citenamefont {Kajueter}(1996)}]{Kajueter96}%
  \BibitemOpen
  \bibfield  {author} {\bibinfo {author} {\bibfnamefont {H.}~\bibnamefont
  {Kajueter}},\ }\href@noop {} {Ph.D. thesis},\ \bibinfo  {school} {Rutgers
  University} (\bibinfo {year} {1996})\BibitemShut {NoStop}%
\bibitem [{\citenamefont {Smith}\ and\ \citenamefont {Si}(2000)}]{Smith00}%
  \BibitemOpen
  \bibfield  {author} {\bibinfo {author} {\bibfnamefont {J.~L.}\ \bibnamefont
  {Smith}}\ and\ \bibinfo {author} {\bibfnamefont {Q.}~\bibnamefont {Si}},\
  }\href {\doibase 10.1103/PhysRevB.61.5184} {\bibfield  {journal} {\bibinfo
  {journal} {Phys. Rev. B}\ }\textbf {\bibinfo {volume} {61}},\ \bibinfo
  {pages} {5184} (\bibinfo {year} {2000})}\BibitemShut {NoStop}%
\bibitem [{\citenamefont {Chitra}\ and\ \citenamefont
  {Kotliar}(2000)}]{Chitra00}%
  \BibitemOpen
  \bibfield  {author} {\bibinfo {author} {\bibfnamefont {R.}~\bibnamefont
  {Chitra}}\ and\ \bibinfo {author} {\bibfnamefont {G.}~\bibnamefont
  {Kotliar}},\ }\href {\doibase 10.1103/PhysRevLett.84.3678} {\bibfield
  {journal} {\bibinfo  {journal} {Phys. Rev. Lett.}\ }\textbf {\bibinfo
  {volume} {84}},\ \bibinfo {pages} {3678} (\bibinfo {year}
  {2000})}\BibitemShut {NoStop}%
\bibitem [{\citenamefont {Chitra}\ and\ \citenamefont
  {Kotliar}(2001)}]{Chitra01}%
  \BibitemOpen
  \bibfield  {author} {\bibinfo {author} {\bibfnamefont {R.}~\bibnamefont
  {Chitra}}\ and\ \bibinfo {author} {\bibfnamefont {G.}~\bibnamefont
  {Kotliar}},\ }\href {\doibase 10.1103/PhysRevB.63.115110} {\bibfield
  {journal} {\bibinfo  {journal} {Phys. Rev. B}\ }\textbf {\bibinfo {volume}
  {63}},\ \bibinfo {pages} {115110} (\bibinfo {year} {2001})}\BibitemShut
  {NoStop}%
\bibitem [{\citenamefont {Sun}\ and\ \citenamefont {Kotliar}(2002)}]{Sun02}%
  \BibitemOpen
  \bibfield  {author} {\bibinfo {author} {\bibfnamefont {P.}~\bibnamefont
  {Sun}}\ and\ \bibinfo {author} {\bibfnamefont {G.}~\bibnamefont {Kotliar}},\
  }\href {\doibase 10.1103/PhysRevB.66.085120} {\bibfield  {journal} {\bibinfo
  {journal} {Phys. Rev. B}\ }\textbf {\bibinfo {volume} {66}},\ \bibinfo
  {pages} {085120} (\bibinfo {year} {2002})}\BibitemShut {NoStop}%
\bibitem [{\citenamefont {Ayral}\ \emph {et~al.}(2012)\citenamefont {Ayral},
  \citenamefont {Werner},\ and\ \citenamefont {Biermann}}]{Ayral12}%
  \BibitemOpen
  \bibfield  {author} {\bibinfo {author} {\bibfnamefont {T.}~\bibnamefont
  {Ayral}}, \bibinfo {author} {\bibfnamefont {P.}~\bibnamefont {Werner}}, \
  and\ \bibinfo {author} {\bibfnamefont {S.}~\bibnamefont {Biermann}},\ }\href
  {\doibase 10.1103/PhysRevLett.109.226401} {\bibfield  {journal} {\bibinfo
  {journal} {Phys. Rev. Lett.}\ }\textbf {\bibinfo {volume} {109}},\ \bibinfo
  {pages} {226401} (\bibinfo {year} {2012})}\BibitemShut {NoStop}%
\bibitem [{\citenamefont {Ayral}\ \emph {et~al.}(2013)\citenamefont {Ayral},
  \citenamefont {Biermann},\ and\ \citenamefont {Werner}}]{Ayral13}%
  \BibitemOpen
  \bibfield  {author} {\bibinfo {author} {\bibfnamefont {T.}~\bibnamefont
  {Ayral}}, \bibinfo {author} {\bibfnamefont {S.}~\bibnamefont {Biermann}}, \
  and\ \bibinfo {author} {\bibfnamefont {P.}~\bibnamefont {Werner}},\ }\href
  {\doibase 10.1103/PhysRevB.87.125149} {\bibfield  {journal} {\bibinfo
  {journal} {Phys. Rev. B}\ }\textbf {\bibinfo {volume} {87}},\ \bibinfo
  {pages} {125149} (\bibinfo {year} {2013})}\BibitemShut {NoStop}%
\bibitem [{\citenamefont {Rubtsov}\ \emph {et~al.}(2012)\citenamefont
  {Rubtsov}, \citenamefont {Katsnelson},\ and\ \citenamefont
  {Lichtenstein}}]{Rubtsov12}%
  \BibitemOpen
  \bibfield  {author} {\bibinfo {author} {\bibfnamefont {A.~N.}\ \bibnamefont
  {Rubtsov}}, \bibinfo {author} {\bibfnamefont {M.~I.}\ \bibnamefont
  {Katsnelson}}, \ and\ \bibinfo {author} {\bibfnamefont {A.~I.}\ \bibnamefont
  {Lichtenstein}},\ }\href {\doibase 10.1016/j.aop.2012.01.002} {\bibfield
  {journal} {\bibinfo  {journal} {Ann. Phys.}\ }\textbf {\bibinfo {volume}
  {327}},\ \bibinfo {pages} {1320} (\bibinfo {year} {2012})}\BibitemShut
  {NoStop}%
\bibitem [{\citenamefont {van Loon}\ \emph {et~al.}(2014)\citenamefont {van
  Loon}, \citenamefont {Hafermann}, \citenamefont {Lichtenstein}, \citenamefont
  {Rubtsov},\ and\ \citenamefont {Katsnelson}}]{vanLoon14}%
  \BibitemOpen
  \bibfield  {author} {\bibinfo {author} {\bibfnamefont {E.~G. C.~P.}\
  \bibnamefont {van Loon}}, \bibinfo {author} {\bibfnamefont {H.}~\bibnamefont
  {Hafermann}}, \bibinfo {author} {\bibfnamefont {A.~I.}\ \bibnamefont
  {Lichtenstein}}, \bibinfo {author} {\bibfnamefont {A.~N.}\ \bibnamefont
  {Rubtsov}}, \ and\ \bibinfo {author} {\bibfnamefont {M.~I.}\ \bibnamefont
  {Katsnelson}},\ }\href {\doibase 10.1103/PhysRevLett.113.246407} {\bibfield
  {journal} {\bibinfo  {journal} {Phys. Rev. Lett.}\ }\textbf {\bibinfo
  {volume} {113}},\ \bibinfo {pages} {246407} (\bibinfo {year}
  {2014})}\BibitemShut {NoStop}%
\bibitem [{\citenamefont {Hafermann}\ \emph {et~al.}(2014)\citenamefont
  {Hafermann}, \citenamefont {van Loon}, \citenamefont {Katsnelson},
  \citenamefont {Lichtenstein},\ and\ \citenamefont
  {Parcollet}}]{Hafermann14-2}%
  \BibitemOpen
  \bibfield  {author} {\bibinfo {author} {\bibfnamefont {H.}~\bibnamefont
  {Hafermann}}, \bibinfo {author} {\bibfnamefont {E.~G. C.~P.}\ \bibnamefont
  {van Loon}}, \bibinfo {author} {\bibfnamefont {M.~I.}\ \bibnamefont
  {Katsnelson}}, \bibinfo {author} {\bibfnamefont {A.~I.}\ \bibnamefont
  {Lichtenstein}}, \ and\ \bibinfo {author} {\bibfnamefont {O.}~\bibnamefont
  {Parcollet}},\ }\href {\doibase 10.1103/PhysRevB.90.235105} {\bibfield
  {journal} {\bibinfo  {journal} {Phys. Rev. B}\ }\textbf {\bibinfo {volume}
  {90}},\ \bibinfo {pages} {235105} (\bibinfo {year} {2014})}\BibitemShut
  {NoStop}%
\bibitem [{\citenamefont {Hafermann}(2010)}]{Hafermannphd}%
  \BibitemOpen
  \bibfield  {author} {\bibinfo {author} {\bibfnamefont {H.}~\bibnamefont
  {Hafermann}},\ }\emph {\bibinfo {title} {Numerical Approaches to Spatial
  Correlations in Strongly Interacting Fermion Systems}},\ \href@noop {} {Ph.D.
  thesis},\ \bibinfo  {school} {University of Hamburg} (\bibinfo {year}
  {2010})\BibitemShut {NoStop}%
\bibitem [{\citenamefont {Hafermann}\ \emph {et~al.}(2012)\citenamefont
  {Hafermann}, \citenamefont {Lechermann}, \citenamefont {Rubtsov},
  \citenamefont {Katsnelson}, \citenamefont {Georges},\ and\ \citenamefont
  {Lichtenstein}}]{Hafermann12-2}%
  \BibitemOpen
  \bibfield  {author} {\bibinfo {author} {\bibfnamefont {H.}~\bibnamefont
  {Hafermann}}, \bibinfo {author} {\bibfnamefont {F.}~\bibnamefont
  {Lechermann}}, \bibinfo {author} {\bibfnamefont {A.~N.}\ \bibnamefont
  {Rubtsov}}, \bibinfo {author} {\bibfnamefont {M.~I.}\ \bibnamefont
  {Katsnelson}}, \bibinfo {author} {\bibfnamefont {A.}~\bibnamefont {Georges}},
  \ and\ \bibinfo {author} {\bibfnamefont {A.~I.}\ \bibnamefont
  {Lichtenstein}},\ }in\ \href {\doibase 10.1007/978-3-642-10449-7_4} {\emph
  {\bibinfo {booktitle} {Modern Theories of Many-Particle Systems in Condensed
  Matter Physics}}},\ \bibinfo {series} {Lecture Notes in Physics}, Vol.\
  \bibinfo {volume} {843},\ \bibinfo {editor} {edited by\ \bibinfo {editor}
  {\bibfnamefont {D.~C.}\ \bibnamefont {Cabra}}, \bibinfo {editor}
  {\bibfnamefont {A.}~\bibnamefont {Honecker}}, \ and\ \bibinfo {editor}
  {\bibfnamefont {P.}~\bibnamefont {Pujol}}}\ (\bibinfo  {publisher} {Springer
  Berlin Heidelberg},\ \bibinfo {year} {2012})\ pp.\ \bibinfo {pages}
  {145--214}\BibitemShut {NoStop}%
\bibitem [{\citenamefont {Hafermann}\ \emph {et~al.}(2009)\citenamefont
  {Hafermann}, \citenamefont {Li}, \citenamefont {Rubtsov}, \citenamefont
  {Katsnelson}, \citenamefont {Lichtenstein},\ and\ \citenamefont
  {Monien}}]{Hafermann09}%
  \BibitemOpen
  \bibfield  {author} {\bibinfo {author} {\bibfnamefont {H.}~\bibnamefont
  {Hafermann}}, \bibinfo {author} {\bibfnamefont {G.}~\bibnamefont {Li}},
  \bibinfo {author} {\bibfnamefont {A.~N.}\ \bibnamefont {Rubtsov}}, \bibinfo
  {author} {\bibfnamefont {M.~I.}\ \bibnamefont {Katsnelson}}, \bibinfo
  {author} {\bibfnamefont {A.~I.}\ \bibnamefont {Lichtenstein}}, \ and\
  \bibinfo {author} {\bibfnamefont {H.}~\bibnamefont {Monien}},\ }\href
  {\doibase 10.1103/PhysRevLett.102.206401} {\bibfield  {journal} {\bibinfo
  {journal} {Phys. Rev. Lett.}\ }\textbf {\bibinfo {volume} {102}},\ \bibinfo
  {pages} {206401} (\bibinfo {year} {2009})}\BibitemShut {NoStop}%
\bibitem [{\citenamefont {Werner}\ \emph {et~al.}(2006)\citenamefont {Werner},
  \citenamefont {Comanac}, \citenamefont {de' Medici}, \citenamefont {Troyer},\
  and\ \citenamefont {Millis}}]{Werner06}%
  \BibitemOpen
  \bibfield  {author} {\bibinfo {author} {\bibfnamefont {P.}~\bibnamefont
  {Werner}}, \bibinfo {author} {\bibfnamefont {A.}~\bibnamefont {Comanac}},
  \bibinfo {author} {\bibfnamefont {L.}~\bibnamefont {de' Medici}}, \bibinfo
  {author} {\bibfnamefont {M.}~\bibnamefont {Troyer}}, \ and\ \bibinfo {author}
  {\bibfnamefont {A.~J.}\ \bibnamefont {Millis}},\ }\href {\doibase
  10.1103/PhysRevLett.97.076405} {\bibfield  {journal} {\bibinfo  {journal}
  {Phys. Rev. Lett.}\ }\textbf {\bibinfo {volume} {97}},\ \bibinfo {pages}
  {076405} (\bibinfo {year} {2006})}\BibitemShut {NoStop}%
\bibitem [{\citenamefont {Hafermann}\ \emph {et~al.}(2013)\citenamefont
  {Hafermann}, \citenamefont {Werner},\ and\ \citenamefont
  {Gull}}]{Hafermann13}%
  \BibitemOpen
  \bibfield  {author} {\bibinfo {author} {\bibfnamefont {H.}~\bibnamefont
  {Hafermann}}, \bibinfo {author} {\bibfnamefont {P.}~\bibnamefont {Werner}}, \
  and\ \bibinfo {author} {\bibfnamefont {E.}~\bibnamefont {Gull}},\ }\href
  {\doibase http://dx.doi.org/10.1016/j.cpc.2012.12.013} {\bibfield  {journal}
  {\bibinfo  {journal} {Computer Physics Communications}\ }\textbf {\bibinfo
  {volume} {184}},\ \bibinfo {pages} {1280 } (\bibinfo {year}
  {2013})}\BibitemShut {NoStop}%
\bibitem [{\citenamefont {Hafermann}(2014)}]{Hafermann14}%
  \BibitemOpen
  \bibfield  {author} {\bibinfo {author} {\bibfnamefont {H.}~\bibnamefont
  {Hafermann}},\ }\href {\doibase 10.1103/PhysRevB.89.235128} {\bibfield
  {journal} {\bibinfo  {journal} {Phys. Rev. B}\ }\textbf {\bibinfo {volume}
  {89}},\ \bibinfo {pages} {235128} (\bibinfo {year} {2014})}\BibitemShut
  {NoStop}%
\bibitem [{\citenamefont {Huang}\ \emph {et~al.}(2014)\citenamefont {Huang},
  \citenamefont {Ayral}, \citenamefont {Biermann},\ and\ \citenamefont
  {Werner}}]{Huang14}%
  \BibitemOpen
  \bibfield  {author} {\bibinfo {author} {\bibfnamefont {L.}~\bibnamefont
  {Huang}}, \bibinfo {author} {\bibfnamefont {T.}~\bibnamefont {Ayral}},
  \bibinfo {author} {\bibfnamefont {S.}~\bibnamefont {Biermann}}, \ and\
  \bibinfo {author} {\bibfnamefont {P.}~\bibnamefont {Werner}},\ }\href
  {\doibase 10.1103/PhysRevB.90.195114} {\bibfield  {journal} {\bibinfo
  {journal} {Phys. Rev. B}\ }\textbf {\bibinfo {volume} {90}},\ \bibinfo
  {pages} {195114} (\bibinfo {year} {2014})}\BibitemShut {NoStop}%
\bibitem [{\citenamefont {Werner}\ and\ \citenamefont
  {Millis}(2010)}]{Werner10}%
  \BibitemOpen
  \bibfield  {author} {\bibinfo {author} {\bibfnamefont {P.}~\bibnamefont
  {Werner}}\ and\ \bibinfo {author} {\bibfnamefont {A.~J.}\ \bibnamefont
  {Millis}},\ }\href {\doibase 10.1103/PhysRevLett.104.146401} {\bibfield
  {journal} {\bibinfo  {journal} {Phys. Rev. Lett.}\ }\textbf {\bibinfo
  {volume} {104}},\ \bibinfo {pages} {146401} (\bibinfo {year}
  {2010})}\BibitemShut {NoStop}%
\bibitem [{\citenamefont {Brener}\ \emph {et~al.}(2008)\citenamefont {Brener},
  \citenamefont {Hafermann}, \citenamefont {Rubtsov}, \citenamefont
  {Katsnelson},\ and\ \citenamefont {Lichtenstein}}]{Brener08}%
  \BibitemOpen
  \bibfield  {author} {\bibinfo {author} {\bibfnamefont {S.}~\bibnamefont
  {Brener}}, \bibinfo {author} {\bibfnamefont {H.}~\bibnamefont {Hafermann}},
  \bibinfo {author} {\bibfnamefont {A.~N.}\ \bibnamefont {Rubtsov}}, \bibinfo
  {author} {\bibfnamefont {M.~I.}\ \bibnamefont {Katsnelson}}, \ and\ \bibinfo
  {author} {\bibfnamefont {A.~I.}\ \bibnamefont {Lichtenstein}},\ }\href
  {\doibase 10.1103/PhysRevB.77.195105} {\bibfield  {journal} {\bibinfo
  {journal} {Phys. Rev. B}\ }\textbf {\bibinfo {volume} {77}},\ \bibinfo {eid}
  {195105} (\bibinfo {year} {2008})}\BibitemShut {NoStop}%
\bibitem [{\citenamefont {Zhang}\ and\ \citenamefont
  {Callaway}(1989)}]{Zhang89}%
  \BibitemOpen
  \bibfield  {author} {\bibinfo {author} {\bibfnamefont {Y.}~\bibnamefont
  {Zhang}}\ and\ \bibinfo {author} {\bibfnamefont {J.}~\bibnamefont
  {Callaway}},\ }\href {\doibase 10.1103/PhysRevB.39.9397} {\bibfield
  {journal} {\bibinfo  {journal} {Phys. Rev. B}\ }\textbf {\bibinfo {volume}
  {39}},\ \bibinfo {pages} {9397} (\bibinfo {year} {1989})}\BibitemShut
  {NoStop}%
\bibitem [{\citenamefont {Davoudi}\ and\ \citenamefont
  {Tremblay}(2007)}]{Davoudi07}%
  \BibitemOpen
  \bibfield  {author} {\bibinfo {author} {\bibfnamefont {B.}~\bibnamefont
  {Davoudi}}\ and\ \bibinfo {author} {\bibfnamefont {A.-M.~S.}\ \bibnamefont
  {Tremblay}},\ }\href {\doibase 10.1103/PhysRevB.76.085115} {\bibfield
  {journal} {\bibinfo  {journal} {Phys. Rev. B}\ }\textbf {\bibinfo {volume}
  {76}},\ \bibinfo {pages} {085115} (\bibinfo {year} {2007})}\BibitemShut
  {NoStop}%
\bibitem [{\citenamefont {Vonsovsky}\ and\ \citenamefont
  {Katsnelson}(1979)}]{Vonsovsky79}%
  \BibitemOpen
  \bibfield  {author} {\bibinfo {author} {\bibfnamefont {S.~V.}\ \bibnamefont
  {Vonsovsky}}\ and\ \bibinfo {author} {\bibfnamefont {M.~I.}\ \bibnamefont
  {Katsnelson}},\ }\href {http://stacks.iop.org/0022-3719/12/i=11/a=015}
  {\bibfield  {journal} {\bibinfo  {journal} {J. Phys. C: Solid State Phys.}\
  }\textbf {\bibinfo {volume} {12}},\ \bibinfo {pages} {2043} (\bibinfo {year}
  {1979})}\BibitemShut {NoStop}%
\bibitem [{\citenamefont {Bauer}\ \emph {et~al.}(2011)\citenamefont {Bauer},
  \citenamefont {Carr}, \citenamefont {Evertz}, \citenamefont {Feiguin},
  \citenamefont {Freire}, \citenamefont {Fuchs}, \citenamefont {Gamper},
  \citenamefont {Gukelberger}, \citenamefont {Gull}, \citenamefont {Guertler},
  \citenamefont {Hehn}, \citenamefont {Igarashi}, \citenamefont {Isakov},
  \citenamefont {Koop}, \citenamefont {Ma}, \citenamefont {Mates},
  \citenamefont {Matsuo}, \citenamefont {Parcollet}, \citenamefont
  {Pawłowski}, \citenamefont {Picon}, \citenamefont {Pollet}, \citenamefont
  {Santos}, \citenamefont {Scarola}, \citenamefont {Schollwöck}, \citenamefont
  {Silva}, \citenamefont {Surer}, \citenamefont {Todo}, \citenamefont {Trebst},
  \citenamefont {Troyer}, \citenamefont {Wall}, \citenamefont {Werner},\ and\
  \citenamefont {Wessel}}]{ALPS2}%
  \BibitemOpen
  \bibfield  {author} {\bibinfo {author} {\bibfnamefont {B.}~\bibnamefont
  {Bauer}}, \bibinfo {author} {\bibfnamefont {L.~D.}\ \bibnamefont {Carr}},
  \bibinfo {author} {\bibfnamefont {H.~G.}\ \bibnamefont {Evertz}}, \bibinfo
  {author} {\bibfnamefont {A.}~\bibnamefont {Feiguin}}, \bibinfo {author}
  {\bibfnamefont {J.}~\bibnamefont {Freire}}, \bibinfo {author} {\bibfnamefont
  {S.}~\bibnamefont {Fuchs}}, \bibinfo {author} {\bibfnamefont
  {L.}~\bibnamefont {Gamper}}, \bibinfo {author} {\bibfnamefont
  {J.}~\bibnamefont {Gukelberger}}, \bibinfo {author} {\bibfnamefont
  {E.}~\bibnamefont {Gull}}, \bibinfo {author} {\bibfnamefont {S.}~\bibnamefont
  {Guertler}}, \bibinfo {author} {\bibfnamefont {A.}~\bibnamefont {Hehn}},
  \bibinfo {author} {\bibfnamefont {R.}~\bibnamefont {Igarashi}}, \bibinfo
  {author} {\bibfnamefont {S.~V.}\ \bibnamefont {Isakov}}, \bibinfo {author}
  {\bibfnamefont {D.}~\bibnamefont {Koop}}, \bibinfo {author} {\bibfnamefont
  {P.~N.}\ \bibnamefont {Ma}}, \bibinfo {author} {\bibfnamefont
  {P.}~\bibnamefont {Mates}}, \bibinfo {author} {\bibfnamefont
  {H.}~\bibnamefont {Matsuo}}, \bibinfo {author} {\bibfnamefont
  {O.}~\bibnamefont {Parcollet}}, \bibinfo {author} {\bibfnamefont
  {G.}~\bibnamefont {Pawłowski}}, \bibinfo {author} {\bibfnamefont {J.~D.}\
  \bibnamefont {Picon}}, \bibinfo {author} {\bibfnamefont {L.}~\bibnamefont
  {Pollet}}, \bibinfo {author} {\bibfnamefont {E.}~\bibnamefont {Santos}},
  \bibinfo {author} {\bibfnamefont {V.~W.}\ \bibnamefont {Scarola}}, \bibinfo
  {author} {\bibfnamefont {U.}~\bibnamefont {Schollwöck}}, \bibinfo {author}
  {\bibfnamefont {C.}~\bibnamefont {Silva}}, \bibinfo {author} {\bibfnamefont
  {B.}~\bibnamefont {Surer}}, \bibinfo {author} {\bibfnamefont
  {S.}~\bibnamefont {Todo}}, \bibinfo {author} {\bibfnamefont {S.}~\bibnamefont
  {Trebst}}, \bibinfo {author} {\bibfnamefont {M.}~\bibnamefont {Troyer}},
  \bibinfo {author} {\bibfnamefont {M.~L.}\ \bibnamefont {Wall}}, \bibinfo
  {author} {\bibfnamefont {P.}~\bibnamefont {Werner}}, \ and\ \bibinfo {author}
  {\bibfnamefont {S.}~\bibnamefont {Wessel}},\ }\href@noop {} {\bibfield
  {journal} {\bibinfo  {journal} {J. Stat. Mech.: Theory Exp.}\ }\textbf
  {\bibinfo {volume} {2011}},\ \bibinfo {pages} {P05001} (\bibinfo {year}
  {2011})}\BibitemShut {NoStop}%
\bibitem [{\citenamefont {Otsuki}(2013)}]{Otsuki13}%
  \BibitemOpen
  \bibfield  {author} {\bibinfo {author} {\bibfnamefont {J.}~\bibnamefont
  {Otsuki}},\ }\href {\doibase 10.1103/PhysRevB.87.125102} {\bibfield
  {journal} {\bibinfo  {journal} {Phys. Rev. B}\ }\textbf {\bibinfo {volume}
  {87}},\ \bibinfo {pages} {125102} (\bibinfo {year} {2013})}\BibitemShut
  {NoStop}%
\bibitem [{\citenamefont {Rohringer}\ \emph {et~al.}(2012)\citenamefont
  {Rohringer}, \citenamefont {Valli},\ and\ \citenamefont
  {Toschi}}]{Rohringer12}%
  \BibitemOpen
  \bibfield  {author} {\bibinfo {author} {\bibfnamefont {G.}~\bibnamefont
  {Rohringer}}, \bibinfo {author} {\bibfnamefont {A.}~\bibnamefont {Valli}}, \
  and\ \bibinfo {author} {\bibfnamefont {A.}~\bibnamefont {Toschi}},\ }\href
  {\doibase 10.1103/PhysRevB.86.125114} {\bibfield  {journal} {\bibinfo
  {journal} {Phys. Rev. B}\ }\textbf {\bibinfo {volume} {86}},\ \bibinfo
  {pages} {125114} (\bibinfo {year} {2012})}\BibitemShut {NoStop}%
\bibitem [{\citenamefont {Brown}(1994)}]{Brown94}%
  \BibitemOpen
  \bibfield  {author} {\bibinfo {author} {\bibfnamefont {L.}~\bibnamefont
  {Brown}},\ }\href@noop {} {\emph {\bibinfo {title} {Quantum Field Theory}}}\
  (\bibinfo  {publisher} {Cambridge University Press},\ \bibinfo {year}
  {1994})\BibitemShut {NoStop}%
\bibitem [{\citenamefont {Jarrell}\ and\ \citenamefont
  {Gubernatis}(1996)}]{Jarrell1996133}%
  \BibitemOpen
  \bibfield  {author} {\bibinfo {author} {\bibfnamefont {M.}~\bibnamefont
  {Jarrell}}\ and\ \bibinfo {author} {\bibfnamefont {J.}~\bibnamefont
  {Gubernatis}},\ }\href {\doibase
  http://dx.doi.org/10.1016/0370-1573(95)00074-7} {\bibfield  {journal}
  {\bibinfo  {journal} {Phys. Rep.}\ }\textbf {\bibinfo {volume} {269}},\
  \bibinfo {pages} {133 } (\bibinfo {year} {1996})}\BibitemShut {NoStop}%
\end{thebibliography}%

\end{document}